\documentclass[sn-aps,iicol]{sn-jnl}

\jyear{2022}%

\usepackage{combelow}
\usepackage{cancel}

\usepackage[utf8]{inputenc}
\usepackage{bm}
\usepackage{bbold}
\usepackage{mathtools}

\usepackage{graphicx}
\usepackage{subfigure}
\usepackage{scalerel,amssymb}
\DeclareMathAlphabet{\pazocal}{OMS}{zplm}{m}{n}

\usepackage{braket}
\usepackage{tensor}
\usepackage{slashed}
\usepackage{xcolor}
\usepackage{xr}

\definecolor{purple}{rgb}{0.8,0,0.6}

\def\lambada{\lambda} 

\def\AVW{{\mathfrak{a}}}

\usepackage{fancyhdr}
\usepackage{lastpage}

\newcommand{\beqn}{\begin{eqnarray}}
\newcommand{\eeqn}{\end{eqnarray}}
\newcommand{\beqs}{\begin{subequations}}
\newcommand{\eeqs}{\end{subequations}\\[-2mm]\noindent}

\begin{document}

\title{Vortical waves in a quantum fluid with vector, axial, and helical charges. I. Non-dissipative transport}

\author[1]{\fnm{Sergio} \sur{Morales-Tejera}}\email{sergio.morales@e-uvt.ro}

\author*[1]{\fnm{Victor E.} \sur{Ambru\cb{s}}}\email{victor.ambrus@e-uvt.ro}

\author*[1,2]{\fnm{Maxim N.} \sur{Chernodub}}\email{maxim.chernodub@univ-tours.fr}

\affil*[1]{
\orgdiv{Department of Physics}, 
\orgname{West University of Timi\cb{s}oara},
\orgaddress{\street{Bd.~Vasile P\^arvan 4}, \city{Timi\cb{s}oara}, \postcode{300223}, \country{Romania}}}

\affil[2]{
\orgdiv{Institut Denis Poisson, CNRS UMR 7013},
\orgname{Universit\'e de Tours}, 
\orgaddress{\city{Tours}, \postcode{37200}, \country{France}}}

\abstract{Due to the spin-orbit coupling, Dirac fermions, submerged in a thermal bath with finite macroscopic vorticity, exhibit a spin polarisation along the direction parallel to the vorticity vector $\bm{\Omega}$. Due to the symmetries of the Lagrangian for free massless Dirac particles, there are three independent and classically conserved currents corresponding to the vector, axial, and helical charges. The constitutive relations for the charge currents and the stress-energy tensor at thermal equilibrium, derived in the framework of quantum field theory at finite temperature, reveal vorticity-induced contributions that deviate from the perfect fluid form. In this paper, we consider the mode structure of the corresponding hydrodynamical theory and derive collective excitations associated with coherent fluctuations of all three charges. We show that the chirally imbalanced rotating fluid should possess non-reciprocal gapless waves that propagate with different velocities along and opposite to the vorticity vector. We also uncover a strictly unidirectional mode, which we call the Axial Vortical Wave, propagating in the background of the axial charge density. The emergence of this wave can be traced back to earlier studies of vortical chiral fluids in a hydrodynamic approach. We also point out an unexpected instability in the limit of degenerate matter and discuss possible solutions when helicity and axial charge non-conservation are taken into account.}


\maketitle

\section{Introduction}\label{sec:intro}

Chiral fluids possess a class of hydrodynamic gapless excitations that emerge due to a coherent interplay of appropriate channels in the transfer and accumulation of (approximately) conserved charges of the system~\cite{Kharzeev:2010gd, Jiang:2015cva, Newman:2005hd}. These hydrodynamic modes appear in the electromagnetic field background and curved (notably, vortical) spacetimes. They are distinguished from other excitations because their very existence is supported by the anomalous breaking of continuous internal symmetries in the system, such as the chiral anomaly or the mixed gauge-gravitational anomaly (for an extensive review, see Ref.~\cite{Kharzeev:2013jha}).

The Chiral Magnetic Wave~\cite{Kharzeev:2010gd} represents a neat example of an anomalous hydrodynamic mode. This excitation is supported by two transport phenomena, both arising due to the axial anomaly: the Chiral Magnetic Effect~\cite{Fukushima:2008xe, Vilenkin:1980fu, Alekseev:1998ds} and the Chiral Separation Effect~\cite{Son:2004tq, Metlitski:2005pr}. The first phenomenon generates a vector current in a region with a nonvanishing chiral charge density. This current, directed along the axis of the background magnetic field, leads to a vector charge accumulation, which serves, in turn, as a source for the chiral current generated now by the second effect. The chiral current transports the chiral charge further along the magnetic field axis, and the process repeats cyclically again, producing the Chiral Magnetic Wave excitation in the vector and axial charge densities of chiral fermions. 

The Chiral Magnetic Wave was hypothesized to emerge in the quark-gluon plasma that possibly has certain experimental signatures~\cite{ALICE:2023weh}. A similar excitation also appears in the vortical backgrounds via chiral vortical effects~\cite{Vilenkin:1978is, Vilenkin:1979ui}, which produce intertwining fluctuations of vector and axial currents and their charges along the axis of rotation~\cite{Jiang:2015cva}. 

In the standard picture described above, chiral fluids are traditionally defined as two-component systems that incorporate vector and axial charges and their currents. However, in addition to these local quantities, ensembles of massless fermions also possess a third class of charges and currents associated with the helicity of massless fermions (see a detailed pedagogical discussion in Ref.~\cite{Pal:2010ih} along with a more recent account in~\cite{Ambrus:2019ayb, Ambrus:2019khr}). In a generic ensemble of fermions, helicity enters as an independent local characteristic of the ensemble, which is distinct from the vector (electric) and axial (chiral) charges. 

Our paper intensively explores the role of the helical degrees of freedom in the hydrodynamic context of vortical chiral fluids. The spectrum of the hydrodynamic modes in rotating fluids contains a multitude of hydrodynamical gapless excitations, even if the helical degree of freedom is not accounted for \cite{Kharzeev:2010gd, Newman:2005hd, Gahramanov:2012wz, Abbasi:2015saa, Yamamoto:2015ria, Jiang:2015cva, Chernodub:2015gxa, Abbasi:2016rds, Kalaydzhyan:2016dyr, Abbasi:2017tea, Gorbar:2017toh}.
The main aim of our work is to identify the physical consequences of the correct incorporation of the helicity property, which must be present in a system possessing more than one chiral fermion, for the hydrodynamics of these fluids. In this paper, we work in the dissipationless limit, when both the axial and helical quantum numbers are strictly conserved.

Helicity is often confusingly identified with chirality, even though these two properties reflect separate physical features of fermions. Its precise nature is highlighted by the simple observation that the chirality property equally applies to particles and anti-particles, while the helicity distinguishes between them. The critical distinction, often misunderstood, lies in the fact that a fermion's chirality matches its helicity, whereas an anti-fermion's chirality is opposite to its helicity. This difference leads to an incorrect generalization, ``chirality equals helicity,'' which does not hold in many-body particle-antiparticle systems. For instance, a fermion--anti-fermion pair can have either combined chirality and zero helicity (if their chiralities align) or zero total chirality but non-zero helicity (if their chiralities are opposite). This striking dissimilarity between chirality and helicity can also be emphasized by the difference in their charge conjugation symmetries: while the chiral charge density is a $C$-even quantity, its helical counterpart has a $C$-odd symmetry with respect to the charge conjugation~\cite{Ambrus:2019ayb, Ambrus:2019khr}. 

The distinct nature of the vector, chiral, and helical quantum numbers is highlighted by the following relation between the charges of an individual massless (anti-)fermion: 
\begin{align}
 q_V q_A &= q_H, & q_A q_H &= q_V, \nonumber\\
 q_H q_V &= q_A, & q_V q_A q_H &= 1,
 \label{eq:charge_products}
\end{align}
where the vector charge $q_V$ discriminates a particle $(q_V = + 1)$ from an anti-particle $(q_V = - 1)$, the helical charge $q_H$ labels the relative orientation of the momentum and spin directions, distinguishing parallel vectors $(q_H = +1)$ and antiparallel vectors $(q_H = -1)$, while the last entry in the triad, the chiral charge $q_A$, shows that the chirality for a particle (anti-particle) coincides with (is opposite to) its helicity $q_H$. The relation~\eqref{eq:charge_products} links tightly the helicity and chirality for a single massless (anti-)particle. However, it is not generally valid for a system of massless fermions, therefore making the total helicity charge of an ensemble disconnected from the total chiral charge of the same ensemble. Thus, the helicity becomes an independent degree of freedom of the system, distinct from chirality. The similarities and differences between chiral and helical degrees of freedom, already emphasised by Eq.~\eqref{eq:charge_products}, are illustrated in detail in Fig.~\ref{fig:CHVE}, discussed in Sec.~\ref{sec:picture}, where the origin of Chiral and Helical Vortical Effects is demonstrated in a simple pictorial form based on a straightforward counting of the vector, axial, and helical degrees of freedom.

In addition, one may argue that the helical degree of freedom is as good (bad) as the chiral charge of massive fermions: neither of them is conserved in physically relevant theories such as massive QED or QCD. In the relaxation time approximation, the pertinence of these quantum numbers for the dynamics of the system is determined by the relative order of magnitude of the axial, $\tau_A$, and helical, $\tau_H$, relaxation times, which show how fast these charges dissolve in the system.  Similarly to the axial (chiral) charge~\cite{Ruggieri:2016asg, Astrakhantsev:2019zkr}, the relaxation of the helical charge might also be a rather fast process~\cite{Ambrus:2019khr}. In addition to the charge relaxation times, the system is also characterized by the kinetic relaxation time $\tau_R$, which encodes how fast fluctuations in the thermodynamic characteristics of the system (such as pressure and energy density) relax towards the thermodynamic equilibrium. In general, all three relaxation times, $\tau_R$, $\tau_A$, and $\tau_H$, are independent of each other. 
We defer the discussion of the consequences of these dissipative effects to the paper representing the second part of this work \cite{Morales-Tejera:2024mtx}. 
In this first part, we will focus on exploring the wide spectrum of coherent excitations that develop at the level of the vector, axial, and helicity currents at both large and small temperatures, with or without background axial or helical imbalance.

Note that there exist other conserved currents that can serve as a measure of polarization, e.g., the associated spin operator studied in \cite{Cotaescu:2022ntp}. Such current could be used in place of helicity to describe the polarisation of the system. The study of the effect of imbalance in the corresponding spin charge requires knowledge of the associated vortical effects, which, to the best of our knowledge, have not been studied so far.

We also mention that in the free theory, there is an infinite number of conserved operators. 
Take, for example, the number operator $\widehat{N}_j = \hat{a}^\dagger_j \hat{a}_j$ associated to particles in mode $j$. This operator commutes with the free Hamiltonian, $\widehat{H} = \sum_j E_j (\hat{a}^\dagger_j \hat{a}_j + \text{c.c.})$ and its expectation value describes the number of particles of type $j$ in a given state. In a realistic system, deviations from the thermodynamic ensemble average will rapidly dissipate, as all interactions involving particles in mode $j$ (e.g., $2$-particle interactions) will lead to changes in the population number of this mode. It is, therefore, unreasonable to associate a chemical potential to model the imbalance in a single mode $j$ within the grand canonical ensemble. The situation for the helicity operator is significantly different, as only a subclass of interactions modifies the helicity charge, namely the helicity-violating pair annihilation (HVPA) processes discussed in Ref.~\cite{Ambrus:2019khr}. A leading-order estimate reveals that the relaxation time associated with such processes can be comparable to the lifetime of the system. We discuss the effects of the dissipation of helicity charge in the companion paper, Ref.~\cite{Morales-Tejera:2024mtx}.

In this paper, we concentrate on helicity-catalyzed waves that emerge in the rotating, also called vortical, fluids. While certain properties of gapless hydrodynamic excitations in the magnetic background field were already analyzed in Ref.~\cite{Ambrus:2023erf}, the vortical background represents a more attractive perspective from the point of view of the experimental verification of these effects in relativistic heavy-ion collisions. Here, the vorticity refers to the local angular momentum or rotational motion of the quark-gluon plasma (QGP), which is generated in a highly vortical state due to the angular momentum conservation and the initial asymmetry in the non-central collision geometry~\cite{Huang:2020dtn}. 

While both vorticity and magnetic fields are generated almost simultaneously during the collision, the key difference between them is their lifetime. Magnetic field decays very rapidly as the expanding plasma cannot maintain it due to relatively low electrical conductivity~\cite{Bloczynski2013, Tuchin2016, Grsoy2018}. In contrast, the vorticity, once established in the QGP, is sustained due to the mechanical conservation of the angular momentum~\cite{Huang:2020dtn}. The emergence of vorticity can be inferred from the spin polarization of emitted particles~\cite{Florkowski:2018fap, Becattini:2021lfq, Becattini:2022zvf}, which is a measure of the alignment of their spins with the (local or global) angular momentum of the QGP~\cite{STAR:2007ccu, STAR:2017ckg}.

The hydrodynamic wave excitations that emerge in vortical plasmas of chiral fermions propagate along the vorticity vector in the plasma. They can leave experimental signatures that resemble the Chiral Magnetic Wave, supposedly imprinted in other observables. Below, we concentrate on the theoretical questions related to the very existence of these modes, their spectrum, and their lifetimes. The discussion of their experimental signatures will be presented elsewhere.

In our paper, we stress that the incorporation of the helical charges is important for the hydrodynamics of the system. In particular, in a system of rotating Dirac fermions at finite temperature and density, the accounting of the helical degree of freedom drastically affects the hydrodynamic spectrum. For example, a new type of excitations, suggested under a collective name ``the helical vortical effects,'' emerges as hydrodynamic modes that intertwine coherent fluctuations of vector, axial, and helical charges~\cite{Ambrus:2019khr}. At the typical parameters of the quark-gluon plasma, the helical modes show a rather profound difference compared to chiral modes, implying, in particular, that the Helical Vortical Wave propagates much faster than the Chiral Vortical Wave~\cite{Ambrus:2019khr}. Notice that the Chiral and Helical Vortical Waves and their various generalizations, considered in this paper, should be distinguished from the waves obtained in the absence of helicity, described in \cite{Gorbar:2017toh}. Similarly, they should not be confused with the chiral density wave~\cite{Nakano:2004cd} and its dual analogue~\cite{Tatsumi:2004dx} (also, in rotating matter~\cite{TabatabaeeMehr:2023tpt, Ghalati:2023npr}) that emerge in the dense matter and are characterized by a static coherent state of spatially varying scalar and pseudoscalar condensates.

We will also show below that in the presence of a non-vanishing axial (chiral) chemical potential, the system of massless fermions hosts the Axial Vortical Wave (AVW). On a background with positive (negative) chiral imbalance, this rather unusual excitation propagates only opposite to (along) the vector of vorticity and not forward (backwards). Notice that the existence of the AVW does not rely on helical degrees of freedom, as this excitation can also propagate in the purely axial sector.
The non-reciprocity of wave propagation is a remarkable and relatively rare phenomenon that needs a particular set of conditions. For example, in the condensed matter setting, one can find a unidirectional propagation of phonons associated with longitudinal vibrations of a twisted crystal lattice of Weyl semimetals that exhibit the axial anomaly~\cite{Chernodub:2019lhw}. The phonons propagate only along the direction of the twist and not backwards. This excitation, dubbed the chiral sound wave, resembles the Axial Vortical Wave because the twist of the crystal can be associated with the vorticity of the underlying ion lattice.

The appearance of the Axial Vortical Wave can be traced back to the very detailed pioneering study of Ref.~\cite{Gorbar:2017toh}, where a linear mode has been obtained in a vortical background of chiral fermions characterized by finite vector and axial densities at finite temperature. Albeit the velocity of the wave of Ref.~\cite{Gorbar:2017toh} has been obtained in an implicit form as a solution of a simple algebraic equation, we will show in our article that it corresponds to the Axial Vortical Wave in the high-temperature limit. We will also show below that this wave has a uni-directional nature that leads to a nonreciprocity of other hydrodynamic modes that are coupled with this wave.

In order to give a brief overview and straightforward description of our findings, here we summarize them in a very short way:

\begin{enumerate}

\item We generalize the existing spectrum of collective excitations in the pair of vector and axial quantum numbers~\cite{Kharzeev:2010gd, Newman:2005hd, Gahramanov:2012wz, Abbasi:2015saa, Yamamoto:2015ria, Jiang:2015cva, Chernodub:2015gxa, Abbasi:2016rds, Kalaydzhyan:2016dyr, Abbasi:2017tea} in a chiral fluid by including fluctuations of the helicity degree of freedom and by taking into account arbitrary vector, axial and helical chemical potentials.

\item We provide an explicit equation for the velocity of the helical vortical wave in an unpolarized ($\mu_A=\mu_H=0$) plasma at arbitrary vector chemical potential.

\item We observe that the coupling between axial imbalance ($\mu_A$) and vorticity ($\Omega$) induces non-reciprocal propagation both for the Helical Vortical Wave and the Axial Vortical Wave. At certain values of the chemical potentials, the propagation becomes uni-directional. We also discuss the emergence of unidirectional propagation from the perspective of discrete symmetries.

\item We discuss that the correct result for the dispersion relation of the sound modes along the vorticity direction involves nontrivial transverse-plane dynamics.

\item We verify that the results obtained in the high-density limit, where the axial and helical degrees of freedom are strongly correlated, reproduce the previous results \cite{Gorbar:2017toh} that do not include fluctuations of helicity.

\end{enumerate}

The structure of the paper is as follows. In Sec.~\ref{sec:2}, we describe the rotating state in thermal equilibrium.  We then formulate the equations obeyed by the fluctuations around this state in the Landau frame in the vicinity of the rotation axis. In our setup, the fluctuations of the charge sector decouple from the energy-momentum sector in the Landau frame. 

As already mentioned above, in this work, we analyze the wave spectrum arising in a rotating V/A/H fluid when all three charges are conserved. 
In Sec.~\ref{sec:largeT}, we discuss the wave spectrum in the large temperature limit. The emergence of the Helical Vortical Wave and the strictly unidirectional Axial Vortical Wave is demonstrated in Subsections~\ref{sec:largeT:HVW} and \ref{sec:largeT:AVW}.
We discuss the case of an unpolarized plasma in Subsec.~\ref{sec:unpolarized}, where the axial and the helical chemical potentials vanish, which can be treated exactly to a high degree.
In Sec.~\ref{sec:largemu}, we discuss the so-called degenerate limit when the vector chemical potential $\mu_V$ is larger than the temperature and the other chemical potentials. This case ---which can be dubbed as a weakly polarized plasma--- corresponds to a realistic high-density matter state ($\lvert \mu_V \rvert \gg T$). At the same time, the smallness of the other chemical potentials ($\mu_A$ and $\mu_H$) emulates the non-conservation of the corresponding charges.

\section{Anomalous transport in vortical fluids}\label{sec:2}

We start this section with a brief review of the vortical effects involving the thermal expectation values of the energy-momentum tensor and of the vector, axial and helicity currents, summarized in Subsec.~\ref{sec:2:quantum}. We then discuss the transition from the natural thermometer ($\beta$) frame to the Landau frame in the limit of slow rotation in Subsec.~\ref{sec:2:Landau}. 
Then, we consider the conservation properties of the vortical/axial/helical ($V/A/H$, for shortness) currents considered in this paper. We leave a discussion of the mechanisms leading to their non-conservation to the companion paper \cite{Morales-Tejera:2024mtx}. Finally, we present the general framework for the analysis of vortical waves in Subsec.~\ref{sec:2:eqs}.

\subsection{Quantum vortical effects}\label{sec:2:quantum}

Due to spin-orbit coupling, a fermionic plasma under rotation develops a spin polarization current along the direction of the local vorticity vector. More specifically, let us consider free massless fermions, described by the Dirac Lagrangian $\mathcal{L} = i \bar{\psi} \slashed{\partial} \psi$. These fermions may be characterized by their vector charge, distinguishing between particles and anti-particles ($\sigma = \pm 1$), axial charge ($\xi = \pm 1$), 
and helical charge ($\lambda = \pm 1/2$). Rotating states may be described in the grand canonical ensemble using the density operator
\begin{equation}
 \hat{\rho} = \exp\left[-\beta(\widehat{H} - \Omega \widehat{J}^z - \vec{\mu} \cdot {\widehat{\vec Q}})\right],
 \label{eq:rho}
\end{equation}
where $\widehat{H}$ and $\widehat{J}^z$ are the system's Hamiltonian and $z$-axis total angular momentum, while $\widehat{\vec{Q}} = (\widehat{Q}_V, \widehat{Q}_A, \widehat{Q}_H)$ collectively denotes the vector, axial and helical charge operators, respectively. The Lagrange multipliers $\beta$, $\Omega$ and $\vec{\mu} = (\mu_V, \mu_A, \mu_H)$ corresponding to these (conserved) operators characterize the system's inverse temperature, angular velocity, and vector/axial/helical chemical potentials, respectively. The statistical operator \eqref{eq:rho} defines a preferred frame, called the thermometer (or $\beta$) frame, characterized by the four-velocity corresponding to rigid rotation:
\begin{equation}
 u^\mu_\Omega \partial_\mu = \Gamma_\Omega(\partial_t + \Omega \partial_\varphi), 
 \quad 
 \Gamma_\Omega = \frac{1}{\sqrt{1- \rho^2 \Omega^2}},
 \label{eq:u_RR}
\end{equation}
where $(t,\rho,\varphi,z)$ represent cylindrical coordinates and $\rho$ is the transverse-plane distance to the rotation axis $z$. 
The velocity profile in Eq.~\eqref{eq:u_RR} allows one to introduce a so-called kinematic tetrad, comprised of the four-velocity $u^\mu$, the kinematic vorticity vector $\omega^\mu = \frac{1}{2} \varepsilon^{\mu\nu\lambda\sigma} u_\nu \partial_\lambda u_\sigma$, acceleration vector $a^\mu = u^\nu \partial_\nu u^\mu$, as well as a fourth vector $\tau^\mu = -\varepsilon^{\mu\nu\lambda\sigma} \omega_\nu a_\lambda u_\sigma$. In the case of rigid rotation, these four vectors are mutually orthogonal. Explicitly, $u^\mu$ is given in Eq.~\eqref{eq:u_RR}, while the other three vectors are given as follows:
\begin{align}
 \omega^\mu_\Omega \partial_\mu & = \Omega \Gamma_\Omega^2 \partial_z, \nonumber\\
 a^\mu_\Omega \partial_\mu & = -\rho \Omega^2 \Gamma_\Omega^2 \partial_\rho, \\
 \tau_\Omega^\mu \partial_\mu & = -\rho \Omega^3 \Gamma_\Omega^5(\rho \Omega \partial_t + \rho^{-1} \partial_\varphi)\,.\nonumber
\end{align}

In the classical Dirac theory, the energy-momentum tensor $T^{\mu\nu}$ reads 
\begin{subequations}
\begin{equation}
 T^{\mu\nu} = \frac{i}{2} \bar{\psi}\gamma^{(\mu} \overleftrightarrow{\partial}^{\nu)} \psi,
\end{equation}
and the charge currents $J^{\mu}_{V/A/H}$ read
\begin{align}
 J^\mu_V & = \bar{\psi} \gamma^\mu \psi, \qquad
 J^\mu_A  = \bar{\psi} \gamma^\mu \gamma^5 \psi, \\
 J^\mu_H & = \bar{\psi} \gamma^\mu h \psi + \overline{h \psi} \gamma^\mu \psi\,.
 \nonumber
\end{align}
\end{subequations}
In thermal field theory, one may compute the thermal expectation value of an operator $\widehat{A}$ via $A = \langle \widehat{A} \rangle \equiv Z^{-1} {\rm tr}(\hat{\rho} \widehat{A})$, where $Z = {\rm tr}(\hat{\rho})$ is the partition function. The expectation values $T^{\mu\nu} \equiv \langle \widehat{T}^{\mu\nu} \rangle$ and $J^\mu_\ell \equiv \langle \widehat{J}^\mu_\ell \rangle$ of, respectively, the energy-momentum tensor and the charge currents can be obtained in quantum field theory \cite{Becattini:2014yxa,Becattini:2015nva,Ambrus:2019cvr}. As is customary in relativistic fluid dynamics, one can decompose these quantities uniquely after defining a so-called hydrodynamic frame that fixes the fluid's four-velocity. In the $\beta$ frame, the four-velocity is given by $u^\mu_\Omega$ and
\begin{align}
 J^\mu_\ell &= Q_{\ell;\beta} u_\Omega^\mu + V^\mu_{\ell;\beta}, \nonumber \\
 T^{\mu\nu} &= E_\beta u_\Omega^\mu u_\Omega^\nu - (P_\beta + \varpi_\beta) \Delta_\Omega^{\mu\nu} + 
 \pi_\beta^{\mu\nu}\nonumber\\ & + W_\beta^\mu u_\Omega^\nu + W_\beta^\nu u_\Omega^\mu,
 \label{eq:macro_RR}
\end{align}
where the subscript $\beta$ reminds us that all quantities appearing on the right-hand sides of the above relations are computed in the $\beta$ frame.
The charge densities $Q_{\ell;\beta}$ ($\ell \in \{V, A, H\}$), energy density $E_\beta$ and thermodynamic pressure $P_\beta$ are terms characteristic of a perfect fluid. 

In Eq.~\eqref{eq:macro_RR}, the tensor $\Delta_\Omega^{\mu\nu} = g^{\mu\nu} - u_\Omega^\mu u_\Omega^\nu$ is a projector on the hypersurface orthogonal to $u_\Omega^\mu$, while the deviations from the perfect fluid form, namely $\varpi_\beta$, $V^\mu_{\ell;\beta}$, $W_\beta^\mu$ and $\pi_\beta^{\mu\nu}$, are by construction traceless and orthogonal to $u_\Omega^\mu$. These terms arise as quantum corrections induced by the local acceleration and vorticity of the fluid. For a conformal fluid, the equation of state $E_\beta = 3P_\beta$ and the tracelessness condition $T^\mu{}_\mu = 0$ imply that the dynamic pressure vanishes, $\varpi_\beta = 0$. The constitutive equations for $V_{\ell;\beta}^\mu$, $W_\beta^\mu$ and $\pi_\beta^{\mu\nu}$ in the $\beta$ frame are given by the vortical effects derived in Ref.~\cite{Ambrus:2019khr}:
\begin{gather}
 V^\mu_{\ell;\beta} = \sigma^\omega_{\ell; \beta} \omega^\mu_\Omega + \sigma^\tau_{\beta;\ell} \tau^\mu_\Omega, \nonumber\\
 W_\beta^\mu = \sigma^\omega_{\varepsilon;\beta} \omega_\Omega^\mu + \sigma^\tau_{\varepsilon;\beta} \tau_\Omega^\mu,\nonumber\\
 \pi_\beta^{\mu\nu} = \pi_{1;\beta}  \left(\tau_\Omega^\mu \tau_\Omega^\nu - \frac{\boldsymbol{\omega}_\Omega^2}{2} a_\Omega^\mu a_\Omega^\nu - \frac{\mathbf{a}_\Omega^2}{2} \omega_\Omega^\mu \omega_\Omega^\nu \right) \nonumber\\ + \pi_{2;\beta} (\omega_\Omega^\mu \tau_\Omega^\nu + \omega_\Omega^\nu \tau_\Omega^\mu),
\end{gather}
where $\sigma^\omega_{\ell;\beta} = (\sigma^\omega_{V;\beta}, \sigma^\omega_{A;\beta}, \sigma^\omega_{H;\beta})$ represent the vector/axial/helical vortical conductivities and $\sigma^\omega_{\varepsilon;\beta}$ is the vortical heat conductivity. Likewise, $\sigma^\tau_{\ell;\beta}$ and $\sigma^\tau_{\varepsilon;\beta}$ represent circular conductivities. Finally, $\pi_{1;\beta}$ and $\pi_{2;\beta}$ represent shear-stress coefficients.

All of the above scalar functions can be calculated in quantum field theory as the sum of a classical term followed by quantum corrections, with the latter being at least quadratic in $\hbar \Omega$. As pointed out in Ref.~\cite{Ambrus:2019khr}, the classical pressure can be calculated on the basis of an ensemble of polarized fermions described by the Fermi-Dirac distribution,
\begin{equation}
 f^{{\rm eq};\sigma}_{\mathbf{p},\lambda} = \left[ 
 \exp\left(\frac{p \cdot u - \mu_{\sigma,\lambda}}{T}\right) 
 + 1 \right]^{-1},
 \label{eq:feq}
\end{equation}
where $p^\mu = (p^0, \mathbf{p})$ is the particle four-momentum ($p^0 = \lvert\mathbf{p}\rvert$ for massless fermions), $\sigma = \pm 1$ distinguishes between particles and anti-particles, while $\lambda = \pm 1/2$ labels the particle helicity. The chemical potential $\mu_{\sigma,\lambda}$ arises due to the vector, axial, and helical imbalances in the system, having the expression
\begin{equation}
    \mu_{\sigma,\lambda} = {\vec q}_{\sigma,\lambda} \cdot {\vec \mu}
    = \sigma \mu_V + 2\lambda \mu_A + 2 \sigma \lambda \mu_H,
\end{equation}
where ${\vec q}_{\sigma,\lambda} = \{\sigma,2\lambda,2 \sigma \lambda\}$ collects the vector, axial and helical charges. In Eq.~\eqref{eq:feq}, $T$ and $\mu_{\sigma,\lambda}$ represent the local temperature and a set of chemical potentials, respectively. 
For example, for the rigidly-rotating system, when $u^\mu \rightarrow u^\mu_\Omega$, the temperature $T$ and chemical potentials $\mu_{\sigma,\lambda}$ are given by the Tolman-Ehrenfest law,
\begin{equation}
 T = \Gamma_\Omega T_0, \qquad \mu_{\sigma,\lambda} = \Gamma_\Omega \mu_{\sigma,\lambda}^0,
\end{equation}
with $(T_0, \mu^0_{\sigma,\lambda})$ being the values of the temperature and chemical potentials on the rotation axis.

The above kinetic theory framework allows the classical contribution to the thermodynamic pressure of the system to be computed as
\begin{align}
 P_{\beta;{\rm cl}} &= \frac{1}{3} \sum_{\sigma,\lambda} \int dP \, E_\mathbf{p}^2 f^{{\rm eq};\sigma}_{\mathbf{p}, \lambda} \nonumber\\&= -\frac{T^4}{\pi^2} \sum_{\sigma,\lambda} {\rm Li}_4(-e^{\mu_{\sigma,\lambda} / T}),
 \label{eq:Pcl}
\end{align}
where $dP = d^3p / [(2\pi)^3 p^0]$ is the Lorentz-invariant momentum space integration measure, $E_{\mathbf{p}} {=} p \cdot u$ is the particle energy in the fluid rest frame and ${\rm Li}_s(z) = \sum_{n = 1}^\infty z^n/n^s$ is the polylogarithm. All other quantities can be computed from $P_{\beta; {\rm cl}}$ via thermodynamic derivatives, see Eq.~(70) in Ref.~\cite{Ambrus:2019khr}:
\begin{align}
 Q_{\ell;\beta; {\rm cl}} & = \frac{\partial P_{\beta;{\rm cl}}}{\partial \mu_\ell}, \qquad 
 \sigma^\omega_{\ell;\beta; {\rm cl}}  = \frac{1}{2} \frac{\partial^2 P_{\beta;{\rm cl}}}{\partial \mu_A \partial \mu_\ell},
 \label{eq:thermocl}\\
 \sigma^\tau_{\ell;\beta;{\rm cl}} &= \frac{1}{12} \frac{\partial^3 P_{\beta;{\rm cl}}}{\partial^2\mu_A \partial \mu_\ell},
 \nonumber
\end{align}
while $\sigma^\omega_{\varepsilon;{\rm cl}} = Q_{A;{\rm cl}}$.
Finally, $\pi_{1;\beta;{\rm cl}} = -2\sigma^\pi_{A;\beta;{\rm cl}} / 27$ and $\pi_{2;\beta;{\rm cl}} = -2 \sigma^\tau_{A;\beta;{\rm cl}}$ [see Eq.~(94) in Ref.~\cite{Ambrus:2019khr}], where $\sigma^\pi_{\ell;\beta;{\rm cl}} = \frac{1}{2} \partial^4P_{\beta;{\rm cl}} / \partial^3\mu_A \partial\mu_\ell$ [cf.~Eq.~(70) in Ref.~\cite{Ambrus:2019khr}].

\subsection{The Landau frame}\label{sec:2:Landau}

In the case when the rotation is slow, one may neglect quadratic terms in $\Omega$ for a region close to the rotation axis. Since the vector $\tau^\mu$ is of third order in $\Omega$, it can be neglected, such that $J^\mu_\ell$ and $T^{\mu\nu}$ simplify to
\begin{align}\label{eq:JT-beta}
 J^\mu_\ell \simeq& Q_{\ell;\beta} u^\mu_\Omega + \sigma^\omega_{\ell;\beta} \omega^\mu_\Omega, \nonumber\\
 T^{\mu\nu} \simeq& (E_\beta+P_\beta) u_\Omega^\mu u_\Omega^\nu - P_\beta g^{\mu\nu} \nonumber\\ &+ \sigma^\omega_{\varepsilon;\beta} (\omega_\Omega^\mu u_\Omega^\nu + u_\Omega^\mu \omega_\Omega^\nu),
\end{align}
where we used the subscript $\beta$ to indicate that the above decomposition is performed with respect to the $\beta$-frame four-velocity, $u^\mu_\Omega$. Neglecting quadratic terms in $\Omega$ also implies that the quantum corrections to the above expressions can be ignored and, e.g., $P_\beta$ can be safely replaced by $P_{\beta;{\rm cl}}$ in Eq.~\eqref{eq:Pcl}. For notational convenience, we shall drop the ``${\rm cl}$'' subscript from all thermodynamic quantities in what follows.

We now seek to reexpress $J^\mu_\ell$ and $T^{\mu\nu}$ with respect to the Landau frame \cite{Abbasi:2016rds}. The latter frame is characterized by the four-velocity $u^\mu_L$ satisfying the eigenvalue equation $T^\mu{}_\nu u^\nu_L = E_L u^\mu_L$, where $E_L$ represents the energy density of the fluid in the Landau frame. A particular feature of the Landau frame is that the heat-flux type of term is completely removed from $T^{\mu\nu}$, giving rise to the decomposition
\begin{align}
 &J^\mu_\ell = Q_{\ell;L} u_L^\mu + V^\mu_{\ell;L}, \nonumber\\&
 T^{\mu\nu} = (E_L + P_L) u_L^\mu u_L^\nu - P_L g^{\mu\nu} + \pi^{\mu\nu}_L.
 \label{eq_Landau_JT}
\end{align}
Working under the assumption of small $\Omega$, we can determine the Landau four-velocity up to first order with respect to $\Omega$ as 
\begin{equation}\label{eq:landau_fourvel}
 u^\mu_L = u^\mu_\Omega + 
 \frac{\sigma^\omega_{\varepsilon;\beta}}{E + P} \omega_\Omega^\mu.
\end{equation}
To linear order in $\Omega$, the energy density, pressure and charge densities remain unchanged, $E_L = E_\beta \equiv E$, $P_L = P_\beta \equiv P$, and $Q_{L;\ell} = Q_{\beta;\ell} \equiv Q_\ell$, while the shear-stress tensor $\pi^{\mu\nu}_L$ is quadratic with respect to $\Omega$ and will, therefore, be disregarded. For notational convenience, we shall remove the $L$ subscript for Landau-frame quantities. The vortical conductivities now absorb the vortical heat conductivity, as follows:
\begin{equation}
 \sigma^\omega_{\ell} = \sigma^\omega_{\ell;\beta} - \frac{\sigma^\omega_{\varepsilon;\beta} Q_{\ell}}{E + P}.
 \label{eq:Landau_sigma}
\end{equation}

 Since we chose to ignore terms that are quadratic with respect to the angular frequency $\Omega$, it is possible to employ one further simplification. We consider the Lorentz boost to the frame where the $z$ component of the four-velocity vector $u^\mu_L$ vanishes:
 \begin{equation}\label{eq_Lorentz}
  L^{\mu\nu} = g^{\mu\nu} + \frac{\sigma^\omega_{\varepsilon;\beta}}{E + P} (u^\mu_\Omega \omega^\nu_\Omega - u^\nu_\Omega \omega^\mu_\Omega).
 \end{equation}
One can easily check that $u^\mu_L \rightarrow L^\mu{}_\nu u^\nu_L = u^\mu_\Omega + O(\Omega^2) \simeq u^\mu_\Omega$. To linear order, the vorticity vector $\omega^\mu_\Omega$ remains unchanged. Thus, in this boosted Landau frame, we have:
 \begin{align}
  &T^{\mu\nu} = (E + P) u^\mu u^\nu - P g^{\mu\nu}, \nonumber\\& 
  J^\mu_\ell = Q_\ell u^\mu + \sigma^\omega_\ell \omega^\mu,
  \label{eq:boosted_dec}
 \end{align}
 where we dropped both the $L$ and the $\Omega$ labels for notational brevity. 
All quantities appearing above are given by the classical expressions in Eqs.~\eqref{eq:Pcl}--\eqref{eq:thermocl}, i.e. 
\begin{align}
 &P = \frac{E}{3} \simeq -\frac{T^4}{\pi^2} \sum_{\sigma,\lambda} {\rm Li}_4(-e^{\mu_{\sigma,\lambda} / T}), \nonumber\\&
 Q_\ell \simeq \frac{\partial P}{\partial \mu_\ell}, \qquad
 \sigma^\omega_{\ell} \simeq 
 \frac{1}{2} \frac{\partial^2 P}{\partial \mu_\ell \partial\mu_A} - 
 \frac{Q_\ell Q_A}{E+P}.
 \label{eq:Landau_cond}
\end{align}

\subsection{Vector/axial/helical anomalous vortical transport as spin-orbit coupling}
\label{sec:picture}

\begin{figure*}
\centering 
\begin{tabular}{l}
    \includegraphics[width=0.85\linewidth]{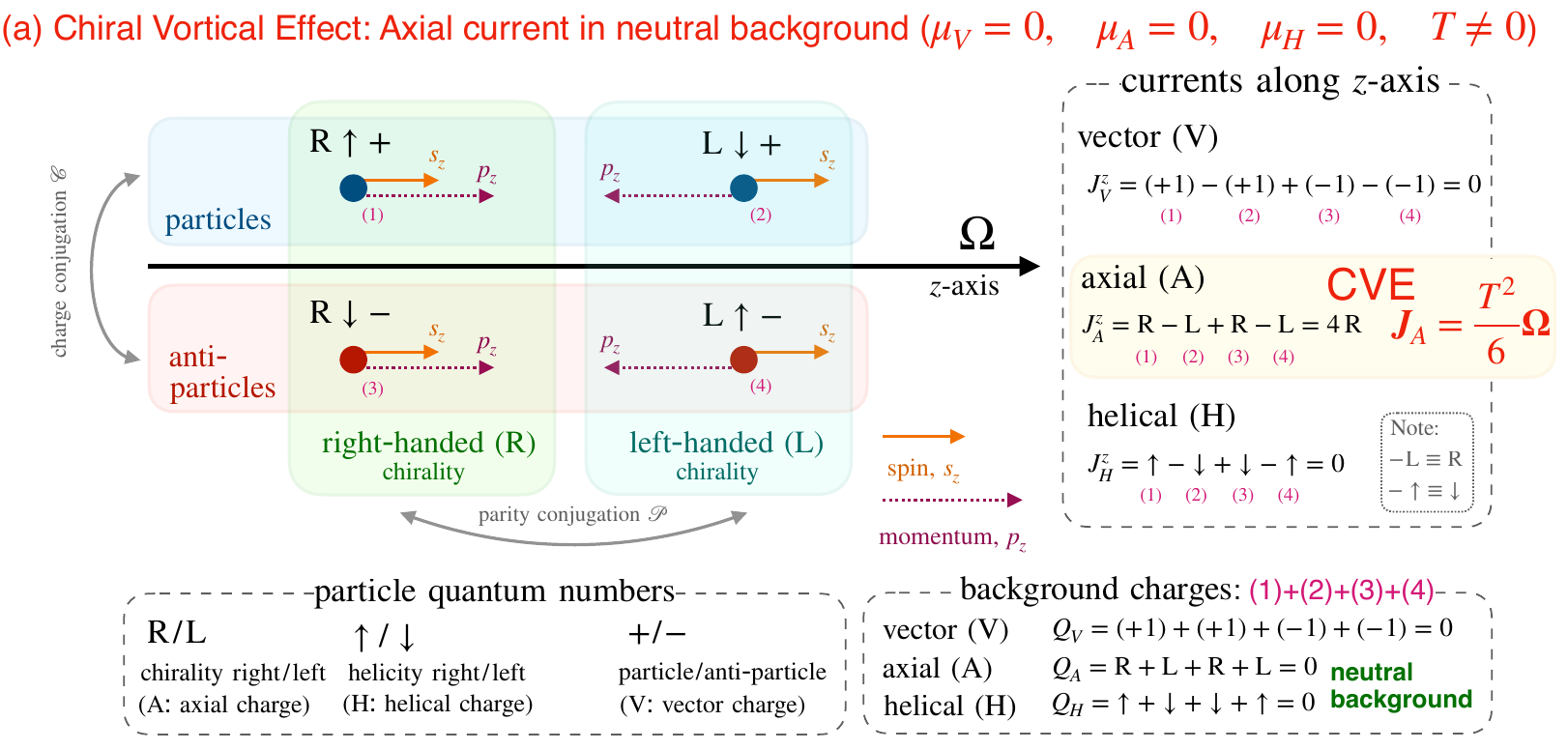} \\[5mm]
    \includegraphics[width=0.85\linewidth]{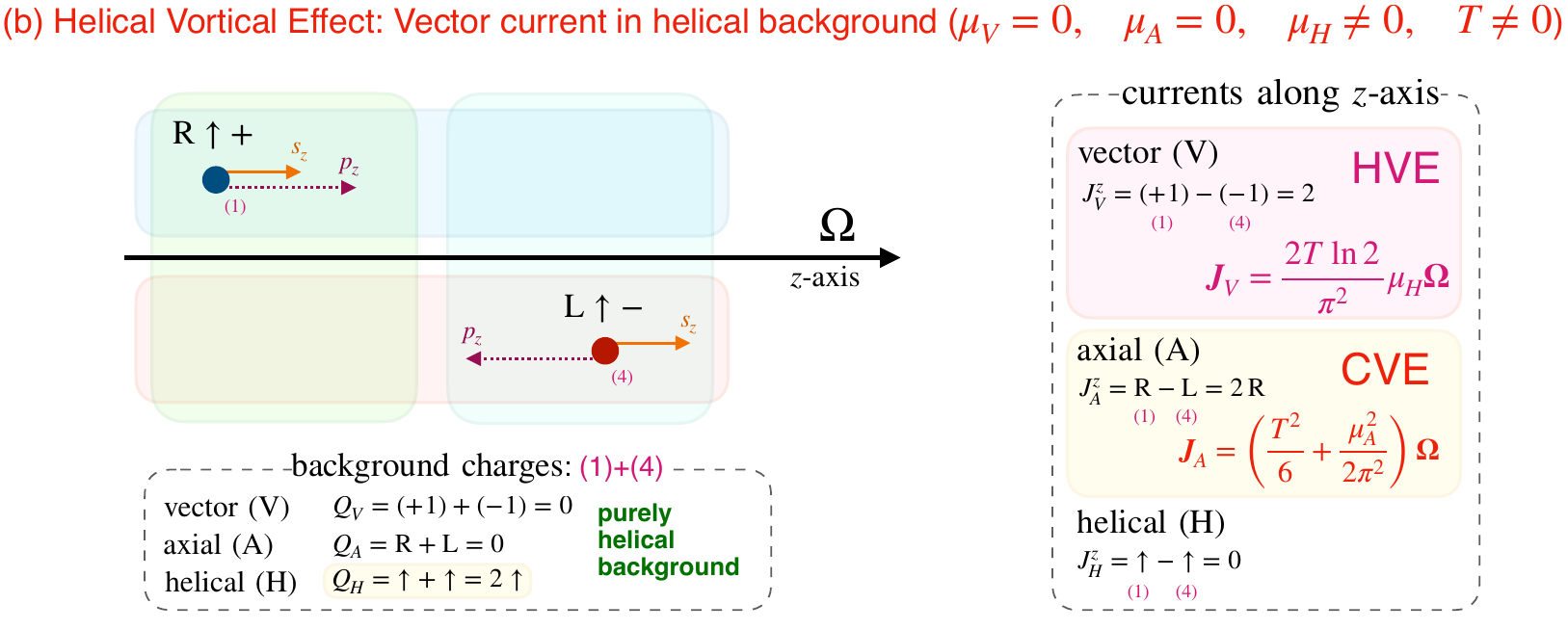} \\[5mm]
    \includegraphics[width=0.85\linewidth]{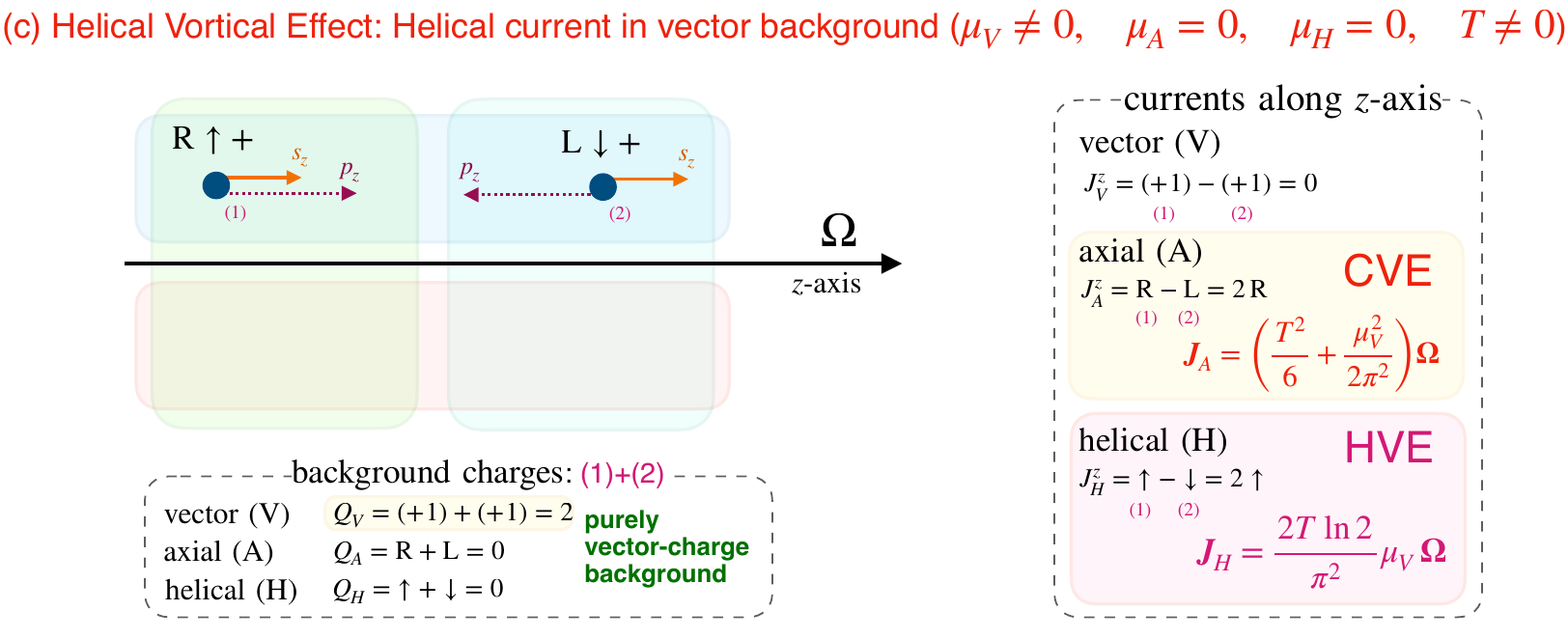} 
\end{tabular}
    \caption{Generation of chiral and helical vortical effects due to spin-orbit coupling in fermion--anti-fermion ensembles rotating with the angular velocity $\boldsymbol \Omega$. An elementary counting of vector, axial, and helical degrees of freedom explains the emergence of (a) the Chiral Vortical Effect~\eqref{eq:sigma:A} in a neutral plasma as well as the Helical Vortical Effects~\eqref{eq:sigma:V} and~\eqref{eq:sigma:H} in the helically imbalanced (b) and degenerate (c) plasma. The chirality of particles is denoted by the letters R/L; their helicity is shown by the arrows $\uparrow/\downarrow$; the vector charge is presented by the signs $+/-$. The counting gives the total charges, $Q_\ell$ and the $z$-axis projection of the currents $J^z_\ell$, with $\ell = V, A, H$ for each case.
    }
    \label{fig:CHVE}
\end{figure*}

Before addressing the technical details of explicit calculations of the hydrodynamic wave spectrum, it is necessary to discuss certain simple and easily understandable examples of anomalous vortical transport. Our calculations involve the helical degree of freedom, which is often misidentified with the chiral property of a fermion. As we highlighted in the Introduction, the chirality and helicity -- that are related, for a single (anti-)fermion, to each other via Eq.~\eqref{eq:charge_products} -- are distinct quantities of fermionic ensembles that constitute different properties of a many-body fermionic system. This distinction also reveals itself at the level of anomalous transport. 

It is worth noticing that the vortical conductivities defined in the Landau frame~\eqref{eq:Landau_sigma} and in the $\beta$ frame differ from each other due to the freedom of interpreting the heat flux as particle diffusion and vice-versa. 
However, in the physically interesting high-temperature limit, the leading contributions to the anomalous conductivities in the $\beta$ frame and in the Landau frame -- which will be given below in Eqs.~\eqref{eq:sigma:VAH} and later in Eqs.~\eqref{eq:VAH_largeT_sigma}, respectively -- are the same for both frames up to subleading $O(T^{0})$ corrections and they agree qualitatively, up to a numerical factor, even in the $O(T^{-1})$ order. Explicitly, the leading anomalous vector, axial and helical conductivities in the $\beta$ frame read, respectively, as follows~\cite{Ambrus:2019khr}:
\begin{subequations}
    \begin{align}
         \sigma^\omega_{\beta;A} & = \frac{T^2}{6} + \frac{1}{2\pi^2} (\mu_V^2 + \mu_A^2 + \mu_H^2) + O(T^{-1})\,, 
         \label{eq:sigma:A} \\
         \sigma^\omega_{\beta;V} & = \frac{2 T \ln 2}{\pi^2} \mu_H + \frac{\mu_V \mu_A}{\pi^2} + O(T^{-1})\,, 
         \label{eq:sigma:V} \\
         \sigma^\omega_{\beta;H} & = \frac{2 T \ln 2}{\pi^2} \mu_V + \frac{\mu_H \mu_A}{\pi^2} + O(T^{-1})\,. 
         \label{eq:sigma:H} 
    \end{align}
 \label{eq:sigma:VAH}
\end{subequations}

In the first term of Eq.~\eqref{eq:sigma:A}, we immediately recognize the Chiral Vortical Effect~\cite{Vilenkin:1978is, Vilenkin:1979ui}, which appears as a result of the axial-gravitational anomaly~\cite{Landsteiner:2011cp}. Despite its anomalous origin, the emergence of the Chiral Vortical Effect~\eqref{eq:sigma:A} can be readily understood on the basis of a simple particle counting of spin degrees of freedom in the system of the chiral fermions polarized in the presence of the vortical (rotating) background. Below, we also demonstrate that the Helical Vortical Effects, given in Eqs.~\eqref{eq:sigma:V} and \eqref{eq:sigma:H}, appear naturally on the very same basis as their famous chiral partner~\eqref{eq:sigma:A}. To this end, we give in Fig.~\ref{fig:CHVE} a pictorial counting for all three effects in Eqs.~\eqref{eq:sigma:VAH} to demonstrate their common origin.

All illustrations in Fig.~\ref{fig:CHVE} are based on the simple fact that the chiralities and helicities are equal (opposite) for particles (antiparticles). The charge conjugation $\mathcal C$ transforms a particle to its antiparticle ($q \to -q$) and, at the same time, flips its helicity ($\uparrow/\downarrow \ \to \ \downarrow/\uparrow$) while leaving unchanged its chirality ($R/L \to R/L$). These properties are consistent with the observation that helicity, similarly to the vector charge of the particle, is a $\mathcal P$--odd quantity while the chirality is a $\mathcal P$-even characteristic of the fermion. On the other hand, the parity transformation $\mathcal P$ does not, expectedly, affect the vector charge, while it naturally flips both helicity ($\uparrow/\downarrow\ \to\ \downarrow/\uparrow$) and chirality ($R/L \to R/L$), which are $\mathcal P$--odd quantities. These properties are consistent with Eq.~\eqref{eq:charge_products} and are summarized in Table~\ref{tbl_symmetries} below.
As seen in Table~\ref{tbl_symmetries}, the $\mathcal{C}$ parities of the spatial currents $\bm{J}_V$, $\bm{J}_A$ and $\bm{J}_H$ are identical with those of the corresponding charges, while their $\mathcal{P}$ and $\mathcal{T}$ parities are reversed. 
The $\mathcal{P}$ and $\mathcal{T}$ parities of the currents $\bm{J}_\ell$ are identically shared by the product $Q_\ell p^z$, where $p^z$ is the $z$-axis momentum of a hypothetical plasma constituent. Therefore, a simple, qualitative explanation of the macroscopic, ensemble-averaged, vortical effects can be visualized pictographically by considering individual constituents and their direction of motion, as indicated in Fig.~\ref{fig:CHVE}.

\begin{table}[t]
\begin{center}
\begin{tabular}{|c||c|c|c||c|c|c||c|c|}
\hline
 &   $Q_V$ & 	$Q_A$ 	& 	$Q_H$	& ${\boldsymbol J}_V$ & 	${\boldsymbol J}_A$ 	& 	${\boldsymbol J}_H$	& $\boldsymbol{v}$ 	&	${\boldsymbol \Omega}$ \\
\hline
$\mathcal C$	&   $-$	&	$+$	&	$-$ &  $-$  &  $+$  &  $-$  &   $+$	&	$+$	\\
\hline
$\mathcal P$	&   $+$	&	$-$	&	$-$ &  $-$  &  $+$  &  $+$ &   	$-$	&	$+$ \\
\hline
$\mathcal T$	&   $+$	&	$+$	&	$+$ &  $-$  &  $-$  &  $-$ &   	$-$	&	$-$	\\
\hline
\end{tabular}
\end{center}
\vskip 3mm
\caption{The charge conjugation $\mathcal C$, the parity inversion $\mathcal P$, and the time reversal $\mathcal T$ symmetries of the vector $Q_V$, axial $Q_A$, and helical $Q_H$ charges, 
their currents, $\bm{J}_\ell$ ($\ell = V,A,H$), as well as the group velocity $\boldsymbol{v} = \partial \omega / \partial \boldsymbol{k}$ and the vorticity $\boldsymbol{\Omega}$. The signs $+/-$ indicate the even/odd nature of these quantities under the corresponding inversions.
}
\label{tbl_symmetries}
\end{table}

The natural appearance of the Chiral Vortical Effect~\eqref{eq:sigma:A} in a system of rotating chiral fermions is illustrated in Fig.~\ref{fig:CHVE}(a). The spin-orbital interaction couples the orbital motion of the fermion--anti-fermion ensemble with the polarization of the spin of each particle. The particle spin $\boldsymbol s$ tends to align with the angular velocity $\boldsymbol \Omega$ regardless of the charge of the particle, as the spin-orbital coupling does not discriminate between particles and antiparticles. In Fig.~\ref{fig:CHVE}(a), we show, for simplicity, a maximally spin-polarized ensemble with the spins of all constituents $\boldsymbol s$ aligned along the angular velocity $\boldsymbol \Omega$. The ensemble is neutral in all possible charges as the total vector, axial and helical numbers are zero. To achieve total neutrality of a non-empty ensemble, one needs at least four constituents: one (chirally) right-handed particle and a left-handed particle, as well as their antiparticles, as shown in the figure. A simple counting, reproduced in this figure as well, shows that the spin polarization in this ensemble generates an axial current along the axis of rotation,  ${\boldsymbol j}_A \propto {\boldsymbol \Omega}$, while the vector and helical currents vanish identically. This simple picture based on degree-counting is perfectly consistent with the prediction of the anomalous transport laws~\eqref{eq:sigma:VAH} for the neutral plasma (with $\mu_V = \mu_A = \mu_H = 0$).

An immediate emergence of one of the Helical Vortical Effects~\eqref{eq:sigma:V} is illustrated in Fig.~\ref{fig:CHVE}(b). Here, we consider a helically nontrivial background where the helical charge is non-zero while all other (vector and axial) charges are vanishing. To achieve this set of quantum numbers, it is sufficient to take a chirally right-handed particle and a chirally left-handed antiparticle. One quickly arrives at the conclusion that in this ensemble, the spin polarization produces the vector current along the angular velocity: ${\boldsymbol J}_V \propto {\boldsymbol \Omega}$. The axial current is also generated in consistency with Eq.~\eqref{eq:sigma:A}.

Figure~\ref{fig:CHVE}(c) shows that in the presence of a vector charge (one right-handed particle and one left-handed particle), the spin-orbit coupling generates the helical current ${\boldsymbol J}_H \propto {\boldsymbol \Omega}$ along the axis of rotation as follows from the other type of the Helical Vortical Effect~\eqref{eq:sigma:H}. The axial current is again generated in consistency with Eq.~\eqref{eq:sigma:A}.

Finally, we consider the purely axial background. Then the anomalous vortical effects~\eqref{eq:sigma:VAH} predict that such an ensemble may only lead to an axial current~\eqref{eq:sigma:A}, while the vector~\eqref{eq:sigma:V} and helical~\eqref{eq:sigma:H} currents should vanish. It is not difficult to check these properties from Fig.~\ref{fig:CHVE}, by considering, for example, an ensemble made of one chirally right-handed particle and its antiparticle. 

Thus, the chirality and helicity are independent numbers that appear naturally in the fermionic systems containing both particles and antiparticles. As we have just seen, the presence of helical charge generates the vector current while the excess of vector charge produces the helical flow along the axis of rotation. Both the helical and vector charge densities produce the axial current. These anomalous transport effects, presented in Eq.~\eqref{eq:sigma:VAH} and qualitatively justified by the counting of Fig.~\ref{fig:CHVE}, intertwine the vector, axial and helical degrees of freedom in the form of hydrodynamic excitations. The rigorous investigation of these hydrodynamic effects is the subject of the present paper.

\subsection{Conservation equations for longitudinal perturbations}\label{sec:2:eqs}

We now consider small perturbations around the rigidly rotating state in terms of fluctuations of velocity, charge densities, and pressure. While doing so, we assume that the constitutive Eqs.~\eqref{eq_Landau_JT} continue to hold, with the quantities $E$, $Q_\ell$ and $\sigma^\omega_\ell$ derived from $P$ as in Eq.~\eqref{eq:Landau_cond}, $u^\mu_L$ being now the perturbed velocity $u^\mu$ and the perturbed vorticity $\omega^\mu$ associated to $u^\mu$.
Perturbations in the chemical potentials $\mu_\ell$ and temperature $T$ are subsequently induced by the relations
\begin{equation}
 \delta P = \frac{\partial P}{\partial T} \delta T + \frac{\partial P}{\partial \mu_\ell} \delta \mu_\ell, \ \ \
 \delta Q_\ell = \frac{\partial Q_{\ell}}{\partial T} \delta T + \frac{\partial Q_{\ell}}{\partial \mu_{\ell'}} \delta\mu_{\ell'},
\end{equation}
with $\partial P / \partial T= s = (E + P - \vec{Q} \cdot \vec{\mu}) / T$ being the entropy density and $\partial P / \partial \mu_\ell = Q_\ell$.
We consider the limit of slow rotation and continue to retain only terms that are linear in the angular frequency $\Omega$. For consistency, we focus on the region around the rotation axis, where $\rho \Omega \ll 1$.

In this paper, we are interested in the study of the propagation of perturbations along the axis of rotation (the $z$ axis).  Decomposing the time-dependent four-velocity $\bar{u}^\mu$, charge densities $\overline{Q}_\ell$ and pressure $\overline{P}$ into a background contribution (denoted without the overline) and a perturbation,
\begin{equation}
 \bar{u}^\mu = u^\mu + \delta \bar{u}^\mu, \quad 
 \overline{Q}_\ell = Q_\ell + \delta \overline{Q}_\ell, \quad 
 \overline{P} = P + \delta \overline{P},
 \label{eq:pert}
\end{equation}
we expand the perturbations $\delta \bar{u}^\mu$, $\delta \overline{Q}_\ell$ and $\delta \overline{P}$ in a Fourier series with respect to $t$ and $z$,
\begin{equation}
 \begin{pmatrix}
  \delta \bar{u}^\mu \\ 
  \delta \overline{Q}_\ell \\
  \delta \overline{P} 
 \end{pmatrix} = \int_{-\infty}^\infty dk\, e^{i k z} \sum_\omega e^{-i \omega t} 
 \begin{pmatrix}
  \delta u^\mu_\omega(k) \\ \delta Q_{\ell;\omega}(k) \\ \delta P_\omega(k)
 \end{pmatrix},
 \label{eq:pert_macro}
\end{equation}
where $\omega \equiv \omega(k)$ is the angular frequency, which is related to the wavenumber $k$ via the dispersion relation, and we have anticipated that the system supports only a discrete set of angular frequencies. Note that the vorticity four-vector, $\omega^\mu$, always carries a Lorentz four-index, whereas its magnitude is denoted by $\Omega$, such that there is no risk of confusion with the angular frequency $\omega$.
The background four-velocity $u^\mu$ is taken to be the Landau four-velocity obtained in Eq. \eqref{eq:landau_fourvel} under the Lorentz boost defined in Eq. \eqref{eq_Lorentz}, which to leading order in $\Omega$ coincides with the four-velicity $u^\mu_\Omega$ introduced in Eq.~\eqref{eq:u_RR} corresponding to that of a rigidly-rotating fluid, while the background pressure and charge densities are assumed to be constant. Note that the perturbation mode amplitudes for the four-velocity, $\delta u_\omega$, charge densities, $\delta Q_{\ell;\omega}$, and pressure, $\delta P_\omega$, can take complex values corresponding to a (relative) phase of these fluctuations.

Both the perturbation amplitudes $\delta A_\omega$ and the angular frequencies $\omega$ will be computed up to first order with respect to the rotational angular frequency $\Omega$:
\begin{align}\label{eq:decomp}
 \delta A_\omega &= \delta A_{\omega;0} + \Omega \delta A_{\omega;1} + \dots,\nonumber\\
 \omega &= \omega_0 + \Omega \omega_1 + \dots.
\end{align}
We take at zeroth order longitudinal perturbations of the velocity, such that $\delta u_{\omega;0}^\mu \partial_\mu = \delta u^z_{\omega;0} \partial_z$. In principle, one is tempted to take all perturbations to be independent on the transverse plane coordinates, $x$ and $y$. However, we will show below that, in the case of the sound modes, at first order with respect to $\Omega$, the amplitudes of the transverse velocity components $\delta u^x_{\omega;1}$ and $\delta u^y_{\omega; 1}$ pick up an $(x,y)$ dependence. Accordingly, we allow for generic $(x,y)$ dependence for $\delta u^x_\omega$ and $\delta u^y_\omega$ while restricting all other perturbations to be independent of the transverse plane coordinates. We will show that these are the minimal requirements to obtain a consistent solution to the conservation equations to linear order. Explicitly, the flucutations of the fluid velocity are give by
\begin{align}
 \delta u^\mu_\omega \partial_\mu = \delta u^t_\omega \partial_t + \delta u^x_\omega \partial_x + \delta u^y_\omega \partial_y + \delta u^z_\omega \partial_z,
\end{align}
where $\delta u^t_\omega = \Omega(-y \delta u^x_\omega + x \delta u^y_\omega)$ is 
fixed demanding that the fluid velocity $u^\mu$ be normalized. 
With respect to $\Omega$, the perturbations have the following form:
\begin{gather}
 \delta u^x_\omega = \Omega \delta u^x_{\omega;1}(x,y), \quad 
 \delta u^y_\omega = \Omega \delta u^y_{\omega;1}(x,y), \nonumber\\
 \delta u^z_\omega = \delta u^z_{\omega;0} + \Omega \delta u^z_{\omega;1},
 \label{eq:pertu}
\end{gather}
where we indicated explicitly the $(x,y)$ dependence for the perturbations that are of first order with respect to $\Omega$. It can be checked that $\delta u^t_\omega \simeq 0$.

These perturbations in the velocity induce a perturbation in the vorticity, 
\begin{multline}
 \bar{\omega}^\mu \partial_\mu = \Omega \partial_z +  \int_{-\infty}^\infty dk\, e^{i k z} 
 \sum_\omega e^{-i \omega t}\delta \omega_\omega^\mu\partial_\mu,
\end{multline}
where
\begin{align}
    \delta \omega_{\omega}^t &= \Omega \delta u^z_{\omega;0},\nonumber\\
    \delta \omega_{\omega}^x &=-\frac{i \Omega}{2}\left(x\omega_0 \delta u_{\omega;0}^z + k\delta u_{\omega;1}^y\right),\nonumber\\
    \delta \omega_{\omega}^y &=-\frac{i\Omega}{2}\left(y\omega_0 \delta u_{\omega;0}^z-k\delta u_{\omega;1}^x\right),\nonumber\\
    \delta \omega_{\omega}^z &= \Omega(\partial_x\delta u_{\omega;1}^y-\partial_y\delta u_{\omega;1}^x).
    \label{eq:pertw}
\end{align}

We are now ready to set up the linear perturbations problem. For this purpose, we assume, as stated in the introduction of this section, that the stress-energy tensor $T^{\mu\nu}$ and the charge currents $J^\mu_\ell$ are given as in Eqs.~\eqref{eq:boosted_dec}. The dynamics of the velocity and pressure perturbations follow by imposing the conservation of energy and momentum, $\partial_\mu T^{\mu\nu} = 0$, leading to 
\begin{subequations}\label{eq_conservation_eqs}
\begin{align}
 D E + (E + P) \theta = 0, \ \ \
 (E + P) Du^\mu - \nabla^\mu P = 0,
 \label{eq_conservation_Tmunu}
\end{align}
where $D = u^\mu \partial_\mu$ represents the comoving derivative, $\nabla^\mu = \Delta^{\mu\nu} \partial_\nu$ is the spatial gradient in the fluid rest frame, while $\theta = \partial_\mu u^\mu$ is the expansion scalar. For simplicity, the overhead bars were dropped in the above equations, however we understand that they will hold also for the perturbed quantities.

Furthermore, we impose the conservation of the charge currents, $\partial_\mu J^\mu_\ell = 0$, with $\ell \in \{V,A,H\}$. All three currents are conserved in the quantum field theory of free (non-interacting) massless fermions. This is certainly true also in the interacting case for the vector current. In a realistic interacting theory, the conservation of the axial current is broken by the axial anomaly \cite{itzykson80,bertlmann96,buzzegoli20phd}. Moreover, interactions mediated by vector bosons (photons in QED or gluons in QCD) break the helicity current conservation through the so-called helicity-violating pair annihilation (HVPA) processes (cf.~Sec.~5.2 in Ref.~\cite{Ambrus:2020oiw}). We address such effects in detail in our companion paper \cite{Morales-Tejera:2024mtx} and instead consider in this paper that all three charge currents are conserved. Then, $\partial_\mu J^\mu_\ell = 0$ leads to:
\begin{align}
 D Q_\ell + Q_\ell \theta + \omega^\mu \partial_\mu \sigma^\omega_\ell + 
 \sigma^\omega_\ell \partial_\mu \omega^\mu &= 0.\label{eq_conservation_Jl}
\end{align}
\end{subequations}

The terms appearing in Eqs.~\eqref{eq_conservation_eqs} above can be computed by going to Fourier space, which amounts to the following substitutions:
\begin{align}
 Df &\rightarrow -i \omega_0  \delta f_{\omega;0}-i\Omega(\omega_0\delta f_{\omega;1}+\omega_1\delta f_{\omega;0}),\nonumber\\
 \theta &\rightarrow i k \delta u_{\omega;0}^z \nonumber\\
 &+ \Omega(ik\delta u_{\omega;1}^z + \partial_x \delta u^x_{\omega;1} + \partial_y \delta u^y_{\omega;1}), \nonumber\\
 \omega^\mu \partial_\mu f &\to ik\Omega \delta f_{\omega;0},
 \nonumber\\ 
 Du^z &\to -i \omega_0  \delta u^z_{\omega;0}-i\Omega(\omega_0\delta u^z_{\omega;1}+\omega_1\delta u^z_{\omega;0}),\nonumber\\
 Du^x &\rightarrow -i\omega_0\Omega\delta u^x_{\omega;1}\quad
 Du^y \rightarrow-i\omega_0\Omega \delta u^y_{\omega;1}   \nonumber\\
 Du^t &\rightarrow 0,\quad \partial_\mu \omega^\mu \rightarrow -2i \omega_0 \Omega \delta u^z_{\omega;0},
 \nonumber\\
 \nabla^\mu P &\rightarrow 
 k \left(\delta P_{\omega;0} + \Omega\delta P_{\omega;1}\right)\delta^\mu_z \nonumber\\
 &-i 
 \omega_0 \Omega (y \delta^\mu_x - x \delta^\mu_y)\delta P_{\omega;0},
 \label{eq:pert_aux2}
\end{align}
where $f$ is a scalar function and the right arrow indicates projecting the quantities to the left onto the $(k, \omega(k))$ Fourier mode. 

It is clear that the stress-energy sector is decoupled from the charge currents sector since it involves only the amplitudes $\delta u_\omega^\mu$ and $\delta P_\omega$.
To zeroth order in $\Omega$, we have
\begin{equation}
 \begin{pmatrix}
  -3\omega_0 & 4 k P \\
  k & -4 P \omega_0
 \end{pmatrix} 
 \begin{pmatrix}
  \delta P_{\omega;0} \\ \delta u^z_{\omega;0}
 \end{pmatrix} = 0.
 \label{eq:sound_det}
\end{equation}
There are two independent solutions to the above equation, which are obtained by the requirement that the determinant of the matrix on the left-hand side of Eq.~\eqref{eq:sound_det} vanishes: 
\begin{equation}
    3\omega_0^2 - k^2 = 0\,. 
\end{equation}
This relation gives rise to the acoustic modes, 
\begin{equation}
 \omega^\pm_{\rm ac.;0} = \pm c_s k, \qquad c_s = 1/\sqrt{3},
\end{equation}
with $c_s$ being the speed of sound in an ultrarelativistic ideal fluid. The leading order amplitude are related by
\begin{equation}\label{eq:ampl-long}
    \delta u^z_{\omega;0} = \dfrac{3\omega_0}{4kP}\delta P_{\omega;0}\,,
\end{equation}
where we have chosen to leave $\delta P_{\omega;0}$ as a free perturbation. Note that $\delta P_{\omega;1}$ can be reabsorbed into $\delta P_{\omega;0}$ and therefore we can set $\delta P_{\omega;1}=0$ without loss of generality. 

We now turn our attention to the conservation equations for the energy-momentum tensor to linear order in $\Omega$.
The energy conservation equation reads:
\begin{subequations}\label{eq:constmunu}
\begin{multline}
 \label{eq:constmunu_E}
 4kP\delta u_{\omega;1}^z-3\omega_0 \delta P_{\omega;1}
 -3\omega_1\delta P_{\omega;0}\\
 -4iP\left(\partial_y\delta u_{\omega;1}^y+\partial_x\delta u_{\omega;1}^x\right) = 0,
\end{multline}
while the equation for the longitudinal perturbation reads:
\begin{equation}
 \label{eq:constmunu_uz}
 4\omega_0 P\delta u_{\omega;1}^z -k \delta P_{\omega;1}+4 P\omega_1\delta u_{\omega;0} = 0.
\end{equation}
In the transverse plane, we uncover:
\begin{align}
    \omega_0 \Omega\left( y\delta P_{\omega;0}-4P\delta u_{\omega;1}^x\right) &=0, \label{eq:constmunu_ux}\\
    -\omega_0 \Omega\left( x\delta P_{\omega;0}+4P\delta u_{\omega;1}^y\right) &=0. \label{eq:constmunu_uy}
\end{align}
\end{subequations}
As anticipated, the transverse equations are solved by $(x,y)$ dependent fluctuations:
\begin{equation}\label{eq:ampl-trans}
    \delta u^x_{\omega;1} = y\dfrac{\delta P_{\omega;0}}{4P},\qquad \delta u^y_{\omega;1} = -x\dfrac{\delta P_{\omega;0}}{4P}.
\end{equation}

Note that the previous solution can be interpreted as a constant fluctuation of the vorticity $\delta\Omega_\omega = \partial_x \delta u^y_\omega -\partial_y \delta u^x_\omega = -\Omega \delta P_{\omega;0}/(4P)$. Substituting the solutions for the amplitudes, shown in Eqs. \eqref{eq:ampl-long} and \eqref{eq:ampl-trans}, onto the conservation equations \eqref{eq:constmunu}, and setting $\delta P_{\omega;1}=0$ without loss of generality, gives
\begin{align}
    \Omega\left( 4k P\delta u^z_{\omega;1}-3\omega_1\delta P_{\omega;0} \right)&=0\,,\nonumber\\
    -\Omega\frac{\omega_0}{k} \left(4kP\delta u^z_{\omega;1}+3\omega_1\delta P_{\omega;0}  \right)&=0\,.
\end{align}
Since $\delta P_{\omega ;0}$ is a free arbitrary perturbation, the only solution to the previous system is 
\begin{equation}
    \omega_1 = 0 \qquad \delta u_{\omega;1}^z=0\,.
\end{equation}
Therefore, the dispersion relation for the acoustic modes traveling along the $z$ direction to leading order in vorticity is given by
\begin{equation}\label{eq:acoustic_vpm}
     \omega^\pm_{\rm ac.} = \pm c_s k
\end{equation}

In Refs.~\cite{Kalaydzhyan:2016dyr,Gorbar:2017toh,Abbasi:2016rds}, the solution of these equations is obtained under the assumption of transverse-plane homogeneity and vanishing transverse velocity amplitudes in the on-axis limit. While it is clear that the transverse perturbations \eqref{eq:ampl-trans} vanish on-axis, the first derivatives remain finite. Neglecting these first derivatives, as implicitly done in the aforementioned references, can affect the dispersion relation for the sound modes, as we show in Appendix \ref{app:cvw_sound}, where we recompute the sound modes in the (unboosted) $\beta$-frame. Note that Eq.~\eqref{eq:acoustic_vpm} coincides with Eq.~(64) of Ref.~\cite{Abbasi:2016rds}, derived also using the boosted Landau frame velocity.

To leading order in $\Omega$, Eqs.~\eqref{eq_conservation_Tmunu} admit also $\delta u_\omega^\mu = \delta P_\omega = 0$ as a trivial solution. These solutions correspond to the charge modes, which are dominated by fluctuations in the charge densities. Setting $\delta u^\mu_\omega = \delta P_\omega =0$, the charge conservation Eqs.~\eqref{eq_conservation_Jl} give us
\begin{equation}
 \bigl(\omega \delta Q_{\ell;\omega} - k \Omega \delta \sigma^\omega_{\ell;\omega} \bigr) {\Bigr\rvert}_{\delta P_\omega =0}
 = 0.
 \label{eq:chargecons_aux}
\end{equation}
The above equation, written with respect to the fluctuations $\delta Q_{\ell;\omega}$ of the charge densities, can be reexpressed in terms of fluctuations in the chemical potentials, $\delta \mu_{\ell;\omega}$, as well as fluctuations in temperature, $\delta T_\omega$. Due to the constraint $\delta P_\omega = 0$, these fluctuations are not mutually independent. In fact, taking into account that $\delta P_\omega = s \delta T_\omega + \sum_\ell Q_\ell \delta \mu_{\ell;\omega}$, one may replace $\delta T_\omega$ via
\begin{equation}
 \delta T_\omega = -\sum_\ell \frac{Q_\ell}{s} \delta \mu_{\ell;\omega}.
\end{equation}
Subsequently, this replacement leads to the following system of equations for the perturbations in the chemical potentials:
\begin{align}
 &\sum_{\ell'} \mathbb{M}_{\ell\ell'} \delta \mu_{\ell';\omega} = 0, \nonumber\\& 
 \mathbb{M}_{\ell\ell'} \equiv \mathbb{M}_{\ell\ell'}(k, \omega) = \frac{\omega}{s} \frac{\partial(P, Q_\ell)}{\partial(T, \mu_{\ell'})} - \frac{k\Omega}{s} \frac{\partial(P, \sigma^\omega_\ell)}{\partial(T, \mu_{\ell'})},
 \label{eq:M_aux}
\end{align}
where it is understood that all quantities appearing in the expression for $\mathbb{M}_{\ell\ell'}$ are evaluated in the background state of the fluid. In the above, we employed the standard notation for the Jacobian,
\begin{equation}
 \frac{\partial(f,g)}{\partial(x,y)} = \frac{\partial f}{\partial x} \frac{\partial g}{\partial y} - \frac{\partial f}{\partial y} \frac{\partial g}{\partial x}.
\end{equation}
Nontrivial solutions for $\omega$ can be obtained by requiring that the determinant of the matrix $\mathbb{M}$ vanishes. Noting that $Q_\ell = \partial P / \partial \mu_\ell$, we can rewrite its temperature derivative as follows:
\begin{equation}
 \frac{\partial Q_\ell}{\partial T} = \frac{\partial s}{\partial \mu_\ell} = \frac{3Q_\ell}{T} - \frac{\vec{\mu}}{T} \cdot \frac{\partial Q_\ell}{\partial \vec{\mu}},
\end{equation}
where we took into account that $s = (E + P - \vec{\mu} \cdot \vec{Q}) / T$, as well as the relation $\partial \vec{Q} / \partial \mu_\ell = \partial Q_\ell / \partial \vec{\mu}$. Similarly, taking into account that $\sigma^\omega_{\ell;\beta} = \frac{1}{2} \partial Q_\ell / \partial \mu_A$ and $\sigma^\omega_\ell = \sigma^\omega_{\ell;\beta} - Q_A Q_\ell / (E + P)$, we have
\begin{equation}
 \frac{\partial \sigma^\omega_{\ell;\beta}}{\partial T} = \frac{2\sigma^\omega_{\ell;\beta}}{T} - \frac{\vec{\mu}}{T} \cdot \frac{\partial \sigma^\omega_{\ell;\beta}}{\partial \vec{\mu}}, \ \ \ 
 \frac{\partial \sigma^\omega_\ell}{\partial T} = \frac{2\sigma^\omega_\ell}{T} - \frac{\vec{\mu}}{T} \cdot \frac{\partial \sigma^\omega_\ell}{\partial \vec{\mu}}.
\end{equation}
Therefore, the matrix elements $\mathbb{M}_{\ell\ell'}$ can be expressed as
\begin{subequations}
 \label{eq:M}
\begin{align}
 &\mathbb{M}_{\ell\ell'} = \omega T^2 \mathbb{M}^\omega_{\ell\ell'} - k\Omega T \mathbb{M}^\Omega_{\ell\ell'}  \nonumber\\
 &=\omega \left(\frac{\partial Q_\ell}{\partial \mu_{\ell'}} - \frac{Q_{\ell'}}{s} \frac{\partial Q_\ell}{\partial T}\right) - k\Omega \left(\frac{\partial \sigma^\omega_\ell}{\partial \mu_{\ell'}} - \frac{Q_{\ell'}}{s} \frac{\partial \sigma^\omega_\ell}{\partial T}\right)\,,
 \label{eq:M_aux2}
\end{align}
where we introduced for later convenience the following notation:
\begin{align}
 \mathbb{M}^\omega_{\ell\ell'} &= \frac{1}{T^2} \left(\frac{\partial Q_\ell}{\partial \mu_{\ell'}} - \frac{3 Q_\ell Q_{\ell'}}{sT} + \frac{Q_{\ell'} \vec{\mu}}{sT} \cdot \frac{\partial Q_\ell}{\partial \vec{\mu}}\right),\label{eq:Mv}\\
 \mathbb{M}^\Omega_{\ell\ell'} &= \frac{1}{T} \left(\frac{\partial \sigma^\omega_\ell}{\partial \mu_{\ell'}} - \frac{2 \sigma^\omega_\ell Q_{\ell'}}{sT} + \frac{Q_{\ell'} \vec{\mu}}{sT} \cdot \frac{\partial \sigma^\omega_\ell}{\partial \vec{\mu}}\right).
 \label{eq:MOmega}
\end{align}
\end{subequations}
The structure of the matrix $\mathbb{M}_{\ell\ell'}$ shows that the angular velocities $\omega(k)$ obey linear dispersion relations,
\begin{equation}
 \omega = k v\,,
\end{equation}
where $v = \omega /k = \partial \omega /\partial k$ represents both the phase and the group velocity of the given excitation mode, being independent of wavenumber $k$.

\subsection{Full space-time solutions from Fourier modes}

Let us consider a function $\bar{f}(t,z)$ characterizing the fluid state, taking the value $f$ in the background fluid state. Under small perturbations, $\delta \bar{f}(t,z) = \bar{f}(t,z) - f$ can be expanded with respect to the Fourier modes considered in Eq.~\eqref{eq:pert_macro} as follows:
\begin{equation}
 \delta \bar{f}(t,z) = \int_{-\infty}^\infty dk\, e^{ikz} \sum_{\omega} e^{-i \omega(k) t} \delta f[k, \omega(k)],
 \label{eq:modes_generic}
\end{equation}
where $\delta f[k, \omega(k)]$ represent the Fourier mode amplitudes.
Let us look at the complex conjugate of $\bar{f}(t,z)$:
\begin{multline}
 \bar{f}^*(t,z) = f^* + \int_{-\infty}^\infty dk\, e^{ikz} \\\times
 \sum_{\omega} e^{i \omega^*(-k) t} \delta f^*[-k, \omega(-k)].
\end{multline}
Imposing that $\bar{f}(t,z)$ is real implies that 
\begin{equation}
 \omega^*(-k) = -\omega(k), \ \
 \delta f^*[-k, \omega(-k)] = \delta f[k, \omega(k)],
 \label{eq:modes_refl}
\end{equation}
while $f^* = f$. This imposes the structure
\begin{equation}
 \omega(k) = k C_\omega(k) - i D_\omega(k),
\end{equation}
where the phase velocity $C_\omega(k) = {\rm Re}[\omega(k)/k]$ and the dissipation rate $D_\omega(k) = -{\rm Im}[\omega(k)]$ are both real and even with respect to $k \rightarrow -k$. The modes allowed by the matrix $\mathbb{M}_{\ell\ell'}$ have the linear dispersion relation $\omega = v k$, hence $C_\omega(k) = v$ and $D_\omega(k) = 0$.
Using the above notation, Eq.~\eqref{eq:modes_generic} can be written as
\begin{align}
 \delta \bar{f}(t,z) &= 2 \int_0^\infty dk\, \sum_\omega\, \{
  \cos(k v t - k z) {\rm Re}[\delta f(k, \omega)] \nonumber\\&
  + \sin(k v t - k z) {\rm Im}[\delta f(k, \omega)] \}.
\end{align}

Frequently, we will consider the case when one of the chemical potentials (the representative one) is initialized according to a simple, harmonic cosine profile, achieved by setting 
\begin{equation}
 \bar{f}(0, z) = f + \delta \bar{f}_0 \cos(kz).
 \label{eq:cos_F}
\end{equation}
This implies that the mode amplitudes $\delta f(k', \omega(k'))$ satisfy
\begin{equation}
 \delta f(k', \omega(k')) = \frac{1}{2} \delta f_\omega [\delta(k' - k) + \delta(k' + k)],
 \label{eq:cos_df}
\end{equation}
which is compatible with Eq.~\eqref{eq:modes_refl}. The constants $\delta f_\omega$ must satisfy 
\begin{equation}
 \sum_\omega \delta f_\omega = \delta \bar{f}_0,
\end{equation}
and the space-time solution reads
\begin{equation}
 \delta\bar{f}(t,z) = \sum_\omega \delta f_\omega \cos(kz - kvt),
 \label{eq:cos_sol}
\end{equation}
with $v = \omega / k$ being independent of $k$.

Finally, we will consider the case of an initial Gaussian profile,
\begin{equation}
 \bar{f}(0, z) = f + \delta \bar{f}_0\, e^{-z^2 / 2 \sigma^2}.
 \label{eq:gauss_F}
\end{equation}
The mode amplitudes then satisfy
\begin{equation}
 \sum_\omega \delta f(k, \omega(k)) = \frac{\sigma}{\sqrt{2\pi}} \delta \bar{f}_0 \, e^{-\sigma^2 k^2 / 2}.
 \label{eq:gauss_fk}
\end{equation}
The exact expression for each mode amplitude $\delta f_k \equiv \delta f(k, \omega(k))$ depends on the conditions imposed at the level of the three independent chemical potentials, $\delta \bar{\mu}_{\ell;0} \equiv \delta \bar{\mu}_{\ell}(0, z)$. Retrieving the space-time solution $\delta \bar{f}(t,z)$ requires the inverse Gaussian integration formula,
\begin{equation}
 \frac{\sigma}{\sqrt{2\pi}} \int_{-\infty}^\infty e^{-i \omega t + ikz} e^{ - \sigma^2 k^2 / 2} dk = e^{-(z - v t)^2 / 2\sigma^2,}
 \label{eq:gauss_inv}
\end{equation}
where we used the property that $\omega =kv$ and $v$ is independent of $k$.

\section{Hydrodynamic waves at high temperature}\label{sec:largeT}

In the large temperature limit, we may assume that $\lvert \mu_\ell \rvert  \ll T$, allowing the polylogarithms in Eq.~\eqref{eq:Pcl} to be expanded in a Taylor series with respect to the dimensionless chemical potentials $\alpha_\ell = \mu_\ell / T$. Using the definition of the polylogarithm, 
\begin{equation}
 {\rm Li}_4(-e^{\mu_{\sigma,\lambda} / T}) = \sum_{n = 1}^\infty \frac{(-1)^n}{n^4} e^{n \mu_{\sigma,\lambda} / T},
\end{equation}
we can expand the exponential in a Taylor series. The sum over $n$ can be performed employing the identity $\sum_{n = 1}^\infty (-1)^n / n^s = -(1 - 2^{1-s}) \zeta(s)$, where $\zeta(s)$ is the Riemann zeta function. This procedure leads us to a number of useful identities:
\begin{gather}
 \sum_{n = 1}^\infty \frac{(-1)^n}{n^4} = -\frac{7\pi^4}{720}, \qquad 
 \sum_{n = 1}^\infty \frac{(-1)^n}{n^3} = -\frac{3\zeta(3)}{4}, \nonumber\\ 
 \sum_{n = 1}^\infty \frac{(-1)^n}{n^2} = -\frac{\pi^2}{12}, \qquad 
 \sum_{n = 1}^\infty \frac{(-1)^n}{n} = -\ln 2, \nonumber\\ 
 \sum_{n = 1}^\infty (-1)^n \rightarrow -\frac{1}{2},
\end{gather}
where the last relation follows only under the assumption of analytical continuation of the $\zeta$ function. The summation over 
the numbers $\sigma = \pm 1$ and $\lambda = \pm 1/2$ 
can be performed using the relations:
\begin{gather}
 \sum_{\sigma,\lambda} 1 = 4, \quad 
 \sum_{\sigma,\lambda} \mu_{\sigma,\lambda} = 0, \nonumber\\
 \sum_{\sigma,\lambda} \mu_{\sigma,\lambda}^2 = 4 {\vec \mu}^2,  \quad
 \sum_{\sigma,\lambda} \mu_{\sigma,\lambda}^3 = 24 \mu_\times^3, \nonumber\\ 
 \sum_{\sigma,\lambda} \mu_{\sigma,\lambda}^4 = 4  \big({\vec \mu}^2\big)^2 + 16 (\mu_V^2 \mu_H^2 + \mu_V^2 \mu_A^2 + \mu_A^2 \mu_H^2),
 \label{eq:largeT_sumsl}
\end{gather}
where we employed Eq.~\eqref{eq:charge_products} to perform the sums and introduced the following notation:
\begin{equation}
 \vec{\mu}^2 = \mu_V^2 + \mu_A^2 + \mu_H^2, \qquad 
 \mu_\times^3 = \mu_V \mu_A \mu_H.
\end{equation}

We therefore obtain for the pressure:
\begin{subequations}\label{eq:VAH_largeT}
\begin{align}
 P &= \frac{7\pi^2 T^4}{180} + \frac{\vec{\mu}^2 T^2}{6}
 + \frac{4 \mu_\times^3 T}{\pi^2} \ln 2
 + \frac{({\vec{\mu}}^2)^2}{12\pi^2} \nonumber\\&
 + \frac{\mu_A^2 \mu_H^2 + \mu_V^2 \mu_H^2 + \mu_V^2 \mu_A^2}{3\pi^2} 
 + O(T^{-1}). \label{eq:VAH_largeT_P}
\end{align}
The charge densities $Q_\ell$ and entropy density $s$ can be found via the following thermodynamic relations~\eqref{eq:thermocl}:
\begin{align}
 Q_\ell &= \frac{\mu_\ell T^2}{3}
 + \frac{4 T \ln 2}{\pi^2} \frac{\partial \mu_\times^3}{\partial \mu_\ell}
 \nonumber\\& + \frac{\mu_\ell(3\vec{\mu}^2 - 2\mu_\ell^2)}{3\pi^2} + O(T^{-1}),\label{eq:VAH_largeT_Q}\\
 s &= \frac{7\pi^2 T^3}{45} + \frac{\vec{\mu}^2 T}{3}
 + \frac{4 \mu_\times^3}{\pi^2} \ln 2
 + O(T^{-2}),\label{eq:VAH_largeT_s}
\end{align}
where it is understood that no summation over $\ell$ is implied in the last term in the above expression for $Q_\ell$.
In order to find the terms appearing in the matrix $\mathbb{M}_{\ell\ell'}$ given in Eq.~\eqref{eq:M}, the derivative of the charge densities with respect to the chemical potential must be computed:
\begin{align}
& \frac{\partial Q_\ell}{\partial \mu_{\ell'}} = \frac{\delta_{\ell\ell'} T^2}{3}
 + \frac{4 T \ln 2}{\pi^2} \frac{\partial^2 \mu_\times^3}{\partial \mu_\ell \partial \mu_{\ell'}}
 \nonumber\\&+ \frac{\delta_{\ell\ell'} (\vec{\mu}^2 - 2\mu_\ell^2) + 2\mu_\ell \mu_{\ell'}}{\pi^2} 
 + O(T^{-1}),\label{eq:VAH_largeT_dQdmu}
\end{align}
where again, there is no summation with respect to $\ell$ in the last term appearing above. The $\beta$-frame vortical conductivities can be obtained from Eq.~\eqref{eq:thermocl} as:
\begin{align}
 &\sigma^\omega_{\beta;\ell} = \frac{\delta_{\ell,A} T^2}{6}
 + \frac{2 T \ln 2}{\pi^2} (\mu_V \delta_{\ell H} + \mu_H \delta_{\ell V}) 
 \nonumber\\&+ \frac{\delta_{\ell A}}{2\pi^2} (\mu_V^2 - \mu_A^2 + \mu_H^2) + \frac{\mu_\ell \mu_A}{\pi^2} +
 O(T^{-1}).
 \label{eq:VAH_largeT_sigmabeta}
\end{align}
Noting that $Q_\ell Q_{\ell'} / (E + P) = 5\mu_\ell \mu_{\ell'} / 7\pi^2 + O(T^{-1})$, the Landau frame vortical conductivities introduced in Eq.~\eqref{eq:Landau_sigma} can be seen to differ from $\sigma^\omega_{\beta;\ell}$ only in the $O(T^0)$ term:
\begin{align}
 &\sigma^\omega_\ell = \frac{\delta_{\ell A} T^2}{6}
 + \frac{2 T \ln 2}{\pi^2} (\mu_V \delta_{\ell H} + \mu_H \delta_{\ell V}) 
 \nonumber\\&+ \frac{\delta_{\ell A}}{2\pi^2} (\mu_V^2 - \mu_A^2 + \mu_H^2) + \frac{2\mu_\ell \mu_A}{7\pi^2} + O(T^{-1}).\label{eq:VAH_largeT_sigma}
\end{align}
Their derivatives with respect to the chemical potentials are
\begin{equation}
\begin{split}
 &\frac{\partial \sigma^\omega_\ell}{\partial \mu_{\ell'}} = \frac{2 T \ln 2}{\pi^2} (\delta_{\ell V} \delta_{\ell' H} + \delta_{\ell H} \delta_{\ell' V}) 
 \nonumber\\&+ \frac{\delta_{\ell A}}{\pi^2}(\mu_{\ell'} - 2\mu_A \delta_{\ell' A}) + \frac{2}{7\pi^2}(\delta_{\ell \ell'} \mu_A + \mu_\ell \delta_{\ell' A}) \nonumber\\&+ O(T^{-1}).
 \label{eq:VAH_largeT_dsigmadmu}
 \end{split}
\end{equation}
\end{subequations}

With the above results, the matrices $\mathbb{M}_\omega$ and $\mathbb{M}_\Omega$ defined in Eq.~\eqref{eq:M} can be expanded as follows:
\begin{align}
 \mathbb{M}_\omega &= \frac{1}{3}\mathbb{I} + 
 \frac{4 \ln 2}{\pi^2 T}
 \begin{pmatrix}
  0 & \mu_H & \mu_A \\
  \mu_H & 0 & \mu_V \\
  \mu_A & \mu_V & 0
 \end{pmatrix} + O(T^{-2}), \nonumber\\
 \mathbb{M}_\Omega &= 
 \frac{2\ln 2}{\pi^2}
 \begin{pmatrix}
  0 &0 & 1 \\ 
  0 & 0 & 0 \\
  1 & 0 & 0
 \end{pmatrix} + \frac{2}{7\pi^2 T}
 \begin{pmatrix}
  \mu_A & \mu_V & 0  \\
  \mu_V & -4\mu_A & \mu_H \\
  0 & \mu_H & \mu_A
 \end{pmatrix} \nonumber\\&+ O(T^{-2}),
 \label{eq:VAH_largeT_Mmu}
\end{align}
where $\mathbb{I}$ is the unit matrix.

The equation ${\rm det}\, (\mathbb{M} / T^2) = 0$ can be solved iteratively, using the formula
\begin{equation}
 {\rm det}(\mathbb{A} + \varepsilon \mathbb{B}) = {\rm det}(\mathbb{A}) [1 + 
 \varepsilon {\rm tr}(\mathbb{A}^{-1} \mathbb{B}) + O(\varepsilon^2)],
 \label{eq:largeT_Mexp}
\end{equation}
which is valid for small $\varepsilon$. We now consider a large-$T$ expansion of the angular frequency~$\omega$,
\begin{equation}
 \omega = \omega_{0} + \frac{\omega_1}{T} + \frac{\omega_2}{T^2} + O(T^{-3}).
 \label{eq:largeT_vexp}
\end{equation}
it is easy to see that the zeroth-order term must vanish since 
\begin{equation}
 {\rm det} [\omega_0 \mathbb{I}] = 0 \Rightarrow \omega_0 = 0.
\end{equation}
Now, taking into account that $\omega = \omega_1 T^{-1} + \dots$, we notice that the leading-order term in $\mathbb{M} / T^2 = \omega \mathbb{M}_\omega - k (\Omega / T) \mathbb{M}_\Omega$ becomes of order $T^{-1}$. 
Substituting Eq.~\eqref{eq:largeT_vexp} into Eq.~\eqref{eq:largeT_Mexp}, we obtain:
\begin{align}
 \frac{1}{T^2} \mathbb{M} =& T^{-1} \mathbb{M}_1 + T^{-2} \mathbb{M}_{2} + O(T^{-3}), \nonumber\\
 \mathbb{M}_1 =& \frac{\omega_1}{3} \mathbb{I} - 
 \frac{2 k \Omega \ln 2}{\pi^2} 
 \begin{pmatrix} 
  0 & 0 & 1 \\ 0 & 0 & 0 \\ 1 & 0 & 0
 \end{pmatrix}, \nonumber\\
 \mathbb{M}_2 =& 
 \frac{\omega_2}{3} \mathbb{I} + 
 \frac{4 \omega_1 \ln 2}{\pi^2} 
 \begin{pmatrix}
  0 & \mu_H & \mu_A \\
  \mu_H & 0 & \mu_V \\
  \mu_A & \mu_V & 0
 \end{pmatrix} \nonumber\\&- 
 \frac{2 k \Omega}{7\pi^2} 
 \begin{pmatrix}
  \mu_A & \mu_V & 0 \\ 
  \mu_V & -4\mu_A & \mu_H \\
  0 & \mu_H & \mu_A
 \end{pmatrix}.
\end{align}
The determinant of $\mathbb{M}_1$ reads
\begin{equation}\label{eq:v1h}
 {\rm det}(\mathbb{M}_1) = \frac{\omega_1 T^2}{27} \left(\frac{\omega_1^2}{T^2} - k^2 c_h^2\right),
\end{equation}
where
\begin{equation}
 c_h = \frac{6 \ln 2}{\pi^2} \frac{\Omega}{T},
 \label{eq_v_HVW_leading}
\end{equation}
and the notation $c_h$ introduced above refers to the propagation speed of the helical vortical wave in a neutral, unpolarized plasma with a conserved helical charge, see Eq.~(116) in Ref.~\cite{Ambrus:2019khr} for details. 
The solution $\omega_1 = 0$ of Eq.~\eqref{eq_v_HVW_leading} corresponds to the so-called \textit{Axial Vortical Wave}, while the solutions $\omega_1 / T = \pm k c_h$ correspond to the \textit{Helical Vortical Waves}. 
Notice that despite the velocity for the Axial Vortical Wave vanishing in the zeroth and first orders, this hydrodynamic mode is still a propagating excitation because the second order brings a nonzero contribution to $\omega$. These properties will be discussed in detail below. Both the axial and helical vortical waves in the high-temperature limit match those found in \cite{Ambrus:2019khr} for a neutral plasma if one neglects the dissipation of helicity in \cite{Ambrus:2019khr}, i.e., in the limit $\tau_H\to \infty$. The results presented in the coming sections constitute a generalisation of the previous results to the presence of non-vanishing chemical potentials for vector, axial and helical charges.

\subsection{Helical Vortical Wave}\label{sec:largeT:HVW}

The solutions $\omega_1/T \rightarrow \omega^{\pm}_{h,1}/T = \pm k c_h $ of Eq.~\eqref{eq:v1h} represent two propagating modes, corresponding to the \textit{Helical Vortical Wave} (HVW). 
In order to compute the second order correction $\omega^{\pm}_{h;2}$ for the angular velocity of the vortical wave, we note that the inverse $(\mathbb{M}_1)^{-1}$ of the matrix $\mathbb{M}_1$ does not exist since we imposed 
${\rm det}(\mathbb{M}_1) = 0$. Nevertheless, the following combination is still finite:
\begin{equation}
 {\rm det}(\mathbb{M}_1) (\mathbb{M}_1)^{-1} = \frac{(\omega^\pm_{h;1})^2}{9} 
 \begin{pmatrix}
  1 & 0 & \pm 1 \\
  0 & 0 & 0 \\
  \pm 1 & 0 & 1
 \end{pmatrix}.
\end{equation}
The leading-order contribution to ${\rm det}(\mathbb{M}/T^2)$ is thus given by $T^{-2} {\rm tr} [{\rm det}(\mathbb{M}_1) \mathbb{M}_1^{-1} \mathbb{M}_2]$, which evaluates to
\begin{equation}
 \frac{2 (\omega^\pm_{h; 1})^2}{27 T^2} \left\{\omega^\pm_{h,2} + 
 \frac{6 k\Omega \mu_A}{7\pi^2} \left[\frac{84}{\pi^2} (\ln 2)^2 - 1\right] \right\} = 0.
\end{equation}
From the above equation, we obtain
\begin{equation}
 \omega^\pm_{h,2} = -\frac{6 k \Omega \mu_A}{7\pi^2} 
 \left[\frac{84}{\pi^2} (\ln 2)^2 - 1\right],
 \label{eq:VAH_largeT_vh12}
\end{equation}
leading to the following extension of Eq.~\eqref{eq_v_HVW_leading} for the velocity of the helical vortical wave:
\begin{align}
 &\frac{\omega^\pm_h}{k} = \pm \frac{6 \ln 2}{\pi^2} \frac{\Omega}{T}
 {-} \frac{6}{7\pi^2} \left[\frac{84}{\pi^2} (\ln 2)^2 {-} 1\right] \frac{\Omega \mu_A}{T^2} \nonumber\\
 &+ O(T^{-3})\,. \qquad {\rm [Helical\ Vortical\ Wave]}
 \label{eq_HVW_speed}
\end{align}
We see that at finite axial chemical potential $\mu_A$, the angular velocity $\omega^\pm_h$ receives a non-reciprocal contribution that distinguishes between the helical vortical waves propagating along and opposite to the direction of vorticity. The non-reciprocal behaviour for the HVW that emerges in the presence of a finite axial chemical potential has not been reported before in the literature.

Let us now consider the relation between the fluctuation amplitudes. Employing a high-temperature expansion similar to that in Eq.~\eqref{eq:largeT_vexp},
\begin{equation}
 \delta \mu^{h;\pm}_{\ell} = \delta \mu^{h;\pm}_{\ell;0} + T^{-1} \delta \mu^{h;\pm}_{\ell;1} + \dots,
\label{eq_delta_mu_T}
\end{equation}
we get a matrix equation for the fluctuations of the chemical potentials, which also involves the first-order corrections. To zeroth order, this equation reads as follows: 
\begin{equation}
 \frac{\omega_1^{h;\pm}}{3}
 \begin{pmatrix}
  \pm 1 & 0 & -1 \\
  0 & \pm 1 & 0\\
  -1 & 0 & \pm 1
 \end{pmatrix} 
 \begin{pmatrix}
  \delta \mu^{h;\pm}_{V; 0} \\ 
  \delta \mu^{h;\pm}_{A; 0} \\ 
  \delta \mu^{h;\pm}_{H; 0}
 \end{pmatrix} = 0,
\end{equation}
giving us
\begin{equation}
 \delta \mu^{h;\pm}_{A; 0} = 0, \qquad
 \delta \mu^{h;\pm}_{H; 0} = \pm \delta \mu^{h;\pm}_{V; 0}.
 \label{eq_HVW_mus_0}
\end{equation}
Therefore, in the leading order in the inverse temperature expansion, the helical vortical wave represents a coherent propagation of helical and vector charges as encoded in the relation~\eqref{eq_HVW_mus_0} between the fluctuations of the corresponding potentials. In the same order, the axial charge content of the helical vortical wave is vanishing. To be explicit, we take the fluctuation amplitude $\delta \mu^{h;\pm}_{V,0} \equiv \delta \mu^{h;\pm}_{V;0}$ of the vector chemical potential to be the reference amplitude scale. Therefore, we 
impose that its higher-order corrections vanish,
$\delta \mu^{h;\pm}_{V;i>0} = 0$.

Now, let us look at the next-to-the-leading order. We have the relation
\begin{equation}
 \mathbb{M}_1 
 \begin{pmatrix}
  \delta \mu^{h;\pm}_{V; 1} \\
  \delta \mu^{h;\pm}_{A; 1} \\
  \delta \mu^{h;\pm}_{H; 1} 
 \end{pmatrix} + \mathbb{M}_2
 \begin{pmatrix}
  1 \\
  0 \\
  \pm 1
 \end{pmatrix} \delta \mu^{h;\pm}_{V;0} = 0,
\end{equation}
which constrains the fluctuations of the chemical potentials as follows:
\begin{align}
 &\delta \mu^{h;\pm}_{H;1} = \pm \delta \mu^{h;\pm}_{V;1} = 0, 
 \nonumber\\&\delta \mu^{h;\pm}_{A;1} = -\frac{\mu_H \pm \mu_V}{7 \ln 2} \left[\frac{84 (\ln 2)^2}{\pi^2} - 1\right] 
 \delta \mu^{h;\pm}_{V;0}.
\end{align}
It can be seen that the first-order corrections to $\delta \mu_{V/H; 1}^{h;\pm}$ are absent, which is consistent with the interpretation that the amplitudes $\delta \mu_{V/H}^\pm$ represent the relevant scale for the induced amplitude of the axial chemical potential. 

Therefore, the HVW is a hydrodynamic excitation in vector and helical charges, with a slight admixture of the axial charge. In the HVW, the vector, axial, and helical chemical potentials are constrained as follows:
\begin{align}
 \delta \mu^{h;\pm}_{H} =& \pm \delta \mu^{h;\pm}_{V}, \qquad\qquad \text{[Helical Vortical Wave]}\nonumber\\
 \delta \mu^{h;\pm}_{A} =& - \frac{1}{7\ln 2} \left[\frac{84 (\ln 2)^2}{\pi^2} - 1\right] 
 \frac{\mu_H \pm \mu_V}{T} \delta \mu^{h;\pm}_{V} \nonumber\\&+ O(T^{-2}).
 \label{eq_HVW_chemicals}
\end{align}

\begin{figure}
\centering 
\begin{tabular}{c}
    \includegraphics[width=0.9\linewidth]{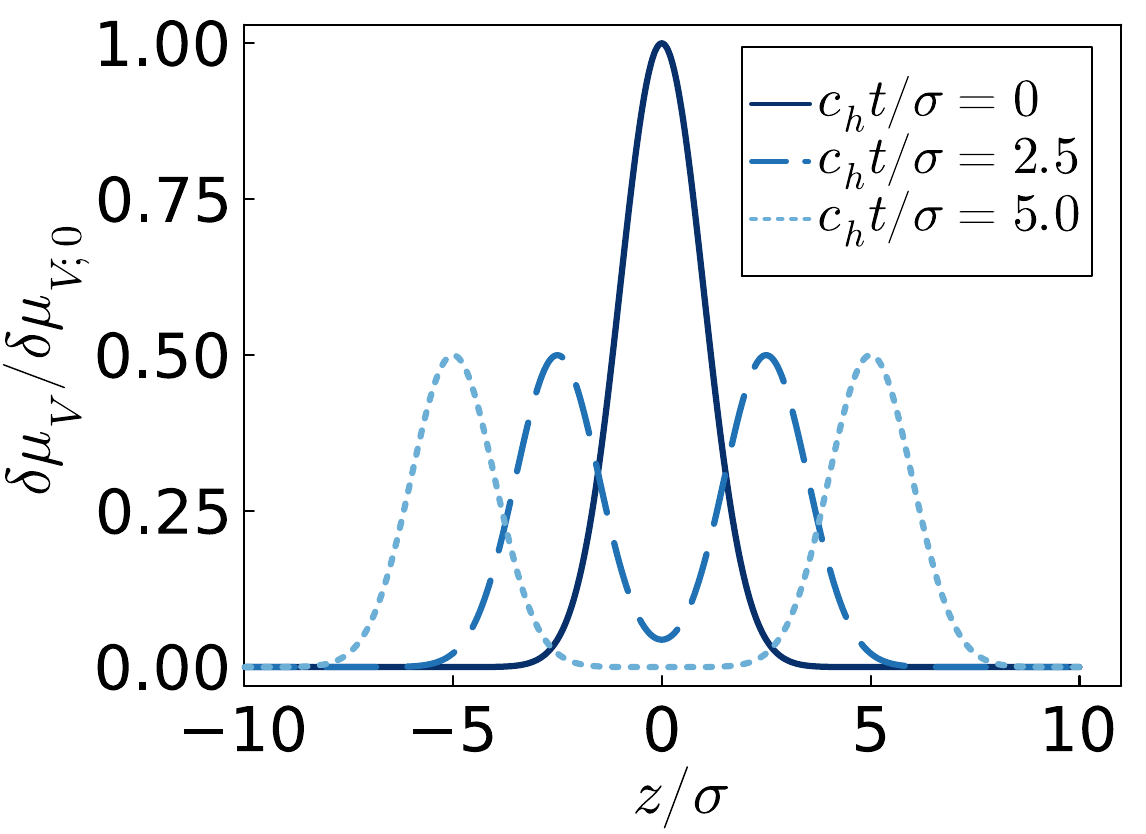}\\
    \includegraphics[width=0.9\linewidth]{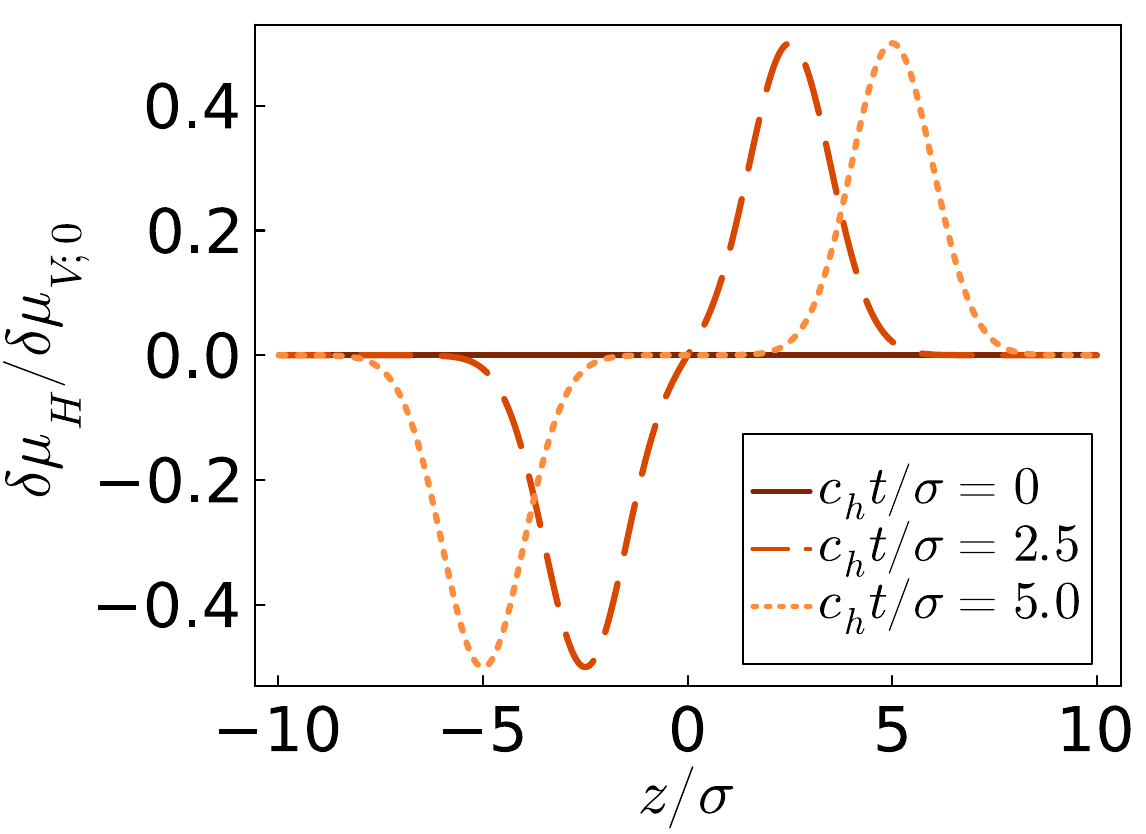}
\end{tabular}
    \caption{Time evolution of the charge perturbations (top) $\delta \bar{\mu}_V(t,z)$ and (bottom) $\delta \bar{\mu}_H(t, z)$, corresponding to the propagation of the helical vortical wave (HVW) through a charge-conserving plasma. The initial conditions are given in Eqs.~\eqref{eq:HVW_gauss_init}. The background state has $\mu_V = \mu_A = \mu_H = 0$, corresponding to the large temperature limit discussed in Sec.~\ref{sec:largeT}. 
    \label{fig:HVW}
    }
\end{figure}

\begin{figure}
\centering 
\begin{tabular}{c}
    \includegraphics[width=0.9\linewidth]{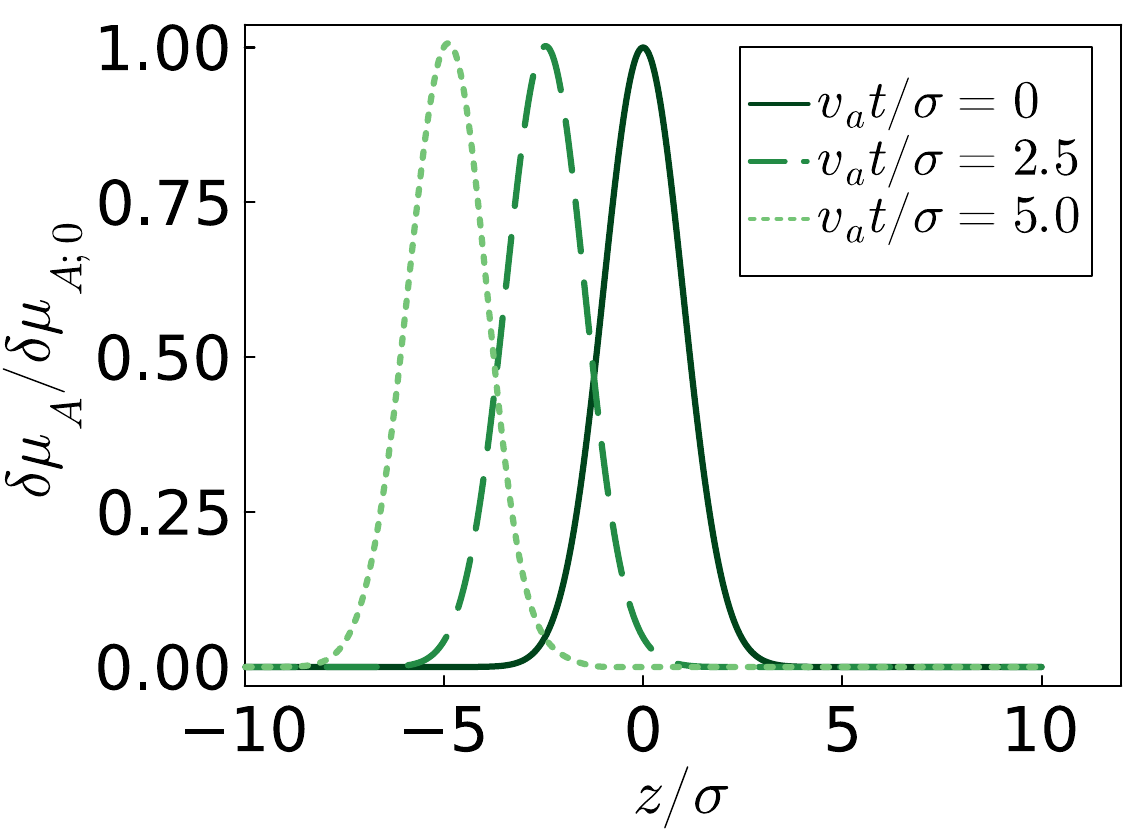}
\end{tabular}
    \caption{Time evolution of the axial charge perturbation $\delta \bar{\mu}_A(t,z)$,  corresponding to the propagation of the axial vortical wave (AVW) through a charge-conserving plasma. The initial conditions are shown in Eq.~\eqref{eq:AVW_gauss_init}. The background state has $\mu_V = \mu_H = 0$, and we took for definiteness $\mu_A / T = 0.1$, corresponding to the large temperature limit discussed in Sec.~\ref{sec:largeT}.
    \label{fig:AVW}
    }
\end{figure}

\subsection{Axial Vortical Wave}
\label{sec:largeT:AVW}

We now focus on the axial vortical wave, for which the zeroth- and first-order terms in the angular frequency vanish, $\omega_0^{\AVW} = \omega_1^\AVW = 0$. Thus, the angular frequency reads $\omega_\AVW = \omega_2^\AVW T^{-2} + O(T^{-3})$. In this case, we have the split
\begin{gather}
 \mathbb{M} = T^2 [ T^{-1} \mathbb{M}_1 + T^{-2} \mathbb{M}_2 + O(T^{-3})], \nonumber\\
 \mathbb{M}_1 = - \frac{2 k\Omega \ln2}{\pi^2} 
 \begin{pmatrix} 
  0 & 0 & 1 \\ 0 & 0 & 0 \\ 1 & 0 & 0
 \end{pmatrix}, \nonumber\\
 \mathbb{M}_2 = 
 \frac{\omega^\AVW_2}{3} 
 \mathbb{I} - \frac{2 k\Omega}{7\pi^2} 
 \begin{pmatrix}
  \mu_A & \mu_V & 0 \\ 
  \mu_V & -4\mu_A & \mu_H \\
  0 & \mu_H & \mu_A
 \end{pmatrix}.
\end{gather}
At the level of the fluctuation amplitudes, the structure of $\mathbb{M}_0$ mandates that 
\begin{equation}
 \delta \mu_{V;0}^\AVW = \delta \mu_{H;0}^\AVW = 0,
\end{equation}
while $\delta \mu_{A;0}^\AVW$ remains arbitrary, and it is the relevant scale for the fluctuations in this axial wave. For the next order, we have
\begin{align}
 &  \left[\frac{\omega_2^\AVW}{3} 
 \begin{pmatrix}
  1 & 0 & 0 \\ 0 & 1 & 0 \\ 0 & 0 & 1
 \end{pmatrix} - \frac{2k \Omega}{7\pi^2} 
 \begin{pmatrix}
  \mu_A & \mu_V & 0 \\ 
  \mu_V & -4\mu_A & \mu_H \\
  0 & \mu_H & \mu_A
 \end{pmatrix}\right]
 \begin{pmatrix}
  0 \\
  \delta \mu_{A;0}^\AVW \\
  0
 \end{pmatrix}  \nonumber\\&  -\frac{2 k\Omega \ln2}{\pi^2} 
 \begin{pmatrix} 
  0 & 0 & 1 \\ 0 & 0 & 0 \\ 1 & 0 & 0
 \end{pmatrix}
 \begin{pmatrix}
  \delta \mu_{V;1}^\AVW \\
  \delta \mu_{A;1}^\AVW \\
  \delta \mu_{H;1}^\AVW 
 \end{pmatrix} = 0.
\end{align}
The equation corresponding to the second line (which contains $\delta \mu_{A;1}^\AVW$) can 
be satisfied only if the leading-order contribution to the velocity is given by
\begin{equation}
 \omega_2^\AVW = -\frac{24 \mu_A k\Omega}{7\pi^2}.\label{eq:VAH_largeT_va12}
\end{equation}
The amplitude $\delta \mu_{A;1}^\AVW$ remains unconstrained (as it should be), and without loss of generality, we set it to $0$. The amplitudes corresponding to the vector and helical chemical potentials satisfy
\begin{equation}
 \delta \mu_{V/H;1}^\AVW = -\frac{\mu_{H/V}}{7\ln 2} \delta \mu_{A;0}^\AVW.
 \label{eq:largeT:AVW}
\end{equation}

Restoring the inverse powers of temperature in the above equations, we arrive at the following physical picture of the Axial Vortical Wave. In the leading order, this hydrodynamic excitation is dominated by the axial chemical potential fluctuations (hence the name) $\delta \mu_A$, which is accompanied by the induced vector and helical charge densities: 
\begin{align}
 &\delta \mu^\AVW_{V/H} = -\frac{\mu_{H/V}}{7T \ln 2} \delta \mu^\AVW_{A} + O(T^{-2}),   \nonumber\\& {\rm [Axial\ Vortical\ Wave]}\,.
 \label{eq_AVW_mu}
\end{align}
In other words, the leading role in this hydrodynamic excitation is taken by the axial charge density. In the presence of the vector (helical) chemical potential, the helical (vector) chemical potential fluctuates in coherence with the axial chemical potential according to Eq.~\eqref{eq_AVW_mu}.

The axial vortical wave slowly propagates along the vorticity vector with the speed
\begin{align}
    &v_\AVW \equiv \frac{\omega_\AVW}{k} = -\frac{24}{7\pi^2} \frac{\mu_A \Omega}{T^2} + O(T^{-3}),
       \nonumber\\& \text{[Axial Vortical Wave]}\,.
    \label{eq_AVW_v}
\end{align}
The Axial Vortical Wave has three notable features. First, it propagates only in the presence of the axial background charge characterized by a nonzero axial chemical potential, $\mu_A \neq 0$. Second, the propagation has a strictly uni-directional nature~\eqref{eq_AVW_v}: depending on the sign of the axial chemical potential, the wave propagates in the direction along (opposite to) the vorticity vector $\boldsymbol{\Omega}$ for $\mu_A < 0$ ($\mu_A > 0$). Third, the Axial Vortical Wave does not generate oscillations in vector and helical chemical potentials if their background values are vanishing, $\mu_V = \mu_H = 0$. In this case, the wave propagates as an oscillation in the axial chemical potential only. 

Our results for the wave spectrum differ from those obtained previously in the literature.  Refs.~\cite{Jiang:2015cva,Kalaydzhyan:2016dyr,Gorbar:2017toh} reported two propagating modes, while we naturally uncover three modes. In Appendix~\ref{app:AVW}, we explicitly compare our results to the study of Ref.~\cite{Gorbar:2017toh} and show that the axial vortical wave velocity \eqref{eq_AVW_v} is the same in both cases, while the HVW has no analogue in the aforementioned works. 

\subsection{Propagation properties}\label{sec:largeT:solutions}

We now consider two concrete examples to illustrate the properties of the HVW and AVW, respectively. In the first case, we take the following initial conditions:
\begin{align}
 &\bar{\mu}_V(t = 0, z) = \mu_V + \delta \bar{\mu}_{V;0} \cos(k z), \nonumber\\& 
 \bar{\mu}_A(t = 0, z) = \bar{\mu}_H(t = 0, z) = 0,
 \label{eq:HVW_init}
\end{align}
where we used the overhead bar $\bar{\mu}_V$ to denote time-dependent quantities. Background values are denoted without an overhead bar (only the vector chemical potential has a non-vanishing background value in the above). We then have three equations for the mode amplitudes,
\begin{align}
 \delta \mu_V^{h;+} + \delta \mu_V^{h;-} + \delta \mu_V^\AVW &= \delta \bar{\mu}_{V;0}, \nonumber\\
 \delta \mu_A^{h;+} + \delta \mu_A^{h;-} + \delta \mu_A^\AVW &= 0, \nonumber\\
 \delta \mu_H^{h;+} + \delta \mu_H^{h;-} + \delta \mu_H^\AVW &= 0.
\end{align}
Using Eq.~\eqref{eq:largeT:AVW}, it can be seen that $\delta \mu_V^\AVW = 0$. Employing the last two of the above relations, it can be shown that $\delta \mu_A^\AVW = 0$, which leads to the solution
\begin{equation}
 \delta \mu_V^{h;+} = \delta \mu_V^{h;-} = \frac{1}{2} \delta \bar{\mu}_{V;0}, 
\end{equation}
while $\delta \mu_H^{h;\pm} = \pm \delta \mu_V^{h;\pm}$ and $\delta \mu_A^{h;\pm} = 0$, with $\delta \mu_V^\AVW = \delta \mu_A^\AVW = \delta \mu_H^\AVW = 0$. The full solution describes the standing helical vortical wave in an unpolarized plasma,
\begin{align}
 &\bar{\mu}_V(t, z) = \mu_V + \delta \bar{\mu}_{V;0} \cos(k c_h t) \cos(k z),\nonumber\\&
 \bar{\mu}_H(t,z) = \delta \bar{\mu}_{V;0} \sin(k c_h t) \sin(k z),
 \label{eq:HVW_ampl}
\end{align}
while $\bar{\mu}_A(t,z) = 0$. 

We now consider the Gaussian example suggested in Eq.~\eqref{eq:gauss_F}. We take 
\begin{align}
 &\bar{\mu}_V(0,z) = \mu_V + \delta \bar{\mu}_{V;0}\, e^{-z^2 / 2\sigma^2}, \nonumber\\& 
 \bar{\mu}_A(0,z) = \bar{\mu}_H(0,z) = 0.
 \label{eq:HVW_gauss_init}
\end{align}
Applying Eq.~\eqref{eq:gauss_fk}, the mode amplitudes can be found as
\begin{equation}
 \delta \mu_V^{h;+}(k) = \delta \mu_V^{h;-}(k) = \frac{\sigma\, \delta \bar{\mu}_{V;0}}{2\sqrt{2\pi}} e^{-\sigma^2 k^2 / 2}, 
\end{equation}
while $\delta \mu_H^{h;\pm}(k) = \pm \delta \mu_V^{h;\pm}(k)$. Reconstituting the time-dependent profile using Eq.~\eqref{eq:gauss_inv}, we find
\begin{align}
 \delta \bar{\mu}_V(t, z) = \frac{\delta \bar{\mu}_{V;0}}{2} \left[e^{-(z - c_h t)^2 / 2\sigma^2} + e^{-(z + c_h t)^2 / 2\sigma^2}\right],\nonumber\\
 \delta \bar{\mu}_H(t, z) = \frac{\delta \bar{\mu}_{V;0}}{2} \left[e^{-(z - c_h t)^2 / 2\sigma^2} - e^{-(z + c_h t)^2 / 2\sigma^2}\right].
 \label{eq:HVW_gauss}
\end{align}
The HVW splits the initial Gaussian in two lumps travelling along and opposite to the vorticity vector. The excess vector charge is carried symmetrically in both directions. The lump travelling in the direction of the vorticity vector has equal vector and helicity charges, while the lump travelling opposite to the vorticity vector presents vector and helicity charges of equal magnitude but opposite sign. Thus, the HVW generates a local helicity imbalance that propagates out towards the system edges. 
This behaviour is illustrated in Fig.~\ref{fig:HVW}.

To illustrate the properties of the axial vortical wave, we set the initial conditions 
\begin{align}
 &\bar{\mu}_A(0, z) = \mu_A + \delta \bar{\mu}_{A;0} \cos(kz), \nonumber\\&
 \bar{\mu}_H(0, z) = -\frac{\mu_V}{7 T \ln 2} \delta \bar{\mu}_{A;0} \cos(k z),
\end{align}
while $\bar{\mu}_V(0, z) = \mu_V$.
Then, the HVW modes vanish, $\delta \mu_\ell^{h;\pm} = 0$, for all $\ell \in \{V, A, H\}$. Moreover, $\delta \mu_V^\AVW = 0$, while $\delta \mu_H^\AVW = -\mu_V \delta \bar{\mu}_{A;0} / (7 T \ln 2)$. The full solution reads
\begin{align}
 &\bar{\mu}_A(t, z) = \mu_A + \delta \bar{\mu}_{A;0} \cos(\omega_\AVW t - kz), \nonumber\\&
 \bar{\mu}_H(t, z) = -\frac{\mu_V \delta \bar{\mu}_{A;0}}{7 T \ln 2} \cos(\omega_\AVW t - kz),
 \label{eq:AVW_ampl}
\end{align}
where $\omega_{\AVW} = k v_\AVW \simeq -24 \mu_A k\Omega / (7 \pi^2 T^2)$. 
Contrary to the solution in Eq.~\eqref{eq:HVW_ampl}, 
the AVW is not a standing wave. It is rather a propagating solution, transferring axial charge in the direction opposite (parallel) to that of the vorticity for a positive (negative) background axial chemical potential. 
This feature is better seen when considering the Gaussian example with the following initial configuration:
\begin{align}
 \bar{\mu}_V(0, z) &= \mu_V, \quad 
 \bar{\mu}_A(0, z) = \mu_A + \delta \bar{\mu}_A\, e^{-z^2 / 2\sigma^2}, \nonumber\\
 \bar{\mu}_H(0, z) &= \delta \bar{\mu}_H(0, z) = -\frac{\mu_V}{7 T \ln 2} \bar{\mu}_A(0, z).\label{eq:AVW_gauss_init}
\end{align}
Since only the axial mode is excited by these initial conditions, the solution trivially reads 
\begin{align}
 \bar{\mu}_A(t, z) = \mu_A + \delta \bar{\mu}_A\, e^{-(z - v_\AVW t)^2 / 2\sigma^2},
 \label{eq:AVW_gauss}
\end{align}
while $\bar{\mu}_V(t, z) = \mu_V$ and $\bar{\mu}_H(t,z) = -\mu_V \delta \bar{\mu}_A(t, z) / (7 T \ln 2)$. The above solution shows clearly that the AVW will lead to uni-directional transport of the excess axial charge in the direction opposite to $\mu_A \boldsymbol{\Omega}$. 
This behaviour is illustrated in Fig.~\ref{fig:AVW}.

\subsection{Symmetries of the helical and axial vortical waves}\label{sec:largeT:symmetries}

Qualitatively, the emergence of the non-reciprocal propagation effects can be understood on the basis of discrete $\mathcal{CPT}$ symmetries of the quantities involved in the process (summarized in Table~\ref{tbl_symmetries}). For a slowly rotating fluid of massless fermions, the leading contribution to the velocity of a hydrodynamic wave excitation in a rotating fluid should be linearly proportional to the magnitude of the corresponding angular velocity, $v \propto \Omega$ (here, we remove all indices for simplicity). One can also notice the collinearity of these vectors ${\boldsymbol{v}} \lvert \lvert  \boldsymbol{\Omega}$ following from the spatial symmetries of the system. One gets for the dispersion relation the following generic expression: 
\begin{align}
 \omega = C(T,\mu_V,\mu_A,\mu_H) \boldsymbol{k} \cdot {\boldsymbol{\Omega}}, 
    \label{eq_v_Omega}
\end{align}
where $C$ is a function of all parameters of the system.

Consider first a neutral fluid with vanishing chemical potentials, $\vec{\mu} = 0$. Since temperature $T$ is the only dimensionful parameter in this case, one has $C \propto 1/T$ for dimensional reasons. The latter statement also implies that $C$ is a $\mathcal P$- and $\mathcal T$-even quantity in a neutral fluid. 

The $\mathcal{T}$ symmetries of the two sides of Eq.~\eqref{eq_v_Omega} are different: while the wave velocity $v$ is a $\mathcal P$-odd quantity, the angular velocity $\Omega$ is $\mathcal{P}$-even ({\it c.f.} Table~\ref{tbl_symmetries}). Equation~\eqref{eq_v_Omega} is preserved under the time reversal $\mathcal T$ and the charge conjugation $\mathcal C$ transformations, while it changes the sign under the parity inversion $\mathcal P$.

Now we notice that in the discussed system, all chemical potentials vanish, thus implying that its $\mathcal P$ (and $\mathcal T$) symmetries are unbroken. Therefore, a $\mathcal P$ transformation applied to the system should give us a system with identical properties. The latter statement is formally inconsistent with the $\mathcal P$-property of Eq.~\eqref{eq_v_Omega} as the parity transform ---corresponding to the inversion of all spatial coordinates--- flips the sign of this equation. Therefore, the wave is either absent ($C = 0$), or there are two identical waves, with $C \to \pm \lvert C \rvert$, propagating in opposite directions with the same velocities. These branches of Eq.~\eqref{eq_v_Omega} are then mapped to each other by the $\mathcal P$ transformation, and the system maintains the invariance under parity transformation. The latter property, indeed, is realized in our case of the Helical Vortical Waves, indicating the reciprocity of the hydrodynamic spectrum of the neutral fluid.

Can the generic law~\eqref{eq_v_Omega} describe non-reciprocal waves? To figure this out, let us consider the system with all three chemical potentials non-vanishing. One can write for the proportionality function~\eqref{eq_v_Omega}
\begin{align}
   & C(T,\vec\mu) = T c_T + \mu_V c_V + \mu_A c_A + \mu_H c_H\nonumber\\& + \mu_V \mu_A c_{VA} + \mu_V \mu_H c_{VH} + \mu_A \mu_H c_{AH}\,,
    \label{eq_v_Omega_gen}
\end{align}
where $c_T$, $c_\ell$ and $c_{\ell,\ell'}$ with $\ell,\ell' = V,A,H$ are $\mathcal{CPT}$-invariant functions of the temperature $T$ and the chemical potentials $\vec\mu = (\mu_V,\mu_A,\mu_H)$.

We require that the $\mathcal C$ symmetry of the system should be unbroken to be consistent with the anticipations in quantum field theory. In our case, this statement means that the spectrum of the hydrodynamic excitations in the system with particles and the identical system made of anti-particles should be the same. Since $\mu_V$ and $\mu_H$ are $\mathcal C$-odd quantities the corresponding coefficients must vanish in Eq.~\eqref{eq_v_Omega_gen}: $c_V = c_H = c_{VA} = c_{AH} = 0$. Therefore, we are only left with two terms in the generic expression for the group velocity:
\begin{align}
    {\boldsymbol{v}} = 
   \frac{d\omega}{d\boldsymbol{k}} =
    c_T T {\boldsymbol{\Omega}} + c_A \mu_A {\boldsymbol{\Omega}} + c_{VH} \mu_V \mu_H {\boldsymbol{\Omega}} \,,
    \label{eq_v_Omega_gen_2}
\end{align}
where we remind that $c_T$, $c_A$ and $c_{VH}$ are $\mathcal{CPT}$-invariant quantities.

As we discussed above, the first term in Eq.~\eqref{eq_v_Omega_gen_2} breaks parity inversion symmetry for the hydrodynamic wave, leading to the existence of two identical counter-propagating branches that preserve the reciprocity of the system. The second and the third terms do not break any symmetry since the combinations $\mu_A {\boldsymbol{\Omega}}$ and $\mu_V \mu_H {\boldsymbol{\Omega}}$ have the same $\mathcal{CPT}$ properties as the velocity ${\boldsymbol{v}}$ ($\mu_A$ and the product $\mu_V \mu_H$ share identical $\mathcal{CPT}$ symmetries). Thus, the last two terms describe a wave excitation that can have no reciprocal partner.

Thus, the non-reciprocity effects appear in the presence of the finite axial chemical potential (for a chirally imbalanced fluid with $\mu_A \neq 0$), as well as in the helically imbalanced $(\mu_H \neq 0)$ dense $(\mu_V \neq 0)$ fluid. In the lowest order in chemical potentials, the second term in Eq.~\eqref{eq_v_Omega_gen_2} provides a leading contribution to the non-reciprocal effects. This term enters the velocities of the helical~\eqref{eq_HVW_speed} and axial~\eqref{eq_AVW_v} vortical waves.

We conclude that in a chirally imbalanced rotating fluid, the non-reciprocal gapless waves can appear in the hydrodynamic spectrum.

\section{Unpolarized Plasma}\label{sec:unpolarized}

We now consider the limit when the background state is unpolarized, such that $\mu_A = \mu_H = 0$. This limit is particularly relevant in realistic plasmas, when the axial and helical charge conservation is broken by interactions (see companion paper \cite{Morales-Tejera:2024mtx}). 
We refer to this case as the \textit{unpolarized} plasma.

Setting $\mu_A = \mu_H = 0$ in Eq.~\eqref{eq:Pcl} leads to
\begin{align}
 P &= -\frac{T^4}{\pi^2} \sum_{\sigma,\lambda} {\rm Li}_4(-e^{\sigma \mu_V / T}) \nonumber\\
 &= \frac{7 \pi^2 T^4}{180} + \frac{\mu_V^2 T^2}{6} + \frac{\mu_V^4}{12\pi^2},
\end{align}
where we used the following property of the polylogarithm:
\begin{equation}
 {\rm Li}_4(-e^\alpha) + {\rm Li}_4(-e^{-\alpha}) = 
 -\frac{7\pi^4}{360} - \frac{\pi^2 \alpha^2}{12} - \frac{\alpha^4}{24}.
 \label{eq:Li4sum}
\end{equation}
The charge densities $Q_\ell$ can be computed by differentiating the pressure $P$ in Eq.~\eqref{eq:Pcl} with respect to the corresponding chemical potential $\mu_\ell$, as shown in Eq.~\eqref{eq:thermocl}, using the property
\begin{equation}
 \frac{\partial}{\partial \mu_\ell} {\rm Li}_n(-e^{\mu_{\sigma,\lambda}/T}) = \frac{q^\ell_{\sigma,\lambda}}{T} {\rm Li}_{n-1}(-e^{\mu_{\sigma,\lambda} / T}).
\end{equation}
This leads to
\begin{align}
 &Q_\ell = -\frac{T^3}{\pi^2} \sum_{\sigma,\lambda} q_{\sigma,\lambda}^\ell {\rm Li}_3(-e^{\sigma \mu_V / T}) 
 \nonumber\\&= \delta_{\ell, V} \left(\frac{\mu_V T^2}{3} + \frac{\mu_V^3}{3\pi^2}\right),
 \label{eq:unpol_Ql}
\end{align}
where we have used the properties $\sum_{\sigma,\lambda} q_{\sigma,\lambda} = 0$ (thus, $Q_A$ and $Q_H$ vanish), as well as 
\begin{equation}
 {\rm Li}_3(-e^\alpha) - {\rm Li}_3(-e^{-\alpha}) = -\frac{\pi^2 \alpha}{6} - \frac{\alpha^3}{6}.\label{eq:Li3sum}
\end{equation}

Since $Q_A = 0$, the Landau-frame vortical conductivities $\sigma^\omega_{\ell} = \sigma^\omega_{\ell;\beta} - Q_A Q_\ell / (E + P)$ agree with the $\beta$-frame ones, $\sigma^\omega_{\ell} = \sigma^\omega_{\ell;\beta} = \frac{1}{2} \partial^2 P / \partial \mu_A \partial \mu_\ell$.
The second derivatives of the pressure with respect to the chemical potentials can be evaluated as
\begin{align}
 &\frac{\partial^2 P}{\partial \mu_\ell \partial \mu_{\ell'}} = \frac{\partial Q_\ell}{\partial \mu_{\ell'}} = -\frac{T^2}{\pi^2} \sum_{\sigma,\lambda} q^\ell_{\sigma,\lambda} q^{\ell'}_{\sigma,\lambda} {\rm Li}_2(-e^{\mu_{\sigma,\lambda}/T}) \nonumber\\&= 2  
 \begin{pmatrix}
  \sigma^\omega_A & 0 & 0 \\
  0 & \sigma^\omega_A & \sigma^\omega_H \\
  0 & \sigma^\omega_H & \sigma^\omega_A
 \end{pmatrix},
\end{align}
where we used Eq.~\eqref{eq:charge_products} and $q_\ell^2 = 1$ to simplify the product $q^\ell_{\sigma,\lambda} q^{\ell'}_{\sigma,\lambda}$ of two charges. The terms $\partial^2P / \partial \mu_V \partial \mu_A$ and $\partial^2P / \partial \mu_V \partial \mu_H$ involve the products $q^V_{\sigma,\lambda} q^A_{\sigma,\lambda} = q^H_{\sigma,\lambda} = 2\lambda \sigma$ and $q^V_{\sigma,\lambda} q^H_{\sigma,\lambda} = q^A_{\sigma,\lambda} = 2\lambda$, which vanish under the summation with respect to $\lambada$. The axial conductivity is given by
\begin{equation}
 \sigma_A^\omega = -\frac{T^2}{2\pi^2} \sum_{\sigma,\lambda} {\rm Li}_2(-e^{\sigma \mu_V/T}) = \frac{T^2}{6} + \frac{\mu_V^2}{2\pi^2},
\end{equation}
where we used the following property of the polylogarithm function:
\begin{equation}
 {\rm Li}_2(-e^\alpha) + {\rm Li}_2(-e^{-\alpha}) = -\frac{\pi^2}{6} - \frac{\alpha^2}{2}.
\end{equation}
The helical vortical conductivity can be expressed as
\begin{equation}
 \sigma_H^\omega = \frac{T^2}{\pi^2} [{\rm Li}_2(-e^{-\mu_V/T}) - {\rm Li}_2(-e^{\mu_V/T})],
\end{equation}
and reduces for small and large $\lvert \mu_V / T \rvert$ to:
\begin{align}
 \sigma_H^\omega(\lvert \mu_V \rvert \ll T) &= \frac{2 \ln 2}{\pi^2} T \mu_V + O(T^{-1}), \nonumber\\
 \sigma_H^\omega(\lvert \mu_V \rvert \gg T)&\simeq {\rm sgn}(\mu_V) \left[\sigma_A^\omega - \frac{2T^2}{\pi^2} e^{-\lvert \mu_V \rvert/T}\right. \nonumber\\&\left. + O(e^{-2\lvert \mu_V \rvert / T})\right].
\end{align}

Finally, we need to consider the derivatives of the vortical charge conductivities $\sigma^\omega_\ell = \sigma^\omega_{\ell;\beta} - Q_A Q_\ell / (E + P)$ with respect to the chemical potential. Here, the second term makes a non-vanishing contribution since $\partial Q_A / \partial \mu_{\ell'} = 2 \sigma^\omega_{\ell'}$:
\begin{equation}
 \frac{\partial}{\partial \mu_{\ell'}}\left(\frac{Q_A Q_\ell}{E + P}\right)_{\mu_A = \mu_H = 0} = \frac{2 Q_V \sigma^\omega_{\ell'}}{E + P} \delta_{\ell V}.
\end{equation}
The derivative of the $\beta$-frame vortical conductivity evaluates to
\begin{equation}
 \frac{\partial \sigma^\omega_{\ell;\beta}}{\partial \mu_{\ell'}} = \frac{1}{2} \frac{\partial^3P}{\partial\mu_{\ell'} \partial \mu_A \partial \mu_\ell} = \frac{1}{\pi^2} 
 \begin{pmatrix}
  0 & \mu_V & T L \\
  \mu_V & 0 & 0 \\
  T L & 0 & 0
 \end{pmatrix},
\end{equation}
where we introduced the notation $L$ based on the relations
\begin{gather}
 {\rm Li}_1(-e^{\mu_V / T}) - 
 {\rm Li}_1(-e^{-\mu_V / T}) = -\frac{\mu_V}{T}, \nonumber\\
 L = -\frac{1}{T} \left[{\rm Li}_1(-e^{\mu_V / T}) + 
 {\rm Li}_1(-e^{-\mu_V / T})\right] \nonumber\\= 2\ln\left(2\cosh \frac{\mu_V}{2T}\right)\,,
\end{gather}
taking into account that ${\rm Li}_1(-e^\alpha) = -\ln(1 + e^\alpha)$. As with $\sigma^\omega_H$, the degenerate limit for $L$ involves an exponentially-decaying function. For convenience, we list below both the high-temperature and the large-chemical potential limits:
\begin{align}
 L (T \gg \lvert \mu_V \rvert) &= 2\ln 2 + \frac{\mu_V^2}{4T^2} - \frac{\mu_V^4}{96T^4} + O \left(\frac{\mu_V^6}{T^6}\right),\nonumber\\
 L(\lvert \mu_V \rvert \gg T) &= \frac{\lvert \mu_V \rvert}{T} + 2e^{-\lvert \mu_V \rvert / T} + \dots,
\end{align}
where we suppressed terms that decay faster than $e^{-\lvert \mu_V \rvert / T}$.
We thus conclude that 
\begin{equation}
 \frac{\partial \sigma^\omega_\ell}{\partial \mu_{\ell'}} =  -\delta_{\ell V} \frac{2 Q_V \sigma^\omega_{\ell'}}{E + P} + 
 \frac{1}{\pi^2} 
 \begin{pmatrix}
  0 & \mu_V & T L \\
  \mu_V & 0 & 0 \\
  T L & 0 & 0
 \end{pmatrix}\,.
\end{equation}

Once again, we have to find the non-trivial solutions of the system \eqref{eq:M_aux}. As in the previous subsection, let us define the dimensionless chemical potential $\alpha_V = \mu_V/T$. Then the matrix $\mathbb{M}$ has the following structure:
\begin{equation}
 \frac{1}{T^2}\mathbb{M} = \omega \mathbb{M}_\omega (\alpha_V) - \frac{k\Omega}{T} \mathbb{M}_\Omega (\alpha_V)\,,
\end{equation}
with $\mathbb{M}_\omega$ and $\mathbb{M}_\Omega$ given by
\begin{align}
 \mathbb{M}_\omega &= \frac{2}{T^2} 
 \begin{pmatrix}
  \sigma^\omega_A - \frac{T^2}{3} \Delta H & 0 & 0 \\
  0 & \sigma^\omega_A & \sigma^\omega_H \\ 
  0 & \sigma^\omega_H & \sigma^\omega_A
 \end{pmatrix}, 
\nonumber\\ \mathbb{M}_\Omega &= 
 \begin{pmatrix}
  0 & \frac{1}{H} A & \frac{1}{H} B\\
  A & 0 & 0 \\ 
  B & 0 & 0
 \end{pmatrix},
 \label{eq:unpol_M}
\end{align}
where $\Delta H = H - 1$ with $H = (E + P) / sT = 1 + \mu_V Q_V / (sT)$, while $A$ and $B$ are defined as
\begin{equation}
 A = \frac{\alpha_V}{\pi^2} - \frac{Q_V}{3 s}, \qquad 
 B = \frac{HL}{\pi^2} - \frac{2 Q_V}{s T^2}  \sigma^\omega_H.
 \label{eq:unpol_AB}
\end{equation}

Since ${\rm det}(\mathbb{M}_\Omega) = 0$, one can conclude that ${\rm det}(\mathbb{M}) = 0$ has $\omega_0 = 0$ as a solution. This non-propagating mode is a fluctuation in a purely axial-helical sector, i.e. $\delta \mu^0_V = 0$, with the axial and helical fluctuations being related as follows:
\begin{equation}
 A \delta \mu^0_A + B \delta \mu^0_H = 0.
\end{equation}

The other two modes correspond to the helical vortical wave. Their angular frequencies satisfy the following equation:
\begin{align}
 &\omega_h^2 \left(\frac{2}{T^2}\right)^2 \left(\sigma^\omega_A - \frac{T^2}{3} \Delta H \right) [(\sigma^\omega_A)^2 - (\sigma^\omega_H)^2]\nonumber\\& -
 \frac{k^2 \Omega^2}{H T^2} [\sigma_A^\omega (A^2 + B^2) - 2 A B \sigma^\omega_H] = 0.
 \label{eq:unpol_detM}
\end{align}
There are two solution branches:
\begin{equation} \label{eq:vhpm}
 \omega_h^\pm = \pm \frac{k \Omega T}{2} \sqrt{\frac{\sigma^\omega_A(A^2 + B^2) - 2 A B \sigma^\omega_H}
 {H(\sigma^\omega_A - \frac{T^2}{3} \Delta H)[(\sigma^\omega_A)^2 - (\sigma^\omega_H)^2]}}.
\end{equation}
The above solution is completely analytical, allowing the large temperature and degenerate limits to be taken explicitly. At high temperature, we have
\begin{gather}
 \sigma_A^\omega = \frac{T^2}{6} + \frac{\mu_V^2}{2\pi^2}, \qquad 
 \sigma^\omega_H = \frac{2\ln 2}{\pi^2} T^2 \alpha_V + O(\alpha_V^3), \nonumber\\
 H = 1 + \frac{15 \alpha_V^2}{7\pi^2} + O(\alpha_V^4), \qquad
 L = 2 \ln 2 + O(\alpha_V^2), \quad \nonumber\\
 A = \frac{2\alpha_V}{7\pi^2} + O(\alpha_V^3),\nonumber\\
 B = \frac{2\ln 2}{\pi^2} + \frac{7\pi^2 - 120 \ln 2}{28\pi^4} \alpha_V^2 + O(\alpha_V^4).
 \label{eq:unpol_largeT}
\end{gather}
It can be seen that the axial vortical conductivity $\sigma_A$ dominates over its helical counterpart $\sigma_H$, such that to leading order, $\omega_h^\pm \simeq \pm (k \Omega T / 2) \times (B/\sigma_A) \simeq \pm k c_h$, with $c_h$ being the speed of the helical vortical wave given in Eq.~\eqref{eq_v_HVW_leading}. Taking into account terms up to next-to-leading order, we arrive at the following expression for the helical vortical wave:
\begin{align}
 &\omega_h^\pm(T \gg \lvert \mu_V \rvert) \nonumber\\&= \pm \frac{k\Omega}{T}\left[\frac{6\ln 2}{\pi^2} - 0.0124 \frac{\mu_V^2}{T^2} + O(\mu_V^4 / T^4)\right].
 \label{eq:unpol_vH_largeT}
\end{align}
In the limit of vanishing chemical potentials for the neutral plasma, Eq.~\eqref{eq:unpol_vH_largeT} coincides with Eq.~\eqref{eq_HVW_speed}.

\begin{figure}
    \centering
    \includegraphics[scale=0.35]{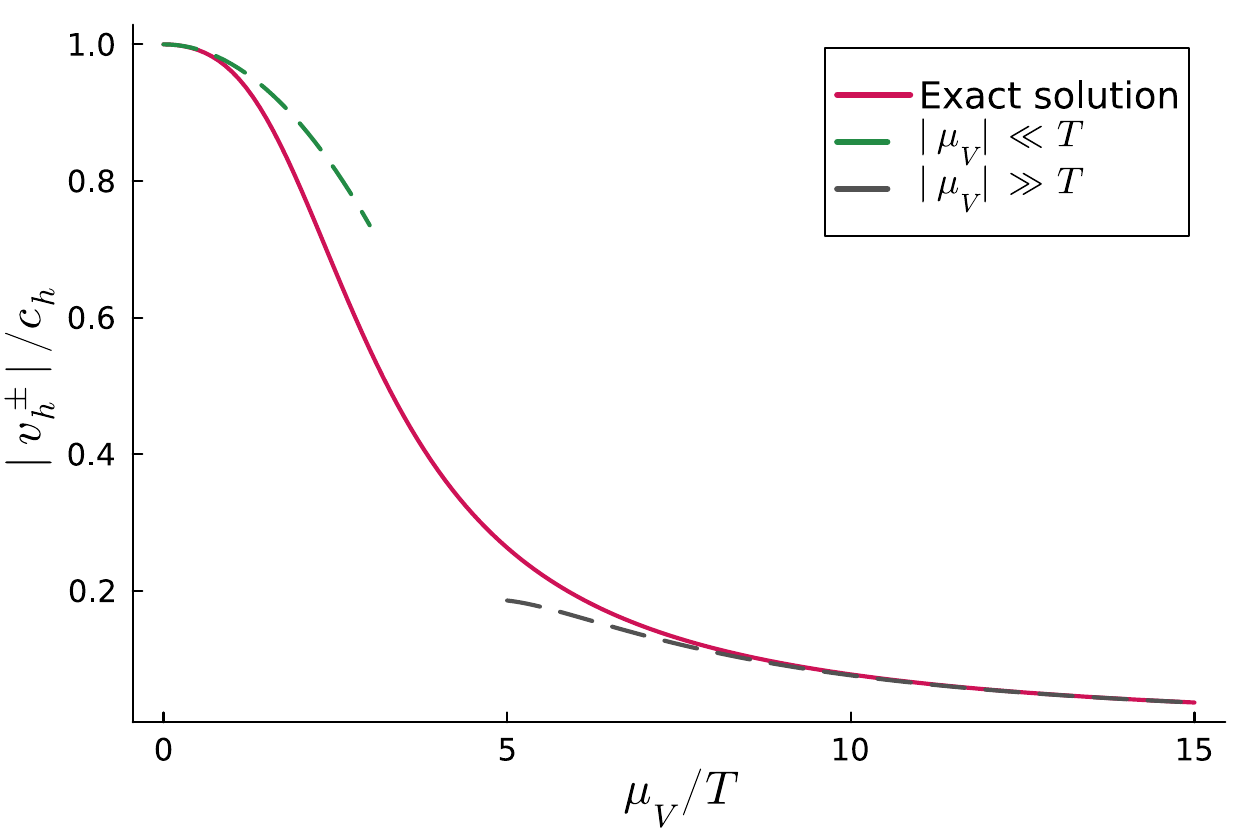}
    \caption{Phase velocity $v_h^\pm = \omega_h^\pm / k$ of the helical vortical wave as a function of $\mu_V/T$, for the unpolarized plasma in which the background axial and helical densities are vanishing ($\mu_A = \mu_H = 0$). The result is normalized to the speed of the HVW in the neutral plasma, $c_h = \dfrac{6\ln 2 }{\pi^2}\dfrac{\Omega}{T}\simeq 0.42 \dfrac{\Omega}{T}$.
    }
    \label{fig:vHVW}
\end{figure}

In the opposite limit of large chemical potential, we have $\sigma^\omega_A - s_V \sigma^\omega_H = (2T^2 / \pi^2) e^{-\lvert \mu_V \rvert/T}$, where we abbreviated $s_V ={\rm sgn}(\mu_V)$, and thus:
\begin{equation}
 (\sigma_A^\omega)^2 - (\sigma_H^\omega)^2 = \frac{4 T^2}{\pi^2} \sigma^\omega_A e^{-\lvert \alpha_V \rvert}.
\end{equation}
The numerator of the fraction appearing under the square root in Eq.~\eqref{eq:vhpm} can be put in the form
\begin{align}
 \sigma^\omega_A(A^2 + B^2) - 2AB \sigma^\omega_H & = \sigma^\omega_A(A - s_V B)^2 \\
 &+ \frac{4s_V T^2}{\pi^2}  A B e^{-\lvert \mu_V \rvert / T}.\nonumber
\end{align}
The first term of the right-hand side of the above equation becomes subleading due to the asymptotic relation
\begin{equation}
 B \simeq s_V A + \frac{2}{\pi^2} \left(H + \frac{2 s_V Q_V}{ s}\right) e^{-\lvert \mu_V \rvert / T}.
\end{equation}
Taking into account that 
\begin{gather}
 H = \frac{\alpha_V^2}{\pi^2} \left(1 + \frac{23\pi^2}{15 \alpha_V^2} + O(\alpha_V^{-4})\right),\nonumber\\
 A = \frac{2\alpha_V}{3\pi^2} \left(1 - \frac{4\pi^2}{15\alpha_V^2} + O(\alpha_V^{-4})\right), \nonumber\\
 \frac{2s_V Q_V}{ s} = \frac{2 s_V}{ \alpha_V} (H - 1) \nonumber\\\simeq \frac{2 s_V \alpha_V}{\pi^2} \left(1 + \frac{8\pi^2}{15 \alpha_V^2} + O(\alpha_V^{-4})\right),
\end{gather}
we finally arrive at
\begin{equation}
 \omega_h^\pm = \pm  \frac{k \Omega A \sqrt{3}}
 {2\sqrt{H \sigma^\omega_A(\frac{3}{T^2} \sigma^\omega_A - \Delta H)}} + O(e^{-\lvert \mu_V \rvert / T}).
 \label{eq:vhpm:deg}
\end{equation}
Expanding the above in powers of $T / \mu_V$, we get
\begin{align}
 &\omega_h^\pm(\lvert \mu_V \rvert \gg T) \nonumber\\&= \pm \frac{2\pi k \Omega T}{\mu_V^2 \sqrt{3}} \left[1 - \frac{7\pi^2 T^2}{6\mu_V^2} + O\left(\frac{T^4}{\mu_V^4}\right)\right].\label{eq:unpol_vH_largemu}
\end{align}
We display the phase velocity of the Helical Vortical Wave $v_h^\pm = \omega_h^\pm / k$ normalized to the neutral plasma velocity $c_h = 6 \Omega \ln 2 / (\pi^2 T)$ in Figure~\ref{fig:vHVW}. The quantity $\lvert v_h^\pm \rvert$ exhibits a monotonically-decreasing dependence on $\alpha_V = \mu_V / T$. We also display using dotted lines the asymptotic limits for large temperature and large chemical potential, shown in Eqs.~\eqref{eq:unpol_vH_largeT} and \eqref{eq:unpol_vH_largemu}, respectively.

The explicit space-time solutions as standing waves or as the evolution of the Gaussian initial state shown in Eqs.~\eqref{eq:HVW_ampl} and \eqref{eq:HVW_gauss} apply straightforwardly to the case considered here by replacing the velocity $\pm c_h$ of the HVW in a neutral plasma with $\omega^\pm_h / k$ given in Eq.~\eqref{eq:vhpm}.

Summarizing, the unpolarized ($\mu_A = \mu_H = 0$) plasma maintains only the Helical Vortical Wave, while its axial vortical analogue is absent since the background axial density vanishes. The finite vector chemical potential $\mu_V$ results in slowing down the propagation of the Helical wave, as indicated in Fig.~\ref{fig:vHVW} and confirmed by the asymptotic results in Eqs.~\eqref{eq:unpol_vH_largeT} and \eqref{eq:unpol_vH_largemu}. The HVW for the unpolarised plasma is of the same nature as the one previously obtained in \cite{Ambrus:2019khr}. The differences come from the fact that Ref.~\cite{Ambrus:2019khr} includes the non-conservation of helicity --we shall address this aspect in our companion paper \cite{Morales-Tejera:2024mtx}-- and in that we generalize the results to non-vanishing vector chemical potential $\mu_V$.

\section{Degenerate limit of high-density matter and non-reciprocity}\label{sec:largemu}

We now consider the case when the fermion gas is strongly degenerate, i.e., when the vector charge density is higher than any other scales in the system so that $\lvert \mu_V \rvert \gg T, \lvert \mu_A \rvert, \lvert \mu_H \rvert$. We analyze this system by rewriting the pressure \eqref{eq:Pcl} as follows:
\begin{multline}
 P = -\frac{T^4}{\pi^2} \sum_{\lambda} \left[
 {\rm Li}_4\left(-e^{\alpha_V + 2\lambda(\alpha_H + \alpha_A)}\right) \right. \\
 \left.+ {\rm Li}_4\left(-e^{-\alpha_V + 2\lambda(\alpha_A - \alpha_H)}\right)\right],
 \label{eq:triad:deg:Pgen}
\end{multline}
where we denoted the dimensionless chemical potentials $\alpha_\ell = \mu_\ell / T$. We now eliminate the polylogarithms using Eq.~\eqref{eq:Li4sum}.
The arguments of the exponentials in the polylogarithms appearing in Eq.~\eqref{eq:triad:deg:Pgen} are, however, not balanced as required in Eq.~\eqref{eq:Li4sum}. To apply this formula, we first rewrite Eq.~\eqref{eq:triad:deg:Pgen} as
\begin{multline}
 P = -\frac{T^4}{\pi^2} \sum_\lambda \left[{\rm Li}_4(-e^{\lvert \alpha_V \rvert + 2\lambda(\alpha_A + s_V \alpha_H)}) \right.\\ \left.+ 
 {\rm Li}_4(-e^{-\lvert \alpha_V \rvert + 2\lambda(\alpha_A - s_V \alpha_H)})\right],
\end{multline}
where $s_V = {\rm sgn}(\alpha_V)$ represents the sign of the vector chemical potential. Taking into account the explicit expression of the polylogarithm,
\begin{equation}
 {\rm Li}_n(z) = \sum_{j = 1}^\infty \frac{z^j}{j^n} = z + \frac{z^2}{2^n} + O(z^3),
 \label{eq:Lin_smallz}
\end{equation}
we seek to eliminate the polylogarithms whose arguments contain the $e^{\lvert \alpha_V \rvert}$ factor by adding and subtracting the term ${\rm Li}_4(-e^{-\lvert \alpha_V \rvert - 2\lambda(\alpha_A + s_V \alpha_H)})$. Taking only the first term in Eq.~\eqref{eq:Lin_smallz}, we see that these added terms behave as $\sim e^{-\lvert \alpha_V \rvert}$ at large $\alpha_V$, and are therefore exponentially suppressed. 

Writing $P = \widetilde{P} + P_e$, we collect in $\widetilde{P}$ the terms obtained by pairing the polylogarithms as in Eq.~\eqref{eq:Li4sum} and in $P_e$ the remaining non-essential terms, which are exponentially suppressed:
\begin{align}
     \widetilde{P} = & -\frac{T^4}{\pi^2} \sum_{\sigma,\lambda} {\rm Li_4}\left(-e^{\sigma \lvert \alpha_V \rvert + 2\lambda \sigma (\mu_A + s_V \mu_H)}\right) \nonumber\\
 = & \frac{7\pi^2 T^4}{180} + \frac{T^2}{6}(\mu_V^2 + \mu_\chi^2) \nonumber\\
 & + \frac{1}{12\pi^2} (\mu_V^4 + 6 \mu_V^2 \mu_\chi^2 + \mu_\chi^4), \nonumber\\
 P_e = & -\frac{T^4}{\pi^2} \sum_{\sigma,\lambda} \sigma {\rm Li}_4(-e^{-\lvert \alpha_V \rvert - 2\lambda s_V \alpha_H + 2\lambda \sigma \alpha_A}) \nonumber\\
 \simeq & -\frac{4T^4}{\pi^2} e^{-\lvert \alpha_V \rvert} \sinh \alpha_A \sinh \alpha_H,
 \label{eq_P_1}
\end{align}
where $\sigma = \pm 1$ was introduced for compactness. It can be seen that the chemical potentials $\mu_A$ and $\mu_H$ enter the dominant part of the pressure $\widetilde{P}$ only through the symmetric combination 
\begin{equation}
\mu_\chi = \mu_A + s_V \mu_H\,.
\label{eq_mu_plus}
\end{equation}
For later convenience, we also introduce the conjugate combination,
\begin{equation}
 \mu_{\tilde{\chi}} = \mu_A - s_V \mu_{H}\,.
\end{equation}

For the purpose of studying the excitations in the degenerate limit, we neglect the exponentially damped contributions coming from $P_e$. We will restore these terms later in this section.
Using the standard machinery described in Sec.~\ref{sec:2:Landau}, we can derive the charge densities, entropy density, and the $\beta$- and Landau-frame vortical conductivities, as follows:
\begin{gather}
 \widetilde{Q}_V = \frac{\mu_V T^2}{3} + \frac{\mu_V^3 + 3 \mu_V \mu_\chi^2}{3\pi^2}, \nonumber\\
 \widetilde{Q}_\chi = \frac{\mu_\chi T^2}{3} + \frac{\mu_\chi^3 + 3 \mu_V^2 \mu_\chi}{3\pi^2}, \nonumber\\
 \tilde{s} = \frac{7\pi^2 T^3}{45} + \frac{T}{3}(\mu_V^2 + \mu_\chi^2),\qquad
 \tilde{\sigma}_{V;\beta}^\omega = \frac{\mu_V \mu_\chi}{\pi^2}, \nonumber\\
 \tilde{\sigma}_{\chi;\beta}^\omega = \frac{T^2}{6} + \frac{\mu_V^2 + \mu_\chi^2}{2\pi^2}, \nonumber\\ 
 \tilde{\sigma}_{V}^\omega = \tilde{\sigma}_{V;\beta}^\omega - \dfrac{\widetilde{Q}_\chi \widetilde{Q}_V}{4\widetilde{P}}, \qquad
 \tilde{\sigma}_\chi^\omega = \tilde{\sigma}_{\chi;\beta}^\omega- \dfrac{\widetilde{Q}_\chi^2}{4\widetilde{P}}\label{eq:deg_expansion}
\end{gather}
where $\widetilde{Q}_\chi = \widetilde{Q}_A = s_V \widetilde{Q}_H$ and $\tilde{\sigma}_{\chi;\beta}^\omega = \tilde{\sigma}_{A;\beta}^\omega = s_V \tilde{\sigma}_{H;\beta}^\omega$.
It can be seen that the axial charge density and vortical conductivity are equal to the helical ones multiplied by the sign $s_V = {\mathrm{sgn}}\, (\mu_V)$ of the vector chemical potential $\mu_V$ because for an ensemble of single-charge particles (as dictated by the high vector potential), the total helical and axial charges are equal to each other up to $s_V$.
Due to this reason, the lines corresponding to $\ell = A$ and $H$, as well as the columns corresponding to $\ell' = A$ and $H$, are proportional to each other at the level of the non-exponential part of the matrix $\widetilde{\mathbb{M}}_{\ell\ell'}$, i.e.
\begin{equation}
 \widetilde{\mathbb{M}}_{A\ell'} = s_V \widetilde{\mathbb{M}}_{H\ell'}, \qquad 
 \widetilde{\mathbb{M}}_{\ell A} = s_V \widetilde{\mathbb{M}}_{\ell H}.
\end{equation}
The above observation implies that, in the strongly degenerate regime, the oscillations in the helical and axial chemical potentials are additive, providing only one degree of freedom in the form $\delta \mu_\chi = \delta \mu_A + s_V \delta \mu_H$. As we just mentioned, this property is to be expected since when the system consists only of particles (that is, there are no antiparticles present), the helicity and chirality are indistinguishable from each other and, therefore, they enter only in the combination~\eqref{eq_mu_plus}. Due to the very same reason, at high vector density, the helical and axial vortical waves form the same hydrodynamic excitation, the Axial-Helical Vortical Wave, which inherits features from both original waves. Finally, since helicity and chirality are indistinguishable in this limit, we expect that the wave spectrum obtained in the degenerate limit can be comparable with the high-density limit reported in earlier works. In Appendix \ref{app:deg} we verify that this is indeed the case.

We now formally introduce a small parameter $\varepsilon$ such that $T \to \varepsilon T$ and $\mu_\chi \to \varepsilon \mu_\chi$. We further consider an expansion of the angular frequency of the form $\tilde{\omega} = \tilde{\omega}_0 + \tilde{\omega}_1 \varepsilon + \dots$. A series expansion with respect to $\varepsilon$ yields:
\begin{align}\label{eq:Matrix_deg}
 \widetilde{\mathbb{M}}_{VV} =& \frac{\tilde{\omega}_0 \mu_V^2}{3\pi^2} + \frac{\varepsilon(4 \mu_\chi k\Omega + \tilde{\omega}_1 \mu_V^2)}{3\pi^2} + O(\varepsilon^2), \nonumber\\
 \widetilde{\mathbb{M}}_{V\chi} =& -\frac{2\varepsilon^2 k \Omega}{\mu_V} \left(\frac{T^2}{3} + \frac{2\mu_\chi^2}{\pi^2}\right) + O(\varepsilon^3), \nonumber\\
 \widetilde{\mathbb{M}}_{\chi V} =& -\frac{2k\Omega \mu_V}{3\pi^2} + \frac{4 \varepsilon \tilde{\omega}_0 \mu_V \mu_\chi}{3\pi^2} + O(\varepsilon^2),\nonumber\\
 \widetilde{\mathbb{M}}_{\chi\chi} =& \frac{\tilde{\omega}_0 \mu_V^2}{\pi^2} + \frac{\varepsilon(6\mu_\chi k\Omega + \tilde{\omega}_1 \mu_V^2)}{\pi^2},
\end{align}
where $\widetilde{\mathbb{M}}_{V \chi} = \widetilde{\mathbb{M}}_{VA} = s_V \widetilde{\mathbb{M}}_{VH}$, 
$\widetilde{\mathbb{M}}_{\chi V} = \widetilde{\mathbb{M}}_{AV} = s_V \widetilde{\mathbb{M}}_{HV}$, and
$\widetilde{\mathbb{M}}_{\chi\chi} = \widetilde{\mathbb{M}}_{AA} = s_V \widetilde{\mathbb{M}}_{AH} = s_V \widetilde{\mathbb{M}}_{HA} = \widetilde{\mathbb{M}}_{HH}$.
Taking the determinant, it can be seen that at order $O(\varepsilon^0)$, we have 
$\tilde{\omega}_0^2 \mu_V^4 / 3\pi^4 = 0$, such that $\tilde{\omega}_0 = 0$ and $\tilde{\omega}$ becomes of first order with respect to $\varepsilon$. The next non-vanishing contribution is of order $\varepsilon^2$, with the corresponding equation given by:
\begin{equation}
 \tilde{\omega}_1^2 \mu_V^4 + 10 \tilde{\omega}_1 \mu_\chi \mu_V^2 k\Omega + 16 \mu_\chi^2 k^2 \Omega^2 - \frac{4\pi^2 T^2}{3} k^2 \Omega^2 = 0.
\end{equation}
Solving the above equation, we obtain the energy dispersion relation of the vortical wave in the degenerate matter:
\begin{align}
 &\tilde{\omega}_\pm = -\frac{5 k\Omega \mu_\chi}{\mu_V^2} \pm \frac{k\Omega}{\mu_V^2} \sqrt{\frac{4\pi^2 T^2}{3} + 9 \mu_\chi^2}
  \nonumber\\& \text{[Axial-Helical Vortical Wave]}.
 \label{eq_v_deg}
\end{align}
The wave appears to couple fluctuations $\widetilde{\delta \mu}_V$ in the vector chemical potential and the fluctuations in the combined axial-helical chemical potentials $\widetilde{\delta\mu}_\chi$ as follows:
\begin{align}
 &\widetilde{\delta \mu}_V^\pm = \frac{3\widetilde{\delta \mu}^\pm_\chi}{2\mu_V} \left(\mu_\chi \pm \sqrt{\frac{4\pi^2 T^2}{3} + 9 \mu_\chi^2}\right)
  \nonumber\\& \text{[Axial-Helical Vortical Wave]}\,.
    \label{eq_mu_deg}
\end{align}
Both the energy dispersion relation~\eqref{eq_v_deg} and the charge density content~\eqref{eq_mu_deg} of the Axial-Helical Vortical Wave depend on the sum $\mu_\chi$ of the axial and helical chemical potentials~\eqref{eq_mu_plus}.

In the presence of a finite axial (or helical) charge density, the propagation of the wave becomes non-reciprocal with respect to the direction of the angular velocity $\boldsymbol \Omega$. In general, the speed of the wave $\tilde{v}_+ = \tilde{\omega}_+/k$ parallel to the direction of the angular velocity $\boldsymbol \Omega$ and the speed $\tilde{v}_-=\tilde{\omega}_-/k$ in the direction anti-parallel to $\boldsymbol \Omega$ are different from each other~\eqref{eq_v_deg}. For example, setting high temperature $T \gg \lvert \mu_\chi \rvert$ (but still maintaining the degenerate limit $\mu_V \gg T$), we clearly see from Eq.~\eqref{eq_v_deg} that the wave propagates in both directions with an offset in the velocities given by the first term. For a positive (negative) sign of the product $\mu_\chi \Omega$, the wave propagates thus faster opposite (along) the vorticity vector $\lvert \tilde{v}_-\rvert > \lvert \tilde{v}_+ \rvert$ ($\lvert \tilde{v}_- \rvert < \lvert \tilde{v}_+ \rvert$). 

Moreover, at the critical temperature
\begin{align}
    T^{\rm deg}_c = \frac{2 \sqrt{3}}{\pi} \lvert \mu_\chi \rvert,
    \label{eq_Tc_deg}
\end{align}
the speed of propagation of the wave along (opposite to) the vorticity vector vanishes if the product $\mu_\chi \Omega$ takes a positive (negative) value. Thus, at the critical temperature~\eqref{eq_Tc_deg}, one of the branches of the axial-helical wave is propagating (anti-)parallel to the vorticity while the other branch represents a static mode. If the temperature is below the critical value~\eqref{eq_Tc_deg}, then both modes propagate in the same direction in a transparent manifestation of non-reciprocity. 

Thus, in the cold and dense rotating matter, the hydrodynamical waves propagate in a non-reciprocal manner with respect to the global angular velocity if the background state is axially or helically imbalanced. Moreover, the relation between the magnitudes of the vector and axial-helical components of the wave differs in the waves propagating in opposite directions~\eqref{eq_mu_deg}. These properties represent a unique feature of hydrodynamic excitations possessing the helical degree of freedom.

Considering now a vector wave constructed as in Eq.~\eqref{eq:cos_F}, namely $\bar{\mu}_V(t = 0, z) = \mu_V + \delta \bar{\mu}_{V;0}\, \cos(k z)$ and $\bar{\mu}_\chi(t = 0, z) = \mu_\chi$, with $\lvert \mu_V \rvert \gg \lvert \mu_\chi \rvert$, we find 
\begin{align}
 \delta \bar{\mu}_V^\pm = \frac{1}{2} \left(1 \pm \frac{\mu_\chi}{\mathfrak{s}}\right) \delta \bar{\mu}_{V;0}, \quad 
 \delta \mu_\chi^\pm =  \pm \frac{\mu_V}{3\mathfrak{s}} \delta \bar{\mu}_{V;0},
\end{align}
with $\mathfrak{s} = \sqrt{4\pi^2 T^2 / 3 + 9 \mu_\chi^2}$. This leads to
\begin{align}
 &\bar{\mu}_V(t, z)= \nonumber\\& \mu_V  + \delta \bar{\mu}_{V;0} \cos\left(kz - \frac{5 \Omega \mu_\chi}{\mu_V^2} k t\right) \cos\left(\frac{\mathfrak{s} \Omega}{\mu_V^2} k t\right) \nonumber\\
 & - 
 \frac{\mu_\chi \delta \bar{\mu}_{V;0}}{\mathfrak{s}} \sin \left(kz - \frac{5 \Omega \mu_\chi}{\mu_V^2} k t\right) \sin\left(\frac{\mathfrak{s} \Omega}{\mu_V^2} k t\right), \nonumber\\
 &\bar{\mu}_\chi(t, z)= \nonumber\\& \mu_\chi - \frac{\mu_V \delta \bar{\mu}_{V;0}}{3\mathfrak{s}} \sin\left(kz - \frac{5 \Omega \mu_\chi}{\mu_V^2} k t\right) \sin\left(\frac{\mathfrak{s} \Omega}{\mu_V^2} k t\right).
\end{align}
In the case of the initial Gaussian distribution in Eq.~\eqref{eq:gauss_F} for the perturbation in the vector chemical potential, $\bar{\mu}_V(0,z) = \mu_V + \delta \bar{\mu}_{V;0}\, e^{-z^2 / 2\sigma^2}$, with constant $\bar{\mu}_\chi(0,z) = \mu_\chi$, we have
\begin{align}
 \bar{\mu}_V(t, z) &= \mu_V + \frac{\delta \bar{\mu}_{V;0}}{2} \left[\left(1 + \frac{\mu_\chi}{\mathfrak{s}}\right) e^{-(z - v_+ t)^2 / 2\sigma^2}\right. \nonumber\\& \left.+ \left(1 - \frac{\mu_\chi}{\mathfrak{s}}\right)  e^{-(z v_- t)^2 / 2\sigma^2} \right],\nonumber\\
 \bar{\mu}_\chi(t, z) &= \mu_\chi+  \frac{\mu_V \delta \bar{\mu}_{V;0}}{3\mathfrak{s}} \left[
 e^{-(z - v_+ t)^2 / 2\sigma^2}\right. \nonumber\\&\left.- e^{-(z - v_- t)^2 / 2\sigma^2} \right],
 \label{eq:deg_gauss}
\end{align}
where $\tilde{v}_\pm = -5 \Omega \mu_\chi / \mu_V^2 \pm \Omega \mathfrak{s} / \mu_V^2$ represent the group velocities corresponding to the two propagating modes.

\begin{figure*}
    \centering
    \begin{tabular}{cc}
    \includegraphics[width=.49\linewidth]{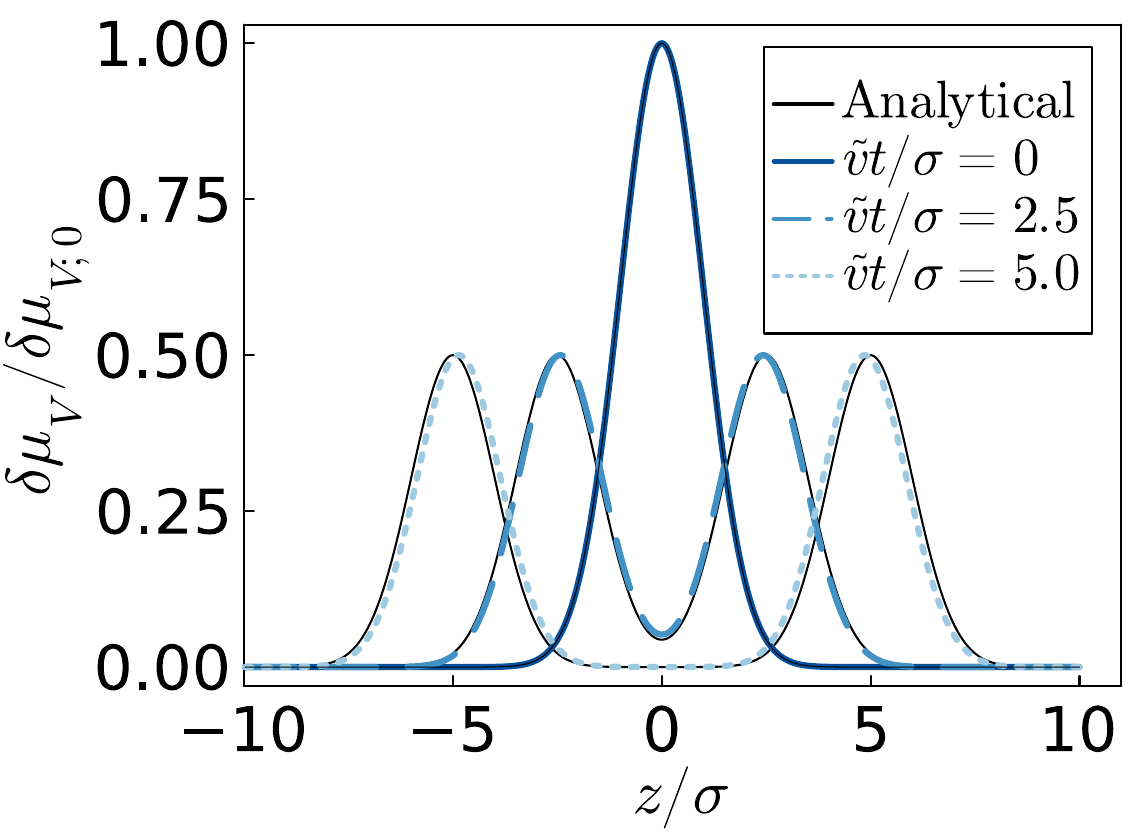} &
    \includegraphics[width=.49\linewidth]{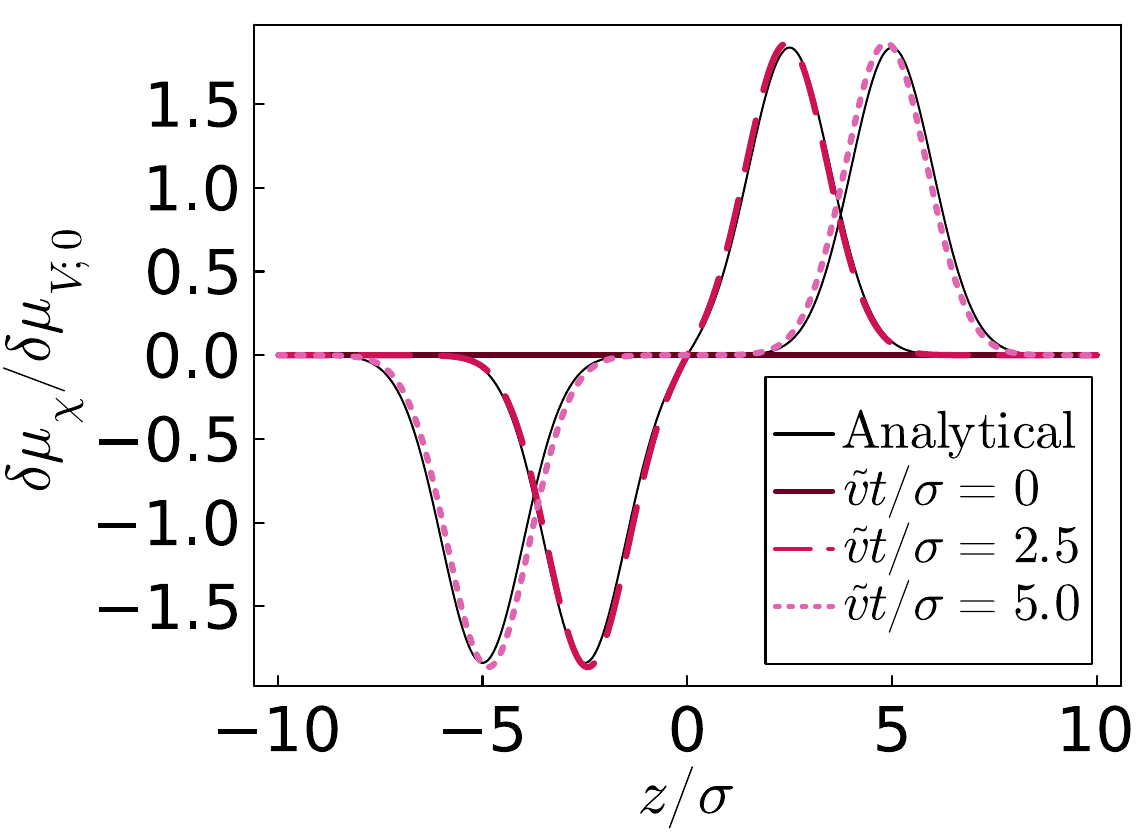} \\
    \multicolumn{2}{c}{\includegraphics[width=.49\linewidth]{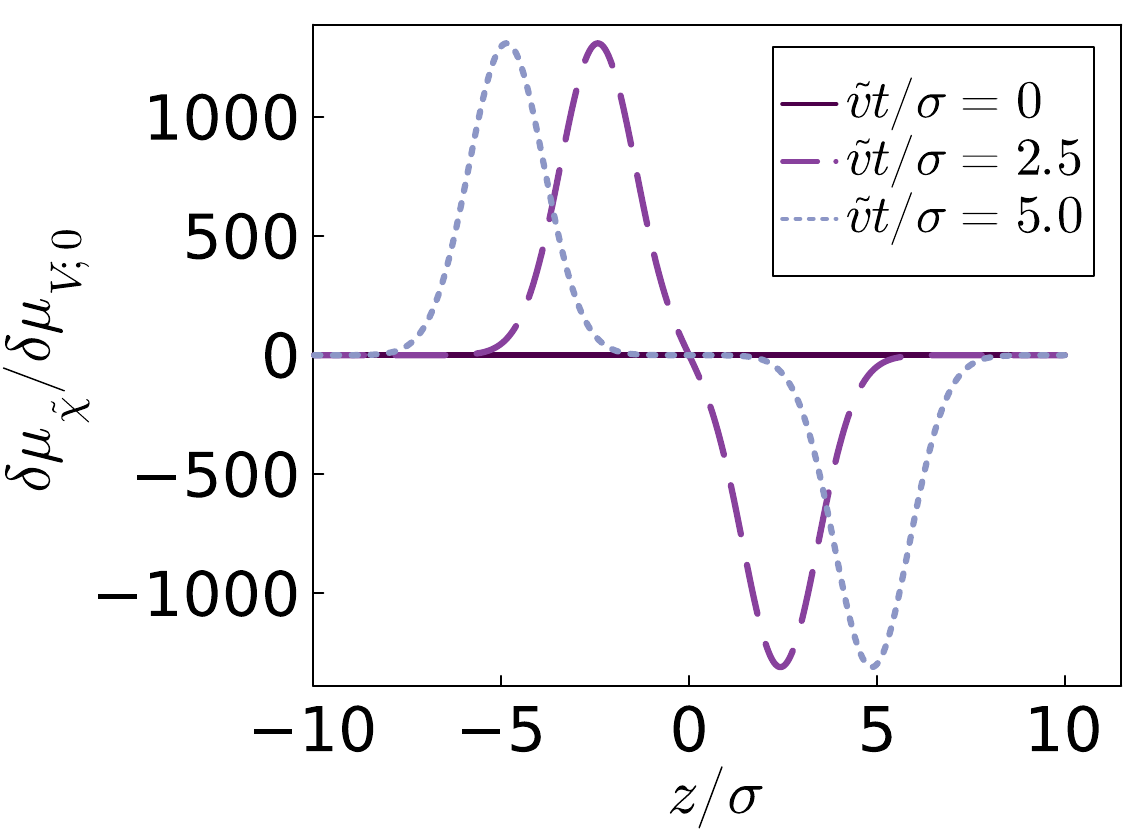}}
    \end{tabular}
    \caption{Time evolution of the charge perturbations (top left) $\delta \bar{\mu}_V(t,z)$, (top right) $\delta \bar{\mu}_{\chi}(t, z)$ and (bottom) $\delta\bar{\mu}_{\tilde\chi}$, corresponding to the propagation of the axial-helical vortical wave through a charge-conserving plasma. The initial conditions are 
    given
    in Eqs.~\eqref{eq:HVW_gauss_init}. The background state is unpolarized ($\mu_A = \mu_H = 0$) and has $\alpha_V = \mu_V / T = 20$, corresponding to the degenerate limit discussed in Sec.~\ref{sec:largemu}.
    In the label, $\tilde{v}$ is the absolute value of the corresponding speed of the wave $\tilde{v}_\pm=\tilde{\omega}_\pm/k$ defined in \ref{eq_v_deg}. The analytical solution in Eq.~\eqref{eq:deg_gauss}, shown in black, is superimposed onto the numerical solutions (blue and pink). 
    \label{fig:HVW_deg}
    }
\end{figure*}

The propagation of the initial Gaussian distribution is considered in Fig.~\ref{fig:HVW_deg}. For definiteness, we took a dimensionless vector chemical potential $\alpha_V = 20$, while the background axial and helical chemical potentials were set to zero. The left and middle panels demonstrate the expected analytical solution derived in Eq.~\eqref{eq:deg_gauss}. Surprisingly, the third panel reveals that the combination of chemical potentials $\delta \bar{\mu}_{\tilde{\chi}} = \delta \bar{\mu}_A - s_V \delta \bar{\mu}_H$ grows to very high values. In the degenerate limit described by the tilde quantities (e.g., $\widetilde{P}$) in Eq.~\eqref{eq:Matrix_deg}, the combination $\delta \bar{\mu}_{\tilde{\chi}}$ is not accessible. 

To understand the behaviour described above, we move back to the unpolarized plasma considered in Sec.~\ref{sec:unpolarized} and rederive the relation between all mode amplitudes before taking the degenerate limit. Considering the exact expression for the matrices $\mathbb{M}_\omega$ and $\mathbb{M}_\Omega$ in Eq.~\eqref{eq:unpol_M}, we can find the relations between the amplitudes corresponding to the trivial mode $\omega_0 = 0$ as follows:
\begin{align}
    &\delta\mu^0_V =0, \nonumber\\&
    \delta\mu^0_{\tilde{\chi}} = -\frac{A+ s_V B}{A - s_V B} \delta\mu^0_\chi \simeq -\frac{2\pi^2}{3\lvert \alpha_V \rvert} e^{\lvert \alpha_V \rvert} \delta\mu^0_\chi.
\end{align}
Clearly, exciting the $\delta \mu^0_\chi$ amplitude leads to an exponentially-larger amplitude $\delta \mu^0_{\tilde{\chi}}$. 

In the case of the non-trivial (helical) modes, when $\omega^\pm_h$ is given by Eq.~\eqref{eq:vhpm}, we can take the amplitude of the vector charge fluctuation, $\delta \mu_V^{h;\pm}$, as the reference amplitude. The amplitudes $\delta \mu^{h;\pm}_\chi$ and $\delta \mu^{h;\pm}_{\tilde{\chi}}$ are then given by
\begin{align}
 \delta\mu_\chi^{h;\pm} &= \frac{k \Omega T}{2\omega^\pm_h} \frac{A + s_V B}{\sigma_A^\omega + s_V \sigma_H^\omega} \delta \mu^{h;\pm}_{V} \simeq \pm \frac{\alpha_V}{\pi \sqrt{3}} \delta \mu_V^{h;\pm},\nonumber\\
 \delta\mu_{\tilde{\chi}}^{h;\pm} &= \frac{k \Omega T}{2\omega^\pm_h} \frac{A - s_V B}{\sigma_A^\omega - s_V \sigma_H^\omega} \delta \mu^{h;\pm}_{V} \nonumber\\&\simeq \pm \frac{s_V \alpha_V^4 \sqrt{3}}{4\pi^3} \delta \mu^{h;\pm}_V,
\end{align}
where we displayed just the leading-order contribution in the degenerate limit. It can be seen that as $\lvert \alpha_V \rvert \rightarrow \infty$, a small excitation in $\delta \mu_V$ in either modes $\delta \mu^{h;\pm}_V$ will lead to excitations in $\delta \mu_{\tilde{\chi}}^{h;\pm}$ that are $\alpha_V^4$ times larger, signalling the breakdown of the linear perturbations ansatz in the degenerate matter. Coming back to Fig.~\ref{fig:HVW_deg} and considering the ratio $\alpha_V = \mu_V / T = 20$, our estimate above indicates $\delta \mu^{h;\pm}_{\tilde{\chi}} / \delta \mu_{V;0} \simeq 1117$ to leading order in $\alpha_V$, which is in line with the results shown in the bottom panel of this figure.

The above discussion indicates that degenerate matter under rotation is unstable under small perturbations. A small excitation of the vector chemical potential induces huge excitations in the axial and helical chemical potentials, invalidating the linear-regime ansatz of this analysis. One may question the physical soundness of these results. In our companion paper \cite{Morales-Tejera:2024mtx}, we show that in a realistic plasma, the non-conservation of the helicity charge prevents the axial and helical mode amplitudes from growing, leading to perfectly reasonable propagation properties.

\section{Summary and Conclusions} \label{sec:conc}

\begin{figure*}[!ht]
    \centering
    \includegraphics[width=.9\linewidth]{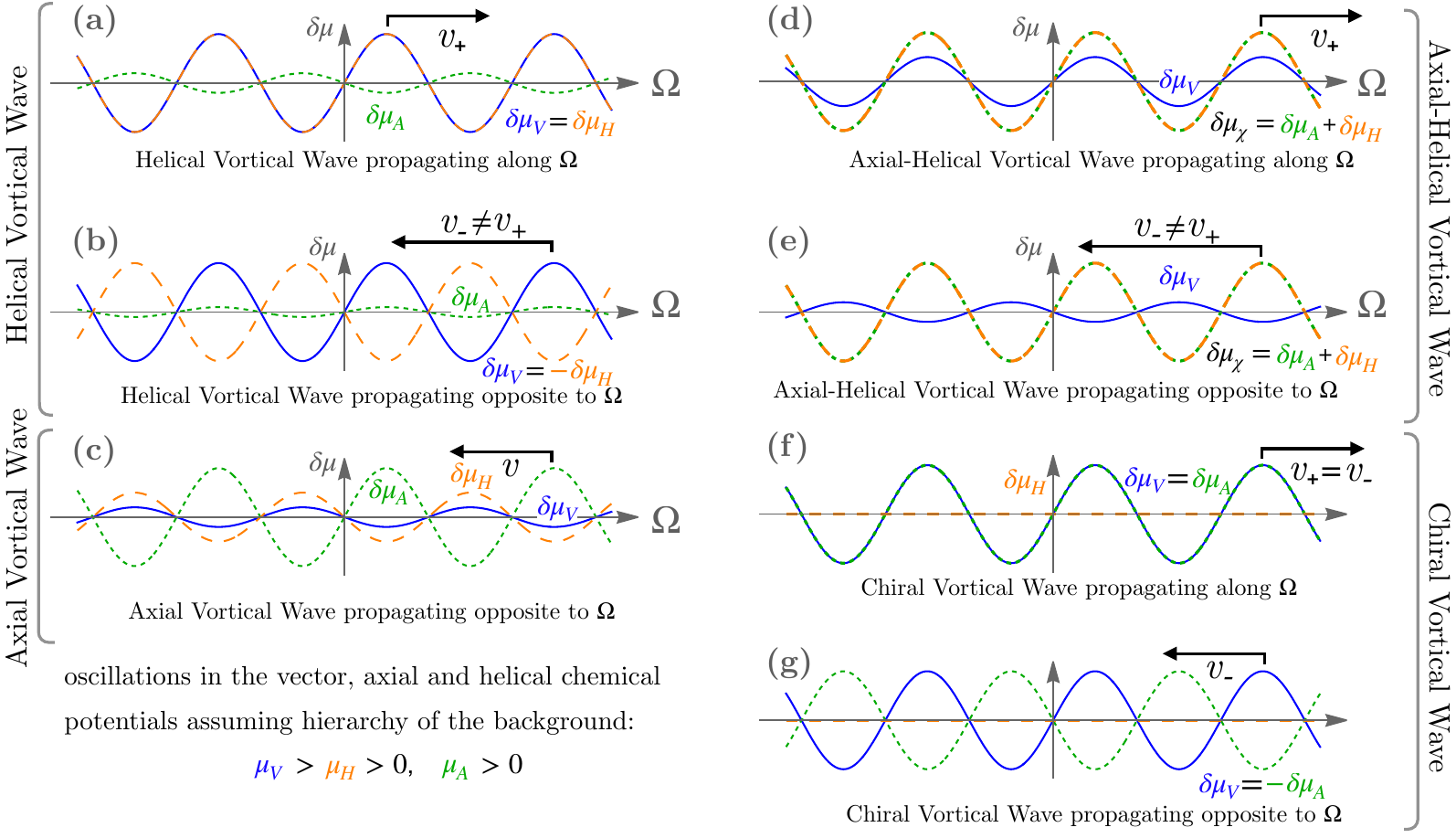}
    \caption{Summarising illustration of four classes of gapless hydrodynamic excitations discussed in the paper for the hierarchy $\mu_V > \mu_H >0$ and $\mu_A > 0$ of the background chemical potentials. For simplicity, a sinusoidal wave is shown only. A detailed description is given in the text of the Conclusions. The results of this paper include the finding of the non-reciprocity of the Helical Vortical Wave, which propagates with different velocities (a) along and (b) opposite to the angular velocity vector; (c) the uni-directional nature of the Axial Vortical Wave, which propagates in one direction only; the merging, in the high-density limit, of these waves into a single Axial-Helical Vortical Wave excitation, again with different velocities (d) along and (e) opposite to the angular velocity vector. Plots (f) and (g) show, for completeness of the presentation, the known Chiral Vortical Wave studied in Refs.~\cite{Jiang:2015cva,Kalaydzhyan:2016dyr,Gorbar:2017toh}, which propagates with the same velocities in both directions.}
    \label{fig:waves}
\end{figure*}

In our work, we challenge the traditional approach to chiral systems, which presumes that the chiral fluids are completely described only by the pair of the vector and axial local charges. We show that the inclusion of the helical degree of freedom enriches the hydrodynamic spectrum of the system. Since we explored various regimes of the vector-axial-helical triad of charges, it is worth summarizing the main features of our findings in a compact set of plots shown in Fig.~\ref{fig:waves} and described concisely in the figure caption. Below, we also summarize the results in a more extended form.

\paragraph{Non-reciprocity of the helical vortical wave} 
In the high-temperature limit of low-density plasmas, where the temperature significantly exceeds the magnitude of any background chemical potential, we confirmed the existence of the Helical Vortical Wave, which corresponds to a gapless hydrodynamic excitation acting mainly in the helical and vector sectors of charge densities. In the next order in the series over the inverse temperature, this wave receives a slight admixture of the axial charge~\eqref{eq_HVW_chemicals}, indicating that all components of the fluid, represented by the vector, helical and axial charges, fluctuate coherently as the wave propagates as it is illustrated in Figs.~\ref{fig:waves}\textcolor{blue}{(a)} and \ref{fig:waves}\textcolor{blue}{(b)}. This next-order-correction gives us also a qualitatively unusual result implying the non-reciprocity of these hydrodynamic excitations: the helical vortical waves propagating along and opposite to the direction of vorticity possess different velocities~\eqref{eq_HVW_speed}:
\begin{multline}
   v^\pm_h = \pm \frac{6 \ln 2}{\pi^2} \frac{\Omega}{T} - \frac{6}{7\pi^2} \left[\frac{84}{\pi^2} (\ln 2)^2 {-} 1\right] \frac{\Omega \mu_A}{T^2} \\ + O\bigl(T^{-3}\bigr)\,. 
\end{multline}

\paragraph{Uni-directionality of the axial vortical wave} 
Our approach also allows us to uncover yet another gapless wave in the system: the Axial Vortical Wave. This wave propagates only in the presence of the axial chemical potential $\mu_A$ since its velocity~\eqref{eq_AVW_v},
\begin{align}
    v_\AVW = -\frac{24}{7\pi^2} \frac{\mu_A \Omega}{T^2} + O(T^{-3})\,,
\end{align}
vanishes if $\mu_A = 0$. In the case when the other two charge densities are vanishing, the wave exhibits oscillations in the axial charge density, which propagate unidirectionally along (opposite to) the vorticity vector $\boldsymbol{\Omega}$ for $\mu_A < 0$ ($\mu_A > 0$). In the presence of the background of vector (helical) charges, the purely axial wave is also accompanied by coherent fluctuations in axial and helical (or vector) charges~\eqref{eq_AVW_mu}, as it is illustrated in Fig. \ref{fig:waves}\textcolor{blue}{(c)}. 

It is noteworthy to stress that the appearance of unidirectional transport in this simple homogeneous system is a fascinating and unexpected phenomenon that is distinct from the processes that usually occur in high-temperature plasmas. For example, in most classical acoustics and hydrodynamics, sound waves typically exhibit reciprocal propagation, meaning that the waves can travel forward and backwards along the same path without change in their propagation characteristics. On the contrary,  the unidirectionality, seen from the point of view of ordinary acoustical waves, emerges as a result of the meticulously engineered breaking of time-reversal symmetry in the medium through which sound waves propagate. For example, in systems that break the time-reversal symmetry, sound waves can be engineered to propagate in one direction while being significantly attenuated in the opposite direction~\cite{Cummer2016}.

The theoretical basis for unidirectional sound transport can, for example, be linked to the concept of topological insulators that represent materials that maintain propagating excitations only along their surface and not through their bulk~\cite{Moore2010}. Analogously, the unidirectional transport of sound waves has been realized through specific mechanisms, such as using exotic active materials with time-varying properties, creating asymmetric structures that induce directional bias, or employing magnetic fields in conjunction with magneto-acoustic materials~\cite{Nash2015}.

On the practical level of acoustic (or, gapless, in a more generic sense) applications, the unidirectional transport includes acoustic diodes or rectifiers, which allow sound to pass through in one direction while blocking it in the opposite direction~\cite{Fleury2015}. These devices could be used in noise reduction systems, acoustic logic devices, and improved ultrasonic imaging technologies~\cite{Mousavi2015}. Unidirectional sound propagation is an unusual phenomenon, and its appearance in a chiral plasma represents a fascinating, unexpected effect of an interplay of anomalous transport effects.

\paragraph{Axial-helical vortical wave at high density}
Coming back to the field-theoretical applications in chiral plasmas, we highlight our observation that at high densities ($\lvert \mu_V \rvert \gg T, \lvert \mu_A \rvert, \lvert \mu_H \rvert$), both the helical and the axial waves merge into a common hydrodynamic excitation, the Axial-Helical Vortical Wave. In this limit of the degenerate Dirac fluid, the magnitudes of helical and axial charge densities of the ensemble of particles are indistinguishable from each other since the helical and axial charges of a given particle (or anti-particle) are the same up to a sign. The Axial-Helical Vortical Wave maintains the property of non-reciprocity reflected in the velocities of its branches~\eqref{eq_v_deg}:
\begin{align}
 v^\pm_{\AVW h} = -\frac{5\Omega \mu_\chi}{\mu_V^2} \pm \frac{\Omega}{\mu_V^2} \sqrt{\frac{4\pi^2 T^2}{3} + 9 \mu_\chi^2}\,,
\end{align}
where $\mu_\chi = \mu_A + \mu_H \, \text{sgn}\,\mu_V$.
Interestingly, there exists a critical temperature~\eqref{eq_Tc_deg}, at which one of the branches of this mixed wave stops propagating while the other one moves parallel to the direction of the vorticity vector. The mixed axial-helical vortical wave is illustrated in Figs. \ref{fig:waves}\textcolor{blue}{(d)} and~\ref{fig:waves}\textcolor{blue}{(e)}.

The discussion in this paper focused on the simplified model where both the axial and the helical charges are conserved. As already discussed in Ref.~\cite{Ambrus:2019khr}, the helicity charge dissipates in a realistic plasma due to helicity-violation pair annihilation (HVPA) processes. Furthermore, the conservation of the axial charge is violated due to interactions via the axial anomaly \cite{bertlmann96}. The latter is dynamically relevant for the crossover region of finite-temperature QCD \cite{Ruggieri:2016asg}. For the sake of completeness, we show in Figs. \ref{fig:waves}\textcolor{blue}{(f)} and \ref{fig:waves}\textcolor{blue}{(g)} the chiral vortical wave \cite{Jiang:2015cva} which emerges in the limit where the helical degree of freedom is frozen due to strong relaxation~\cite{Morales-Tejera:2024mtx}. We explore the consequences of the axial and helical charge non-conservation on the excitations spectrum revealed here in the companion paper~\cite{Morales-Tejera:2024mtx}.

{\bf Acknowledgments.}
We thank P. Aasha for fruitful discussions. We also thank an anonymous referee for helping us improve the connections to existing literature.
This work was funded by the EU’s NextGenerationEU instrument through the National Recovery and Resilience Plan of Romania - Pillar III-C9-I8, managed by the Ministry of Research, Innovation and Digitization, within the project entitled ``Facets of Rotating Quark-Gluon Plasma'' (FORQ), contract no. 760079/23.05.2023 code CF 103/15.11.2022. 
V.E.A.~gratefully acknowledges partial support through a grant of the Ministry of Research, Innovation and Digitization, CNCS - UEFISCDI, project number PN-III-P1-1.1-TE-2021-1707, within PNCDI III.

\appendix 

\section{Comparison with the chiral vortical wave in the un-boosted frame.}
\label{app:cvw}

In this appendix, we provide a direct comparison with the results for the chiral vortical wave studied in Ref.~\cite{Gorbar:2017toh}. Note that, in setting up the study of the collective waves in this work, we have taken one preliminary step that modifies the velocities of the collective modes of the system: we performed a Lorentz boost. We take into account this modification prior to comparing the two results.

\subsection{Sound modes}\label{app:cvw_sound}

In Sec. \ref{sec:2:eqs} we have claimed that the sound modes obtained in previous works differ from our results owing to the fact that the spatial gradients of transverse perturbations were neglected. In this section we recompute the sound modes in the unboosted (laboratory) $\beta$-frame and show explicitly how these gradients are required to have a consistent solution.

We consider perturbations $\delta \overline{A}$ around the background state $A$, as described in Eqs.~\eqref{eq:pert}--\eqref{eq:pertw}. We take as background velocity that of rigid rotation, $u^\mu \partial_\mu \simeq \partial_t + \Omega \partial_\varphi$, while the velocity perturbations are similar to those discussed in Eq.~\eqref{eq:pertu}.
Similarly to what was done in the main text, we shall assume that only the transverse fluctuations $\delta \bar{u}^x$ and $\delta \bar{u}^y$ depend on the transverse plane coordinates $(x,y)$, and are of order $O(\Omega)$. The longitudinal perturbations $\delta u^z$ are of zeroth order with respect to $\Omega$ and are considered to be independent of $x$ and $y$. The vorticity $\bar{\omega}^\mu = \omega^\mu + \delta\bar{\omega}^\mu$ has the background value $\omega^\mu = \Omega \delta^\mu_z$ and receives perturbation induced by the four-velocity shown in Eq.~\eqref{eq:pertw}.

The energy momentum tensor in the $\beta$-frame is given in Eq.\eqref{eq:JT-beta}:
\begin{align}
    &T^{\mu\nu} = (E+P)u^\mu u^\nu -P g^{\mu\nu}+ Q_A(\omega^\mu u^\nu + \omega^\nu u^\mu)\,,
\end{align}
where we have identified $Q_A = \sigma^\omega_\varepsilon$ and we have dropped the $\beta$ subindex. The conservation equation $\partial_\mu T^{\mu\nu}=0$ become
\begin{align}
%
 DE+ (E+P)\theta + \omega^\mu\partial_\mu Q_A \nonumber\\
 + Q_A(\partial_\mu\omega^\mu-\omega_\mu Du^\mu) &=0,\nonumber\\
 (E+P)Du^\nu - \nabla^\nu P+\omega^\nu DQ_A + Q_A\theta \omega^\nu\nonumber\\
 + Q_A\left(\omega^\mu\partial_\mu u^\nu  +D\omega^\nu + u^\nu \omega_\mu Du^\mu\right)&= 0.
\end{align}

We now list the necessary relations to evaluate the conservation equations above upon projection to the Fourier space:
\begin{align}
    Df &\to -i \omega_0 \delta f_{\omega;0}-i\Omega(\omega_0\delta f_{\omega;1}+\omega_1\delta f_{\omega;0}),\nonumber\\
    \theta &\rightarrow i k \delta u_{\omega;0}^z + \Omega(ik\delta u_{\omega;1}^z + \partial_x \delta u^x_{\omega;1} \nonumber\\
    &+ \partial_y \delta u^y_{\omega;1}), \nonumber\\
    \omega^\mu\partial_\mu f &\to i k \Omega \delta f_{\omega;0}, \quad \partial_\mu \omega^\mu\to -2i\Omega \omega_0 \delta u^z_{\omega;0},\nonumber\\
    Du^x&\to -i\Omega \omega_0\delta u_{\omega;1}^x, \quad 
    Du^y\to -i\Omega \omega_0\delta u_{\omega;1}^y, \nonumber\\
     Du^z&\to -i\omega_0\delta u_{\omega;0}^z -i\Omega(\omega_0\delta u_{\omega;1}^z + \omega_1 \delta u_{\omega;0}^z) ,\nonumber\\
     Du^t &\to 0,     \qquad
     \omega^\mu\partial_\mu u^\nu \to i k \Omega \delta u_{\omega;0}^z \delta^\nu_z ,\nonumber\\
\nabla^\nu P&\to -i\Omega \omega_0 \delta P_{\omega;0}\left(y\delta^\nu_x-x\delta^\nu_y\right)\nonumber\\
&-ik(\delta P_{\omega;0}+\Omega\delta P_{\omega;1})\delta^\nu_z,\nonumber\\
\omega_\mu Du^\mu&\to i\Omega \omega_0 \delta u_{\omega;0}^z,\qquad D\omega^t\to0,\nonumber\\
D\omega^x&\to -\frac{1}{2}\omega_0 \Omega (x\omega_0\delta u_{\omega;0}^z  + k \delta u_{\omega;1}^y),\nonumber\\
D\omega^y&\to -\frac{1}{2}\omega_0 \Omega (y\omega_0\delta u_{\omega;0}^z  - k \delta u_{\omega;1}^y),\nonumber\\
D\omega^z&\to -\frac{i}{2}\omega_0 \Omega (\partial_y\delta u_{\omega;1}^x  - \partial_x\delta u_{\omega;1}^y),
\end{align}
where $f$ is a scalar function. 

Upon substitution of the relations above, the conservation equations for the energy momentum tensor to zeroth order in $\Omega$ read:
\begin{equation}
 \begin{pmatrix}
  -3\omega_0 & 4 k P \\
  k & -4 P \omega_0
 \end{pmatrix} 
 \begin{pmatrix}
  \delta P_{\omega;0} \\ \delta u^z_{\omega;0}
 \end{pmatrix} = 0,
\end{equation}
for which the non-trivial solutions are the sound modes
\begin{equation}
    \omega_0^\pm = \pm \dfrac{k}{\sqrt{3}},\qquad \delta u^z_{\pm;0} = \dfrac{3\omega_0^\pm}{4kP}\delta P_{\pm;0}\,.
\end{equation}

Taking the amplitude of the pressure perturbations as the reference amplitude, we set $\delta P_{\omega;1} = 0$ without loss of generality. Then, at first order in $\Omega$, the conservation of energy and momentum entails:
\begin{multline}\label{eq:Tmunu-beta-t}
    k\delta Q^A_{\omega;0} -3Q_A\omega_0\delta u^z_{\omega;0}-3\omega_1\delta P_{\omega;0}+4k P \delta u^z_{\omega;1}  \\ 
    - 4i P (\partial_x \delta u^x_{\omega;1}+\partial_y \delta u^y_{\omega;1}) =0,
\end{multline}
\begin{multline}\label{eq:Tmunu-beta-x}
    i\left(\delta u^x_{\omega;1} -\frac{y \delta P_{\omega;0}}{4P}\right) +\frac{Q_A}{8P}\left(k\delta u^y_{\omega;1}+x\omega_0\delta u^z_{\omega;0}\right)\\
    =0,
\end{multline}
\begin{multline}\label{eq:Tmunu-beta-y}
    i\left(\delta u^y_{\omega;1} +\frac{x \delta P_{\omega;0}}{4P}\right) - \frac{Q_A}{8P} \left(k\delta u^x_{\omega;1}-y\omega_0\delta u^z_{\omega;0}\right)\\
    =0,
\end{multline}
\begin{multline}\label{eq:Tmunu-beta-z}
    -\omega_0\delta Q^A_{\omega;0} + (2kQ_A-4P\omega_1)\delta u^z_{\omega;0}-4\omega_0 P \delta u^z_{\omega;1} \\  +\dfrac{1}{2}Q_A\omega_0 (\partial_y \delta u^x_{\omega;1}-\partial_x \delta u^y_{\omega;1})=0.
\end{multline}
Eqs. \eqref{eq:Tmunu-beta-x} and \eqref{eq:Tmunu-beta-y} can be solved algebraically for $\delta u^{x,y}_{\omega;1}$:

\begin{equation}
    \delta u^x_{\omega;1} = y\dfrac{\delta P_{\omega;0}}{4P}\,,\quad \delta u^y_{\omega;1} = -x\dfrac{\delta P_{\omega;0}}{4P}\,.
    \label{eq:beta_dut}
\end{equation}
The result agrees with the solution obtained in Eq. \eqref{eq:ampl-trans}, since the boost \eqref{eq_Lorentz} is trivial in the transverse plane. Note that the non-vanishing derivatives of the transverse perturbations contribute non-trivially to Eq. \eqref{eq:Tmunu-beta-z}. Now we can solve Eq.\eqref{eq:Tmunu-beta-t} algebraically for $\delta u^z_{\omega;1}$,

\begin{multline}
    \delta u^z_{\omega;1} = \dfrac{1}{4kP}(3Q_A\omega_0\delta u^z_{\omega;0}+3\omega_1\delta P_{\omega;0} \\ 
    -k\delta Q_{\omega;0}^A)\,.
\end{multline}
Substituting the above into Eq.\eqref{eq:Tmunu-beta-z} gives:
\begin{multline}\label{eq:beta-omega1}
    \dfrac{3i\omega_0}{4kP}\left(kQ_A-8P\omega_1\right)\delta P_{\omega;0} \\ 
    + \dfrac{1}{2}i Q_A\omega_0\left(\partial_y \delta u^x_{\omega;1}-\partial_x \delta u^y_{\omega;1}\right) = 0.
\end{multline}
Using Eq.~\eqref{eq:beta_dut} to compute the transverse gradients, we arrive at
\begin{align}
    \dfrac{i\omega_0}{kP}\left(kQ_A-6P\omega_1\right)\delta P_{\omega;0} = 0,
\end{align}
which gives the leading correction to the dispersion relation:
\begin{equation}
    \omega_1 = k\dfrac{Q_A}{6P}\,.
\end{equation}
From Eq. \eqref{eq:beta-omega1}, it is clear that dismissing the spatial gradients of the transverse fluctuations gives a different dispersion relation, as obtained previously in the literature (see e.g. Eq. (58) in Ref.  \cite{Gorbar:2017toh}):
\begin{equation}
 \text{Ref.~\cite{Gorbar:2017toh}:} \qquad 
 \tilde{\omega}_{\rm ac.}^{\pm} = \pm \frac{k}{\sqrt{3}} + k\dfrac{Q_A\Omega}{8P}.
\end{equation}

We conclude that the dispersion relation for the sound modes traveling along the $z$ direction is given by
\begin{equation}\label{eq:ac_unboosted}
    \omega^\pm_{\textrm{ac.}} = \pm\dfrac{k}{\sqrt{3}} + k\dfrac{Q_A}{6P}\Omega+O(\Omega^2)\,.
\end{equation}

In order to compare the result in Eq.~\eqref{eq:ac_unboosted} with the result in Eq.~\eqref{eq:acoustic_vpm} of the main text, we rewrite Eq.~\eqref{eq:ac_unboosted} in terms of the boosted components of $k^\mu$, which we denote by $k^\mu_\Lambda =\Lambda^\mu{}_\nu k^\nu$. Applying the boost in Eq.~\eqref{eq_Lorentz}, we find
\begin{align}
    \omega_\Lambda &= \omega - k\dfrac{Q_A}{4P}\Omega\,,\nonumber \\
    k_\Lambda &= k - \omega \dfrac{Q_A}{4P}\Omega\,,
    \label{eq:boosted}
\end{align}
where we have denoted the Lorentz-transformed quantities with the label $\Lambda$. Substituting the previous relations into Eq.~\eqref{eq:ac_unboosted}, we arrive at
\begin{align}
    \omega^{\pm;\Lambda}_{\rm ac.} =& \left(\pm\dfrac{1}{\sqrt{3}}+\dfrac{Q_A \Omega}{6P}\right)\left( k_\Lambda + \omega_{\textrm{ac.}}\dfrac{Q_A \Omega}{4P}\right) \nonumber\\
    &-  \dfrac{Q_A \Omega}{4P} k_\Lambda \nonumber\\
    =& \pm\dfrac{1}{\sqrt{3}} k_\Lambda + O(\Omega^2)\,,
\end{align}
where we have kept only the linear terms in $\Omega$. These modes match our result in Eq. \eqref{eq:acoustic_vpm}.

\subsection{High temperature Axial Vortical Wave}\label{app:AVW}

In Sec. \ref{sec:largeT:AVW} we uncovered the Axial Vortical Wave. We showed that, in the absence of background vector and helical potentials ($\mu_V=\mu_H=0$), the AVW propagates as an oscillation in the axial chemical potential only. Consequently, such a wave should also be present in the absence of helicity, as it has been considered in previous works.
For instance, the chiral vortical wave angular frequencies introduced in Eq.~(59) of Ref.~\cite{Gorbar:2017toh} read as
\begin{equation}
 \omega_{1,2} = v_{1,2} k,
 \label{eq:gorbartoh}
\end{equation}
where we neglected dissipative effects and absorbed the angular velocity $\Omega$ in the linear velocities $v_{1,2}$. The velocities $v_{1,2}$ are the roots of the quadratic equation $a v^2 + b \Omega v + \Omega^2 c = 0$, where the $\Omega$-independent thermodynamic coefficients $a$, $b$ and $c$ are listed in Eqs.~(60)--(62) of Ref.~\cite{Gorbar:2017toh}. 
In the high-temperature limit,  $v_1$ and $v_2$ reduce to
\begin{subequations}
\begin{align}
    v_1 & = -\dfrac{9}{7\pi^2}\frac{\mu_A \Omega}{T^2} + O(T^{-3})\,,\\ 
    v_2 & = \dfrac{3}{\pi^2 }\frac{\mu_A \Omega}{T^2}+ O(T^{-3}).
\end{align}
\label{eq_v21_modes}
\end{subequations}
In order to compare the above modes with our results, we perform the Lorentz boost as shown in Eq.~\eqref{eq:boosted} and obtain, in the large-$T$ limit when $Q_A / 4P \simeq 15 \mu_A / 7\pi^2 T^2$, the following results:
\begin{subequations}
    \begin{align}
    v_1^\Lambda & = -\dfrac{24}{7\pi^2}\frac{\mu_A \Omega}{T^2} + O(T^{-3})\,, \\ v_2^\Lambda & = \dfrac{6}{7\pi^2}\frac{\mu_A \Omega}{T^2} + O(T^{-3})\,.
    \end{align}
\end{subequations}
Thus, $v_1^\Lambda$ coincides with the axial mode given in Eq.~\eqref{eq_AVW_v}. The second velocity, $v_2^\Lambda$, has no analogue with our results. However, one may find a resemblance between $v_2^\Lambda$ above and a part of the non-reciprocal correction to the helical vortical wave, shown in Eq.~\eqref{eq_HVW_speed}. It is worth pointing out that, as opposed to the HVW mode in Eq.~\eqref{eq_HVW_speed}, none of the $v_i$ modes in Eq.~\eqref{eq_v21_modes} propagate in a neutral plasma (where $\mu_V = \mu_A = 0$).

\subsection{Degenerate limit}
\label{app:deg}

In Sec. \ref{sec:largemu}, we observed that the helical and axial degrees of freedom are indistinguishable from each other, up to exponentially small corrections. The coincidence of axial and helical degrees of freedom in this limit suggests that the same type of excitations should have appeared in the previous studies where helicity was not taken into account, e.g. in Ref.~\cite{Gorbar:2017toh}. We now seek to express $v_{1,2}$ from Eq.~\eqref{eq:gorbartoh} by taking the high-density limit. The pair of the modes $v_1$ and $v_2$ attain the following form
\begin{equation}
    v_{1,2} = -\frac{2 \Omega \mu_A}{\mu_V^2} \pm \frac{\Omega}{\mu_V^2} \sqrt{\frac{4\pi^2 T^2}{3} + 9 \mu_A^2}\,.
\end{equation}
As before, we express $v_i$ in the boosted frame using Eq.~\eqref{eq:boosted}:
\begin{equation}
     v_i^{\Lambda} = v_i - \dfrac{Q_A}{4 P}\Omega \simeq v_i -\dfrac{3}{\mu_V^2}\frac{\mu_A \Omega}{T^2},
\end{equation}
where in the last step, we have made use of the high-density expansion of $P$ and $Q_\chi$ given in Eq.~\eqref{eq:deg_expansion}. Note that the connection with Ref.~\cite{Gorbar:2017toh} is made at $\mu_H=0$, and so $Q_A=Q_\chi$ and $\mu_A = \mu_\chi$. All in all, we obtain
\begin{equation}
    v_{1,2}^\Lambda = -\frac{5 \Omega \mu_A}{\mu_V^2} \pm \frac{\Omega}{\mu_V^2} \sqrt{\frac{4\pi^2 T^2}{3} + 9 \mu_A^2}\,,
\end{equation}
in agreement with our result in Eq.~\eqref{eq_v_deg}.

\bibliography{hvw}

\end{document}